# The *DL-Lite* Family and Relations

**Alessandro Artale**                                    ARTALE@INF.UNIBZ.IT
**Diego Calvanese**                                  CALVANESE@INF.UNIBZ.IT
*KRDB Research Centre*
*Free University of Bozen-Bolzano*
*Piazza Domenicani, 3 I-39100 Bolzano, Italy*

**Roman Kontchakov**                              ROMAN@DCS.BBK.AC.UK
**Michael Zakharyaschev**                       MICHAEL@DCS.BBK.AC.UK
*Department of Computer Science and Information Systems*
*Birkbeck College*
*Malet Street, London WC1E 7HX, U.K.*

## Abstract

The recently introduced series of description logics under the common moniker '*DL-Lite*' has attracted attention of the description logic and semantic web communities due to the low computational complexity of inference, on the one hand, and the ability to represent conceptual modeling formalisms, on the other. The main aim of this article is to carry out a thorough and systematic investigation of inference in extensions of the original *DL-Lite* logics along five axes: by (i) adding the Boolean connectives and (ii) number restrictions to concept constructs, (iii) allowing role hierarchies, (iv) allowing role disjointness, symmetry, asymmetry, reflexivity, irreflexivity and transitivity constraints, and (v) adopting or dropping the unique name assumption. We analyze the combined complexity of satisfiability for the resulting logics, as well as the data complexity of instance checking and answering positive existential queries. Our approach is based on embedding *DL-Lite* logics in suitable fragments of the one-variable first-order logic, which provides useful insights into their properties and, in particular, computational behavior.

## 1. Introduction

Description Logic (cf. Baader, Calvanese, McGuinness, Nardi, & Patel-Schneider, 2003 and references therein) is a family of knowledge representation formalisms developed over the past three decades and, in recent years, widely used in various application areas such as:

- *conceptual modeling* (Bergamaschi & Sartori, 1992; Calvanese et al., 1998b, 1999; McGuinness & Wright, 1998; Franconi & Ng, 2000; Borgida & Brachman, 2003; Berardi, Calvanese, & De Giacomo, 2005; Artale et al., 1996, 2007, 2007b),

- *information and data integration* (Beeri, Levy, & Rousset, 1997; Levy & Rousset, 1998; Goasdoue, Lattes, & Rousset, 2000; Calvanese et al., 1998a, 2002a, 2002b, 2008; Noy, 2004; Meyer, Lee, & Booth, 2005),

- *ontology-based data access* (Dolby et al., 2008; Poggi et al., 2008a; Heymans et al., 2008),

- *the Semantic Web* (Heflin & Hendler, 2001; Horrocks, Patel-Schneider, & van Harmelen, 2003).

Description logics (DLs, for short) underlie the standard Web Ontology Language OWL,[1] which is now in the process of being standardized by the W3C in its second edition, OWL 2.

The widespread use of DLs as flexible modeling languages stems from the fact that, similarly to more traditional modeling formalisms, they structure the domain of interest into classes (or concepts, in the DL parlance) of objects with common properties. Properties are associated with objects by means of binary relationships (or roles) to other objects. Constraints available in standard DLs also resemble those used in conceptual modeling formalisms for structuring information: is-a hierarchies (i.e., inclusions) and disjointness for concepts and roles, domain and range constraints for roles, mandatory participation in roles, functionality and more general numeric restrictions for roles, covering within concept hierarchies, etc. In a DL knowledge base (KB), these constraints are combined to form a TBox asserting intensional knowledge, while an ABox collects extensional knowledge about individual objects, such as whether an object is an instance of a concept, or two objects are connected by a role. The standard reasoning services over a DL KB include checking its consistency (or satisfiability), instance checking (whether a certain individual is an instance of a concept), and logic entailment (whether a certain constraint is logically implied by the KB). More sophisticated services are emerging that can support modular development of ontologies by checking, for example, whether one ontology is a conservative extension of another with respect to a certain vocabulary (see, e.g., Ghilardi, Lutz, & Wolter, 2006; Cuenca Grau, Horrocks, Kazakov, & Sattler, 2008; Kontchakov, Wolter, & Zakharyaschev, 2008; Kontchakov, Pulina, Sattler, Schneider, Selmer, Wolter, & Zakharyaschev, 2009).

Description logics have recently been used to provide access to large amounts of data through a high-level conceptual interface, which is of relevance to both data integration and ontology-based data access. In this setting, the TBox constitutes the conceptual, high-level view of the information managed by the system, and the ABox is physically stored in a relational database and accessed using the standard relational database technology (Poggi et al., 2008a; Calvanese et al., 2008). The fundamental inference service in this case is answering queries to the ABox with the constraints in the TBox taken into account. The kind of queries that have most often been considered are first-order *conjunctive queries*, which correspond to the commonly used Select-Project-Join SQL queries. The key properties for such an approach to be viable in practice are (i) efficiency of query evaluation, with the ideal target being traditional database query processing, and (ii) that query evaluation can be done by leveraging the relational technology already used for storing the data.

With these objectives in mind, a series of description logics—*the DL-Lite family*—has recently been proposed and investigated by Calvanese, De Giacomo, Lembo, Lenzerini, and Rosati (2005, 2006, 2008a), and later extended by Artale, Calvanese, Kontchakov, and Zakharyaschev (2007a), Poggi, Lembo, Calvanese, De Giacomo, Lenzerini, and Rosati (2008a). Most logics of the family meet the requirements above and, at the same time, are capable of representing many important types of constraints used in conceptual modeling. In particular, inference in various *DL-Lite* logics can be done efficiently both in the size of the data (data complexity) and in the overall size of the KB (combined complexity): it was shown that KB satisfiability in these logics is polynomial for combined complexity, while answering queries is in $AC^0$ for data complexity—which, roughly, means that, given a conjunctive query over a KB, the query and the TBox can be rewritten (independently of the ABox) into a union of conjunctive queries over the ABox alone. (It is to be emphasized that the data complexity measure is very important in the application context of the *DL-Lite* logics, since one can reasonably assume that the size of the data largely dominates the size of the TBox.) Query rewriting techniques have been implemented in various systems such as QuOnto[2] (Acciarri, Calvanese, De Giacomo, Lembo, Lenzerini, Palmieri, & Rosati, 2005; Poggi, Rodriguez, & Ruzzi, 2008b), ROWLKit (Corona, Ruzzi, & Savo, 2009), Owlgres (Stocker & Smith, 2008) and REQUIEM (Pérez-Urbina, Motik, & Horrocks, 2009). It has also been demonstrated (Kontchakov et al., 2008) that developing, analyzing and re-using *DL-Lite* ontologies (TBoxes) can be supported by efficient tools capable of checking various types of entailment between such ontologies with respect to given vocabularies, in particular, by minimal module extraction tools (Kontchakov et al., 2009)—which do not yet exist for richer languages.

---

[1]. http://www.w3.org/2007/OWL/

The significance of the *DL-Lite* family is testified by the fact that it forms the basis of OWL 2 QL, one of the three profiles of OWL 2.[3] The OWL 2 profiles are fragments of the full OWL 2 language that have been designed and standardized for specific application requirements. According to (the current version of) the official W3C profiles document, the purpose of OWL 2 QL is to be the language of choice for applications that use very large amounts of data and where query answering is the most important reasoning task.

The common denominator of the *DL-Lite* logics constructed so far is as follows: (i) quantification over roles and their inverses is not qualified (in other words, in concepts of the form $\exists R.C$ we must have $C = \top$) and (ii) the TBox axioms are concept inclusions that cannot represent any kind of disjunctive information (say, that two concepts cover the whole domain). The other *DL-Lite*-related dialects were designed—with the aim of capturing more conceptual modeling constraints, but in a somewhat *ad hoc* manner—by extending this 'core' language with a number of constructs such as global functionality constraints, role inclusions and restricted Boolean operators on concepts (see Section 4 for details). Although some attempts have been made (Calvanese et al., 2006; Artale et al., 2007a; Kontchakov & Zakharyaschev, 2008) to put the original *DL-Lite* logics into a more general perspective and investigate their extensions with a variety of DL constructs required for conceptual modeling, the resulting picture still remains rather fragmentary and far from comprehensive. A systematic investigation of the *DL-Lite* family and relatives has become even more urgent and challenging in view of the choice of the constructs to be included in the specification of the OWL 2 QL profile[4] (in particular, because OWL does not make the unique name assumption, UNA, which was usually adopted in *DL-Lite*, and uses equalities and inequalities between object names instead).

The main aim of this article is to fill in this gap and provide a thorough and comprehensive understanding of the interaction between various *DL-Lite* constructs and their impact on the computational complexity of reasoning. To achieve this goal, we consider a spectrum of logics, classified according to five mutually orthogonal features:

(1) the presence or absence of role inclusions;

---

(2) the form of the allowed concept inclusions, where we consider four classes, called *core*, *Krom*, *Horn*, and *Bool*, that exhibit different computational properties;

(3) the form of the allowed numeric constraints, ranging from none, to global functionality constraints only, and to arbitrary number restrictions;

(4) the presence or absence of the unique name assumption (and the equalities and inequalities between object names, if this assumption is dropped); and

(5) the presence or absence of standard role constraints such as disjointness, symmetry, asymmetry, reflexivity, irreflexivity, and transitivity.

For all the resulting cases, we investigate the combined and data complexity of KB satisfiability and instance checking, as well as the data complexity of query answering. The obtained *tight* complexity results are summarized in Section 3.4 (Table 2 and Remark 3.1).

As already mentioned, the original motivation and distinguishing feature for the logics in the *DL-Lite* family was their 'lite'-ness in the sense of low computational complexity of the reasoning tasks (query answering in $AC^0$ for data complexity and tractable KB satisfiability for combined complexity). In the broader perspective we take here, not all of our logics meet this requirement, in particular, those with Krom or Bool concept inclusions[5] However, we identify another distinguishing feature that can be regarded as the natural logic-based characterization of the *DL-Lite* family: embeddability into the one-variable fragment of first-order logic without equality and function symbols. This allows us to relate the complexity of *DL-Lite* logics to the complexity of the corresponding fragments of first-order logic, and thus to obtain a deep insight into the underlying logical properties of each *DL-Lite* variant. For example, most upper complexity bounds established below follow from this embedding and well-known results on the classical decision problem (see, e.g., Börger, Grädel, & Gurevich, 1997) and descriptive complexity (see, e.g., Immerman, 1999).

One of the most interesting findings in this article is that number restrictions, even expressed locally, instead of global role functionality, can be added to the original *DL-Lite* logics (under the UNA and without role inclusions) 'for free,' that is, without changing their computational complexity. The first-order approach shows that in most cases we can also extend the *DL-Lite* logics with the role constraints mentioned above, again keeping the same complexity. It also gives a framework to analyze the effect of adopting or dropping the UNA and using (in)equalities between object names. For example, we observe that if equality is allowed in the language of *DL-Lite* (which only makes sense without the UNA) then query answering becomes LOGSPACE-complete for data complexity, and therefore not first-order rewritable. It also turns out that dropping the UNA results in P-hardness of reasoning (for both combined and data complexity) in the presence of functionality constraints (NLOGSPACE-hardness was shown by Calvanese et al., 2008), and in NP-hardness if arbitrary number restrictions are allowed.

Another interesting finding is the dramatic impact of role inclusions, when combined with number restrictions (or even functionality constraints), on the computational complexity of reasoning. As was already observed by Calvanese et al. (2006), such a combination increases data complexity of instance checking from membership in LOGSPACE to NLogSpace-hardness. We show here that the situation is actually even worse: for data complexity, instance checking turns out to be P-complete in the case of core and Horn logics and coNP-complete in the case of Krom and Bool logics; moreover, KB satisfiability, which is NLogSpace-complete for combined complexity in the simplest 'core' case—i.e., efficiently tractable, when role inclusions or number restrictions are used separately—becomes ExpTime-complete—i.e., provably intractable, when they are used together.

---

5. Note, by the way, that logics with Bool concept inclusions turn out to be quite useful in conceptual modeling and reasonably manageable computationally (Kontchakov et al., 2008).

To retain both role inclusions and functionality constraints in the language and keep complexity within the required limits, Poggi et al. (2008a) introduced another *DL-Lite* dialect, called *DL-Lite$_\mathcal{A}$*, which restricts the interaction between role inclusions and functionality constraints. Here we extend this result by showing that the *DL-Lite* logics with such a limited interaction between role inclusions and number restrictions can still be embedded into the one-variable fragment of first-order logic, and so exhibit the same behavior as their fragments with only role inclusions or only number restrictions.

The article is structured in the following way. In Section 2, we introduce the logics of the extended *DL-Lite* family and illustrate their features as conceptual modeling formalisms. In Section 3, we discuss the reasoning services and the complexity measures to be analyzed in what follows, and give an overview of the obtained complexity results. In Section 4, we place the introduced *DL-Lite* logics in the context of the original *DL-Lite* family, and discuss its relationship with OWL 2. In Section 5, we study the combined complexity of KB satisfiability and instance checking, while in Section 6, we consider the data complexity of these problems. In Section 7, we study the data complexity of query answering. In Section 8, we analyze the impact of dropping the UNA and adding (in)equalities between object names on the complexity of reasoning. Section 9 concludes the article.

## 2. The Extended *DL-Lite* Family of Description Logics

*Description Logic* (Baader et al., 2003) is a family of logics that have been studied and used in knowledge representation and reasoning since the 1980s. In DLs, the elements of the domain of interest are structured into *concepts* (unary predicates), and their properties are specified by means of *roles* (binary predicates). Complex concept and role expressions (or simply concepts and roles) are constructed, starting from a set of concept and role names, by applying suitable constructs, where the set of available constructs depends on the specific description logic. Concepts and roles can then be used in a *knowledge base* to assert knowledge, both at the intensional level, in a so-called *TBox* ('T' for terminological), and at the extensional level, in a so-called *ABox* ('A' for assertional). A TBox typically consists of a set of axioms stating the inclusion between concepts and roles. In an ABox, one can assert membership of objects (i.e., constants) in concepts, or that a pair of objects is connected by a role. DLs are supported by reasoning services, such as satisfiability checking and query answering, that rely on their logic-based semantics.

### 2.1 Syntax and Semantics of the Logics in the *DL-Lite* Family

We introduce now the (extended) *DL-Lite* family of description logics, which was initially proposed with the aim of capturing typical conceptual modeling formalisms, such as UML class diagrams and ER models (see Section 2.2 for details), while maintaining good computational properties of standard DL reasoning tasks (Calvanese et al., 2005). We begin by defining the logic $DL\text{-}Lite_{bool}^{\mathcal{HN}}$, which can be regarded as the *supremum* of the original *DL-Lite* family (Calvanese et al., 2005, 2006, 2007b) in the lattice of description logics.

**$DL\text{-}Lite_{bool}^{\mathcal{HN}}$.** The language of $DL\text{-}Lite_{bool}^{\mathcal{HN}}$ contains *object names* $a_0, a_1, \ldots$, *concept names* $A_0, A_1, \ldots$, and *role names* $P_0, P_1, \ldots$. Complex *roles* $R$ and *concepts* $C$ of this language are defined as follows:

$$
\begin{aligned}
R &\quad ::= \quad P_k \quad | \quad P_k^-, \\
B &\quad ::= \quad \bot \quad | \quad A_k \quad | \quad \geq q\, R, \\
C &\quad ::= \quad B \quad | \quad \neg C \quad | \quad C_1 \sqcap C_2,
\end{aligned}
$$

where $q$ is a positive integer. The concepts of the form $B$ will be called *basic*.

A $DL\text{-}Lite_{bool}^{\mathcal{HN}}$ *TBox*, $\mathcal{T}$, is a finite set of *concept* and *role inclusion axioms* (or simply *concept* and *role inclusions*) of the form:

$$
C_1 \sqsubseteq C_2 \qquad \text{and} \qquad R_1 \sqsubseteq R_2,
$$

and an *ABox*, $\mathcal{A}$, is a finite set of assertions of the form:

$$
A_k(a_i), \qquad \neg A_k(a_i), \qquad P_k(a_i, a_j) \qquad \text{and} \qquad \neg P_k(a_i, a_j).
$$

Taken together, $\mathcal{T}$ and $\mathcal{A}$ constitute the $DL\text{-}Lite_{bool}^{\mathcal{HN}}$ *knowledge base* $\mathcal{K} = (\mathcal{T}, \mathcal{A})$. In the following, we denote by $role(\mathcal{K})$ the set of role names occurring in $\mathcal{T}$ and $\mathcal{A}$, by $role^{\pm}(\mathcal{K})$ the set $\{P_k, P_k^- \mid P_k \in role(\mathcal{K})\}$, and by $ob(\mathcal{A})$ the set of object names in $\mathcal{A}$. For a role $R$, we set:

$$
inv(R) = \begin{cases} P_k^-, & \text{if } R = P_k, \\ P_k, & \text{if } R = P_k^-. \end{cases}
$$

As usual in description logic, an *interpretation*, $\mathcal{I} = (\Delta^{\mathcal{I}}, \cdot^{\mathcal{I}})$, consists of a nonempty *domain* $\Delta^{\mathcal{I}}$ and an interpretation function $\cdot^{\mathcal{I}}$ that assigns to each object name $a_i$ an element $a_i^{\mathcal{I}} \in \Delta^{\mathcal{I}}$, to each concept name $A_k$ a subset $A_k^{\mathcal{I}} \subseteq \Delta^{\mathcal{I}}$ of the domain, and to each role name $P_k$ a binary relation $P_k^{\mathcal{I}} \subseteq \Delta^{\mathcal{I}} \times \Delta^{\mathcal{I}}$ over the domain. Unless otherwise stated, we adopt here the *unique name assumption* (UNA):

$$
a_i^{\mathcal{I}} \neq a_j^{\mathcal{I}} \quad \text{for all} \quad i \neq j. \tag{UNA}
$$

However, we shall always indicate which of our results depend on the UNA and which do not, and when they do depend on this assumption, we discuss also the consequences of dropping it (see also Sections 4 and 8).

The role and concept constructs are interpreted in $\mathcal{I}$ in the standard way:

$$
\begin{aligned}
(P_k^-)^{\mathcal{I}} &= \{(y, x) \in \Delta^{\mathcal{I}} \times \Delta^{\mathcal{I}} \mid (x, y) \in P_k^{\mathcal{I}}\}, &&\text{(inverse role)} \\
\bot^{\mathcal{I}} &= \emptyset, &&\text{(the empty set)} \\
(\geq q\, R)^{\mathcal{I}} &= \{x \in \Delta^{\mathcal{I}} \mid \sharp\{y \in \Delta^{\mathcal{I}} \mid (x, y) \in R^{\mathcal{I}}\} \geq q\}, &&\text{(at least $q$ $R$-successors)} \\
(\neg C)^{\mathcal{I}} &= \Delta^{\mathcal{I}} \setminus C^{\mathcal{I}}, &&\text{(not in $C$)} \\
(C_1 \sqcap C_2)^{\mathcal{I}} &= C_1^{\mathcal{I}} \cap C_2^{\mathcal{I}}, &&\text{(both in $C_1$ and in $C_2$)}
\end{aligned}
$$

where $\sharp X$ denotes the cardinality of $X$. We will use standard abbreviations such as

$$C_1 \sqcup C_2 = \neg(\neg C_1 \sqcap \neg C_2), \qquad \top = \neg\bot, \qquad \exists R = (\geq 1\, R), \qquad \leq q\, R = \neg(\geq q+1\, R).$$

Concepts of the form $\leq q\, R$ and $\geq q\, R$ are called *number restrictions*, and those of the form $\exists R$ are called *existential concepts*.

The *satisfaction relation* $\models$ is also standard:

$$\mathcal{I} \models C_1 \sqsubseteq C_2 \quad \text{iff} \quad C_1^{\mathcal{I}} \subseteq C_2^{\mathcal{I}}, \qquad \mathcal{I} \models R_1 \sqsubseteq R_2 \quad \text{iff} \quad R_1^{\mathcal{I}} \subseteq R_2^{\mathcal{I}},$$
$$\mathcal{I} \models A_k(a_i) \quad \text{iff} \quad a_i^{\mathcal{I}} \in A_k^{\mathcal{I}}, \qquad \mathcal{I} \models P_k(a_i, a_j) \quad \text{iff} \quad (a_i^{\mathcal{I}}, a_j^{\mathcal{I}}) \in P_k^{\mathcal{I}},$$
$$\mathcal{I} \models \neg A_k(a_i) \quad \text{iff} \quad a_i^{\mathcal{I}} \notin A_k^{\mathcal{I}}, \qquad \mathcal{I} \models \neg P_k(a_i, a_j) \quad \text{iff} \quad (a_i^{\mathcal{I}}, a_j^{\mathcal{I}}) \notin P_k^{\mathcal{I}}.$$

A knowledge base $\mathcal{K} = (\mathcal{T}, \mathcal{A})$ is said to be *satisfiable* (or *consistent*) if there is an interpretation, $\mathcal{I}$, satisfying all the members of $\mathcal{T}$ and $\mathcal{A}$. In this case we write $\mathcal{I} \models \mathcal{K}$ (as well as $\mathcal{I} \models \mathcal{T}$ and $\mathcal{I} \models \mathcal{A}$) and say that $\mathcal{I}$ is a *model* of $\mathcal{K}$ (and of $\mathcal{T}$ and $\mathcal{A}$).

The languages of the *DL-Lite* family we investigate in this article are obtained by restricting the language of $DL\text{-}Lite_{bool}^{\mathcal{HN}}$ along three axes: (i) the Boolean operators ($_{bool}$) on concepts, (ii) the number restrictions ($\mathcal{N}$) and (iii) the role inclusions, or hierarchies ($\mathcal{H}$).

Similarly to classical logic, we adopt the following definitions. A $DL\text{-}Lite_{bool}^{\mathcal{HN}}$ TBox $\mathcal{T}$ will be called a *Krom TBox*[6] if its concept inclusions are restricted to:

$$B_1 \sqsubseteq B_2, \qquad B_1 \sqsubseteq \neg B_2 \qquad \text{or} \qquad \neg B_1 \sqsubseteq B_2 \tag{Krom}$$

(here and below all the $B_i$ and $B$ are basic concepts). $\mathcal{T}$ will be called a *Horn TBox* if its concept inclusions are restricted to:

$$\bigsqcap_k B_k \sqsubseteq B \tag{Horn}$$

(by definition, the empty conjunction is $\top$). Finally, we will call $\mathcal{T}$ a *core TBox* if its concept inclusions are restricted to:

$$B_1 \sqsubseteq B_2 \qquad \text{or} \qquad B_1 \sqsubseteq \neg B_2. \tag{core}$$

As $B_1 \sqsubseteq \neg B_2$ is equivalent to $B_1 \sqcap B_2 \sqsubseteq \bot$, core TBoxes can be regarded as sitting in the intersection of Krom and Horn TBoxes.

**Remark 2.1** We will sometimes use conjunctions on the right-hand side of concept inclusions in these restricted languages: $C \sqsubseteq \bigsqcap_k B_k$. Clearly, this 'syntactic sugar' does not add any extra expressive power.

**$DL\text{-}Lite_{krom}^{\mathcal{HN}}$, $DL\text{-}Lite_{horn}^{\mathcal{HN}}$ and $DL\text{-}Lite_{core}^{\mathcal{HN}}$.** The fragments of $DL\text{-}Lite_{bool}^{\mathcal{HN}}$ with Krom, Horn, and core TBoxes will be denoted by $DL\text{-}Lite_{krom}^{\mathcal{HN}}$, $DL\text{-}Lite_{horn}^{\mathcal{HN}}$ and $DL\text{-}Lite_{core}^{\mathcal{HN}}$, respectively. Other fragments are obtained by limiting the use of number restrictions and role inclusions.

---

6. The Krom fragment of first-order logic consists of all formulas in prenex normal form whose quantifier-free part is a conjunction of binary clauses.

**DL-Lite$_\alpha^{\mathcal{H}}$.** The fragment of *DL-Lite$_\alpha^{\mathcal{HN}}$*, $\alpha \in \{core, krom, horn, bool\}$, without number restrictions $\geq q\,R$, for $q \geq 2$, (but with role inclusions) will be denoted by *DL-Lite$_\alpha^{\mathcal{H}}$*. Note that, in *DL-Lite$_\alpha^{\mathcal{H}}$*, we can still use existential concepts $\exists R$ (that is, $\geq 1\,R$).

**DL-Lite$_\alpha^{\mathcal{HF}}$.** Denote by *DL-Lite$_\alpha^{\mathcal{HF}}$* the fragment of *DL-Lite$_\alpha^{\mathcal{HN}}$* in which of all number restrictions $\geq q\,R$, we have existential concepts (with $q = 1$) and only those with $q = 2$ that occur in concept inclusions of the form $\geq 2\,R \sqsubseteq \bot$. Such an inclusion is called a *global functionality constraint* because it states that role $R$ is *functional* (more precisely, if $\mathcal{I} \models (\geq 2\,R \sqsubseteq \bot)$ and both $(x, y) \in R^{\mathcal{I}}$ and $(x, z) \in R^{\mathcal{I}}$, then $y = z$).

**DL-Lite$_\alpha^{\mathcal{N}}$, DL-Lite$_\alpha^{\mathcal{F}}$ and DL-Lite$_\alpha$.** If role inclusions are excluded from the language, then for each $\alpha \in \{core, krom, horn, bool\}$ we obtain three fragments: *DL-Lite$_\alpha^{\mathcal{N}}$* (with arbitrary number restrictions), *DL-Lite$_\alpha^{\mathcal{F}}$* (with functionality constraints and existential concepts $\exists R$), and *DL-Lite$_\alpha$* (without number restrictions different from $\exists R$).

As we shall see later on in this article, the logics of the form *DL-Lite$_\alpha^{\mathcal{HF}}$* and *DL-Lite$_\alpha^{\mathcal{HN}}$*, even for $\alpha = core$, turn out to be computationally rather costly because of the interaction between role inclusions and functionality constraints (or, more generally, number restrictions). On the other hand, for the purpose of conceptual modeling one may need both of these constructs; cf. the example in Section 2.2. A compromise can be found by artificially limiting the interplay between role inclusions and number restrictions in a way similar to the logic *DL-Lite$_\mathcal{A}$* proposed by Poggi et al. (2008a).

For a TBox $\mathcal{T}$, let $\sqsubseteq_\mathcal{T}^*$ denote the reflexive and transitive closure of the relation

$$\{(R, R'), (inv(R), inv(R')) \mid R \sqsubseteq R' \in \mathcal{T}\}$$

and let $R \equiv_\mathcal{T}^* R'$ iff $R \sqsubseteq_\mathcal{T}^* R'$ and $R' \sqsubseteq_\mathcal{T}^* R$. Say that $R'$ is a *proper sub-role of* $R$ in $\mathcal{T}$ if $R' \sqsubseteq_\mathcal{T}^* R$ and $R' \not\equiv_\mathcal{T}^* R$.

**DL-Lite$_\alpha^{(\mathcal{HN})}$.** We now introduce the logics *DL-Lite$_\alpha^{(\mathcal{HN})}$*, $\alpha \in \{core, krom, horn, bool\}$, which, on the one hand, restrict the logics *DL-Lite$_\alpha^{\mathcal{HN}}$* by limiting the interaction between role inclusions and number restrictions in order to reduce complexity of reasoning, and, on the other hand, include additional constructs, such as limited qualified existential quantifiers, role disjointness, (a)symmetry and (ir)reflexivity constraints, which increase the expressive power of the logics but do not affect their computational properties.

*DL-Lite$_\alpha^{(\mathcal{HN})}$* TBoxes $\mathcal{T}$ must satisfy the conditions **(A$_1$)**–**(A$_3$)** below. (We remind the reader that an occurrence of a concept on the right-hand (left-hand) side of a concept inclusion is called *negative* if it is in the scope of an odd (even) number of negations $\neg$; otherwise the occurrence is called *positive*.)

**(A$_1$)** $\mathcal{T}$ may contain only *positive occurrences* of qualified number restrictions $\geq q\,R.C$, where $C$ is a conjunction of concepts allowed on the right-hand side of $\alpha$-concept inclusions;

**(A$_2$)** if $\geq q\,R.C$ occurs in $\mathcal{T}$, then $\mathcal{T}$ does not contain *negative occurrences* of number restrictions $\geq q'\,R$ or $\geq q'\,inv(R)$ with $q' \geq 2$;

**(A$_3$)** if $R$ has a proper sub-role in $\mathcal{T}$, then $\mathcal{T}$ does not contain *negative occurrences* of $\geq q\,R$ or $\geq q\,inv(R)$ with $q \geq 2$.

| role constraints | role inclusions | number restrictions | concept inclusions | | | |
|---|---|---|---|---|---|---|
| | | | core | Krom | Horn | Bool |
| no | no | $\exists R$ | $DL\text{-}Lite_{core}$ | $DL\text{-}Lite_{krom}$ | $DL\text{-}Lite_{horn}$ | $DL\text{-}Lite_{bool}$ |
| | | $\exists R/\text{funct.}$ | $DL\text{-}Lite_{core}^{\mathcal{F}}$ | $DL\text{-}Lite_{krom}^{\mathcal{F}}$ | $DL\text{-}Lite_{horn}^{\mathcal{F}}$ | $DL\text{-}Lite_{bool}^{\mathcal{F}}$ |
| | | $\geq q\,R$ | $DL\text{-}Lite_{core}^{\mathcal{N}}$ | $DL\text{-}Lite_{krom}^{\mathcal{N}}$ | $DL\text{-}Lite_{horn}^{\mathcal{N}}$ | $DL\text{-}Lite_{bool}^{\mathcal{N}}$ |
| no | yes | $\exists R$ | $DL\text{-}Lite_{core}^{\mathcal{H}}$ | $DL\text{-}Lite_{krom}^{\mathcal{H}}$ | $DL\text{-}Lite_{horn}^{\mathcal{H}}$ | $DL\text{-}Lite_{bool}^{\mathcal{H}}$ |
| | | $\exists R/\text{funct.}$ | $DL\text{-}Lite_{core}^{\mathcal{HF}}$ | $DL\text{-}Lite_{krom}^{\mathcal{HF}}$ | $DL\text{-}Lite_{horn}^{\mathcal{HF}}$ | $DL\text{-}Lite_{bool}^{\mathcal{HF}}$ |
| | | $\geq q\,R$ | $DL\text{-}Lite_{core}^{\mathcal{HN}}$ | $DL\text{-}Lite_{krom}^{\mathcal{HN}}$ | $DL\text{-}Lite_{horn}^{\mathcal{HN}}$ | $DL\text{-}Lite_{bool}^{\mathcal{HN}}$ |
| disj. (a)sym. (ir)ref. | yes | $\exists R.C/\text{funct.}^{\text{a)}}$ | $DL\text{-}Lite_{core}^{(\mathcal{HF})}$ | $DL\text{-}Lite_{krom}^{(\mathcal{HF})}$ | $DL\text{-}Lite_{horn}^{(\mathcal{HF})}$ | $DL\text{-}Lite_{bool}^{(\mathcal{HF})}$ |
| | | $\geq q\,R.C^{\text{a)}}$ | $DL\text{-}Lite_{core}^{(\mathcal{HN})}$ | $DL\text{-}Lite_{krom}^{(\mathcal{HN})}$ | $DL\text{-}Lite_{horn}^{(\mathcal{HN})}$ | $DL\text{-}Lite_{bool}^{(\mathcal{HN})}$ |
| disj. (a)sym. (ir)ref. tran. | yes | $\exists R.C/\text{funct.}^{\text{a)}}$ | $DL\text{-}Lite_{core}^{(\mathcal{HF})^+}$ | $DL\text{-}Lite_{krom}^{(\mathcal{HF})^+}$ | $DL\text{-}Lite_{horn}^{(\mathcal{HF})^+}$ | $DL\text{-}Lite_{bool}^{(\mathcal{HF})^+}$ |
| | | $\geq q\,R.C^{\text{a)}}$ | $DL\text{-}Lite_{core}^{(\mathcal{HN})^+}$ | $DL\text{-}Lite_{krom}^{(\mathcal{HN})^+}$ | $DL\text{-}Lite_{horn}^{(\mathcal{HN})^+}$ | $DL\text{-}Lite_{bool}^{(\mathcal{HN})^+}$ |

a) restricted by **(A1)**–**(A3)**.

Table 1: The extended *DL-Lite* family.

(It follows that no $DL\text{-}Lite_{\alpha}^{(\mathcal{HN})}$ TBox can contain both, say, a functionality constraint $\geq 2\,R \sqsubseteq \bot$ and an occurrence of $\geq q\,R.C$, for any $q \geq 1$.)

Additionally, $DL\text{-}Lite_{\alpha}^{(\mathcal{HN})}$ TBoxes can contain *role constraints* (or *axioms*) of the form:

$$\mathsf{Dis}(R_1, R_2), \qquad \mathsf{Asym}(P_k), \qquad \mathsf{Sym}(P_k), \qquad \mathsf{Irr}(P_k), \qquad \text{and} \qquad \mathsf{Ref}(P_k).$$

The meaning of these new constructs is defined as usual: for an interpretation $\mathcal{I} = (\Delta^{\mathcal{I}}, \cdot^{\mathcal{I}})$,

- $(\geq q\,R.C)^{\mathcal{I}} = \big\{x \in \Delta^{\mathcal{I}} \mid \sharp\{y \in C^{\mathcal{I}} \mid (x,y) \in R^{\mathcal{I}}\} \geq q\big\}$;

- $\mathcal{I} \models \mathsf{Dis}(R_1, R_2)$  iff  $R_1^{\mathcal{I}} \cap R_2^{\mathcal{I}} = \emptyset$  (roles $R_1$ and $R_2$ are *disjoint*);

- $\mathcal{I} \models \mathsf{Asym}(P_k)$  iff  $P_k^{\mathcal{I}} \cap (P_k^-)^{\mathcal{I}} = \emptyset$  (role $P_k$ is *asymmetric*);

- $\mathcal{I} \models \mathsf{Sym}(P_k)$  iff  $P_k^{\mathcal{I}} = (P_k^-)^{\mathcal{I}}$  ($P_k$ is *symmetric*);

- $\mathcal{I} \models \mathsf{Irr}(P_k)$  iff  $(x,x) \notin P_k^{\mathcal{I}}$ for all $x \in \Delta^{\mathcal{I}}$  ($P_k$ is *irreflexive*);

- $\mathcal{I} \models \mathsf{Ref}(P_k)$  iff  $(x,x) \in P_k^{\mathcal{I}}$ for all $x \in \Delta^{\mathcal{I}}$  ($P_k$ is *reflexive*).

It is to be emphasized that these extra constructs are often used in conceptual modeling and their introduction in $DL\text{-}Lite_{\alpha}^{(\mathcal{HN})}$ is motivated by the OWL 2 QL proposal. (Note that $DL\text{-}Lite_{\alpha}^{(\mathcal{HN})}$ contains both $DL\text{-}Lite_{\alpha}^{\mathcal{H}}$ and $DL\text{-}Lite_{\alpha}^{\mathcal{N}}$ as its proper fragments.)

**$DL\text{-}Lite_{\alpha}^{(\mathcal{HN})^+}$.** For $\alpha \in \{bool, horn, krom, core\}$, denote by $DL\text{-}Lite_{\alpha}^{(\mathcal{HN})^+}$ the extension of $DL\text{-}Lite_{\alpha}^{(\mathcal{HN})}$ with *role transitivity constraints* of the form $\mathsf{Tra}(P_k)$, the meaning of which is as expected:

- $\mathcal{I} \models \mathsf{Tra}(P_k)$  iff  $(x,y) \in P_k^{\mathcal{I}}$ and $(y,z) \in P_k^{\mathcal{I}}$ imply $(x,z) \in P_k^{\mathcal{I}}$, for all $x,y,z \in \Delta^{\mathcal{I}}$  ($P_k$ is *transitive*).

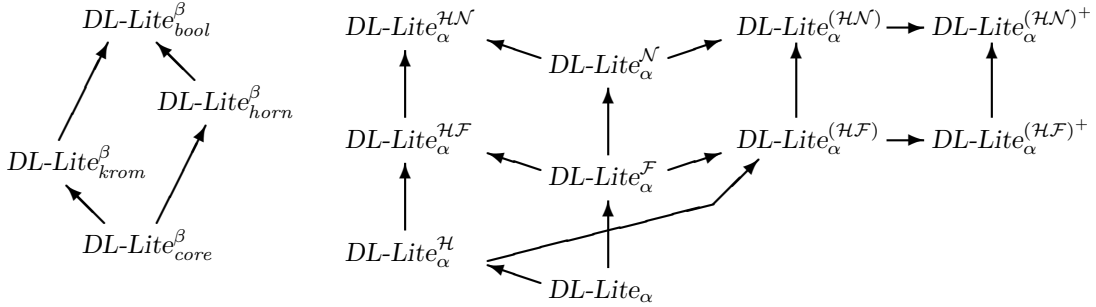

Figure 1: Language inclusions in the extended *DL-Lite* family.

We remind the reader of the standard restriction limiting the use of transitive roles in DLs (see, e.g., Horrocks, Sattler, & Tobies, 2000):

- only *simple* roles $R$ are allowed in concepts of the form $\geq q\,R$, for $q \geq 2$,

where by a *simple role* in a given TBox $\mathcal{T}$ we understand a role without transitive sub-roles (including itself). In particular, if $\mathcal{T}$ contains $\mathsf{Tra}(P)$ then $P$ and $P^-$ are not simple, and so $\mathcal{T}$ cannot contain occurrences of concepts of the form $\geq q\,P$ and $\geq q\,P^-$, for $q \geq 2$.

**$DL\text{-}Lite_\alpha^{(\mathcal{HF})}$ and $DL\text{-}Lite_\alpha^{(\mathcal{HF})^+}$.** We also define languages $DL\text{-}Lite_\alpha^{(\mathcal{HF})}$ as sub-languages of $DL\text{-}Lite_\alpha^{(\mathcal{HN})}$, in which only number restrictions of the form $\exists R$, $\exists R.C$ and functionality constraints $\geq 2\,R \sqsubseteq \bot$ are allowed—provided, of course, that they satisfy $(\mathbf{A_1})$–$(\mathbf{A_3})$; in particular, $\exists R.C$ is not allowed if $R$ is functional. As before, $DL\text{-}Lite_\alpha^{(\mathcal{HF})^+}$ are the extensions of $DL\text{-}Lite_\alpha^{(\mathcal{HF})}$ with role transitivity constraints (satisfying the restriction above).

Thus, the extended *DL-Lite* family we consider in this article consists of 40 different logics collected in Table 1. The inclusions between these logics are shown in Figure 1. They are obtained by taking the product of the left- and right-hand parts of the picture, where the subscript $\alpha$ on the right-hand part ranges over $\{core, krom, horn, bool\}$, i.e., the subscripts on the left-hand part, and similarly, the superscript $\beta$ on the left-hand part ranges over $\{\_, \mathcal{F}, \mathcal{N}, \mathcal{H}, \mathcal{HF}, \mathcal{HN}, (\mathcal{HF}), (\mathcal{HN}), (\mathcal{HF})^+, (\mathcal{HN})^+\}$, i.e., the superscripts on the right-hand part.

The position of these logics relative to other *DL-Lite* logics known in the literature and the OWL 2 QL profile will be discussed in Section 4. And starting from Section 5, we begin a thorough investigation of the computational properties of the logics in the extended *DL-Lite* family, both with and without the UNA. But before that we illustrate the expressive power of the *DL-Lite* logics by a concrete example.

## 2.2 *DL-Lite* for Conceptual Modeling

A tight correspondence between conceptual modeling formalisms, such as the ER model and UML class diagrams, and various description logics has been pointed out in various papers (e.g., Calvanese et al., 1998b, 1999; Borgida & Brachman, 2003; Berardi et al., 2005). Here we give an example showing how *DL-Lite* logics can be used for conceptual modeling purposes; for more details see the work by Artale et al. (2007b).

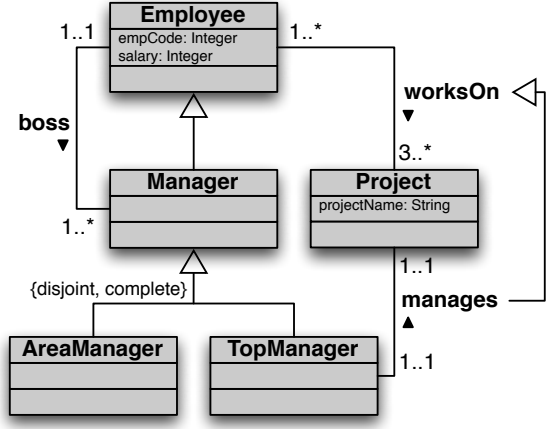

Figure 2: A UML class diagram.

Let us consider the UML class diagram depicted in Figure 2 and representing (a portion of) a company information system. According to the diagram, all managers are employees and are partitioned into area managers and top managers. This information can be represented by means of the following concept inclusions (where in brackets we specify the minimal *DL-Lite* language the inclusion belongs to):

$$Manager \sqsubseteq Employee \qquad (DL\text{-}Lite_{core})$$
$$AreaManager \sqsubseteq Manager \qquad (DL\text{-}Lite_{core})$$
$$TopManager \sqsubseteq Manager \qquad (DL\text{-}Lite_{core})$$
$$AreaManager \sqsubseteq \neg TopManager \qquad (DL\text{-}Lite_{core})$$
$$Manager \sqsubseteq AreaManager \sqcup TopManager \qquad (DL\text{-}Lite_{bool})$$

Each employee has two functional attributes, *empCode* and *salary*, with integer values. Unlike OWL, here we do not distinguish between abstract objects and data values. Hence we model a datatype, such as *Integer*, by means of a concept, and an attribute, such as employee's salary, by means of a role. Thus, *salary* can be represented as follows:

$$Employee \sqsubseteq \exists salary \qquad (DL\text{-}Lite_{core})$$
$$\exists salary^- \sqsubseteq Integer \qquad (DL\text{-}Lite_{core})$$
$$\geq 2\, salary \sqsubseteq \bot \qquad (DL\text{-}Lite_{core}^{\mathcal{F}})$$

The functional attribute *empCode* with values in *Integer* is represented in the same way. The binary relationship *worksOn* has *Employee* as its domain and *Project* as its range:

$$\exists worksOn \sqsubseteq Employee \qquad (DL\text{-}Lite_{core})$$
$$\exists worksOn^- \sqsubseteq Project \qquad (DL\text{-}Lite_{core})$$

The binary relationship *boss* with domain *Employee* and range *Manager* is treated analogously. Each employee works on a project and has exactly one boss, while a project must involve at least three employees:

$$Employee \ \sqsubseteq \ \exists worksOn \qquad (DL\text{-}Lite_{core})$$
$$Employee \ \sqsubseteq \ \exists boss \qquad (DL\text{-}Lite_{core})$$
$$\geq 2 \ boss \ \sqsubseteq \ \bot \qquad (DL\text{-}Lite_{core}^{\mathcal{F}})$$
$$Project \ \sqsubseteq \ \geq 3 \ worksOn^- \qquad (DL\text{-}Lite_{core}^{\mathcal{N}})$$

A top manager manages exactly one project and also works on that project, while a project is managed by exactly one top manager:

$$\exists manages \ \sqsubseteq \ TopManager \qquad (DL\text{-}Lite_{core})$$
$$\exists manages^- \ \sqsubseteq \ Project \qquad (DL\text{-}Lite_{core})$$
$$TopManager \ \sqsubseteq \ \exists manages \qquad (DL\text{-}Lite_{core})$$
$$Project \ \sqsubseteq \ \exists manages^- \qquad (DL\text{-}Lite_{core})$$
$$\geq 2 \ manages \ \sqsubseteq \ \bot \qquad (DL\text{-}Lite_{core}^{\mathcal{F}})$$
$$\geq 2 \ manages^- \ \sqsubseteq \ \bot \qquad (DL\text{-}Lite_{core}^{\mathcal{F}})$$
$$manages \ \sqsubseteq \ worksOn \qquad (DL\text{-}Lite_{core}^{\mathcal{H}})$$

All in all, the only languages in the extended $DL\text{-}Lite$ family capable of representing the UML class diagram in Figure 2 are $DL\text{-}Lite_{bool}^{\mathcal{HN}}$ and $DL\text{-}Lite_{bool}^{(\mathcal{HN})}$. Note, however, that except for the covering constraint, $Manager \sqsubseteq AreaManager \sqcup TopManager$, all other concept inclusions in the $DL\text{-}Lite$ translation of the UML class diagram belong to variants of the 'core' fragments $DL\text{-}Lite_{core}^{\mathcal{HN}}$ and $DL\text{-}Lite_{core}^{(\mathcal{HN})}$. It is not hard to imagine a situation where one needs Horn concept inclusions to represent integrity constraints over UML class diagrams, for example, to express (together with the above axioms) that 'no chief executive officer may work on five projects and be a manager of one of them:'

$$CEO \sqcap (\geq 5 \ worksOn) \sqcap \exists manages \ \sqsubseteq \ \bot \qquad (DL\text{-}Lite_{horn}^{\mathcal{N}})$$

In the context of UML class diagrams, the Krom fragment $DL\text{-}Lite_{krom}$ (with its variants) seems to be useless: it extends $DL\text{-}Lite_{core}$ with concept inclusions of the form $\neg B_1 \sqsubseteq B_2$ or, equivalently, $\top \sqsubseteq B_1 \sqcup B_2$, which are rarely used in conceptual modeling. Indeed, this would correspond to partitioning the whole domain of interest in just two parts, while more general and useful covering constraints of the form $B \sqsubseteq B_1 \sqcup \cdots \sqcup B_k$ require the full Bool language. On the other hand, the Krom fragments are important for pinpointing the borderlines of various complexity classes over the description logics of the $DL\text{-}Lite$ family and their extensions; see Table 2.

## 3. Reasoning in *DL-Lite* Logics

We discuss now the reasoning problems we consider in this article, their mutual relationships, and the complexity measures we adopt. We also provide an overview of the complexity results for $DL\text{-}Lite$ logics obtained in this article.

### 3.1 Reasoning Problems

We will concentrate on three fundamental and standard reasoning tasks for description logics: satisfiability (or consistency), instance checking, and query answering.

For a DL $\mathcal{L}$ in the extended *DL-Lite* family, we define an *$\mathcal{L}$-concept inclusion* as any concept inclusion allowed in $\mathcal{L}$. Similarly, we define the notions of *$\mathcal{L}$-KB* and *$\mathcal{L}$-TBox*. Finally, define an *$\mathcal{L}$-concept* as any concept that can occur on the right-hand side of an *$\mathcal{L}$-concept inclusion* or a conjunction of such concepts.

**Satisfiability.** The *KB satisfiability problem* is to check, given an $\mathcal{L}$-KB $\mathcal{K}$, whether there is a model of $\mathcal{K}$. Clearly, satisfiability is the minimal requirement for any ontology. As is well known in DL (Baader et al., 2003), many other reasoning tasks for description logics are reducible to the satisfiability problem. Consider, for example, the *subsumption problem*: given an $\mathcal{L}$-TBox $\mathcal{T}$ and an $\mathcal{L}$-concept inclusion $C_1 \sqsubseteq C_2$, decide whether $\mathcal{T} \models C_1 \sqsubseteq C_2$, that is, $C_1^{\mathcal{I}} \subseteq C_2^{\mathcal{I}}$, for every model $\mathcal{I}$ of $\mathcal{T}$. To reduce this problem to (un)satisfiability, take a fresh concept name $A$, and set $\mathcal{K} = (\mathcal{T}', \mathcal{A})$, where

$$\mathcal{T}' = \mathcal{T} \cup \{A \sqsubseteq C_1, \ A \sqsubseteq \neg C_2\} \quad \text{and} \quad \mathcal{A} = \{A(a)\}.$$

It is easy to see that $\mathcal{T} \models C_1 \sqsubseteq C_2$ iff $\mathcal{K}$ is *not* satisfiable. For core, Krom and Horn KBs, if $C_2 = \sqcap_k D_k$, where each $D_k$ is a (possibly negated) basic concept, checking unsatisfiability of $\mathcal{K}$ amounts to checking unsatisfiability of each of the KBs $\mathcal{K}_k = (\mathcal{T}_k, \mathcal{A})$, where $\mathcal{T}_k = \mathcal{T} \cup \{A \sqsubseteq C_1, \ A \sqsubseteq \neg D_k\}$ (for Horn KBs, replace $A \sqsubseteq \neg B$ with the equivalent $A \sqcap B \sqsubseteq \bot$).

The *concept satisfiability problem*—given an $\mathcal{L}$-TBox $\mathcal{T}$ and an $\mathcal{L}$-concept $C$, decide whether $C^{\mathcal{I}} \neq \emptyset$ in a model $\mathcal{I}$ of $\mathcal{T}$—is also easily reducible to KB satisfiability. Indeed, take a fresh concept name $A$, a fresh object name $a$, and set $\mathcal{K} = (\mathcal{T}', \mathcal{A})$, where

$$\mathcal{T}' = \mathcal{T} \cup \{A \sqsubseteq C\} \quad \text{and} \quad \mathcal{A} = \{A(a)\}.$$

Then $C$ is satisfiable with respect to $\mathcal{T}$ iff $\mathcal{K}$ is satisfiable.

**Instance checking.** The *instance checking problem* is to decide, given an object name $a$, an $\mathcal{L}$-concept $C$ and an $\mathcal{L}$-KB $\mathcal{K} = (\mathcal{T}, \mathcal{A})$, whether $\mathcal{K} \models C(a)$, that is, $a^{\mathcal{I}} \in C^{\mathcal{I}}$, for every model $\mathcal{I}$ of $\mathcal{K}$. Instance checking is also reducible to (un)satisfiability: an object $a$ is an instance of an $\mathcal{L}$-concept $C$ in every model of $\mathcal{K} = (\mathcal{T}, \mathcal{A})$ iff the KB $\mathcal{K}' = (\mathcal{T}', \mathcal{A}')$, with

$$\mathcal{T}' = \mathcal{T} \cup \{A \sqsubseteq \neg C\} \qquad \text{and} \qquad \mathcal{A}' = \mathcal{A} \cup \{A(a)\},$$

is not satisfiable, where $A$ is a fresh concept name. For core, Krom and Horn KBs, if $C = \sqcap_k D_k$, where each $D_k$ is a (possibly negated) basic concept, we can proceed as for subsumption: checking the unsatisfiability of $\mathcal{K}'$ amounts to checking the unsatisfiability of each KB $\mathcal{K}'_k = (\mathcal{T}'_k, \mathcal{A}')$ with $\mathcal{T}'_k = \mathcal{T} \cup \{A \sqsubseteq \neg D_k\}$.

Conversely, KB satisfiability is reducible to the complement of instance checking: $\mathcal{K}$ is satisfiable iff $\mathcal{K} \not\models A(a)$, for a fresh concept name $A$ and a fresh object $a$.

**Query answering.** A *positive existential query* $\mathsf{q}(x_1, \ldots, x_n)$ is any first-order formula $\varphi(x_1, \ldots, x_n)$ constructed by means of conjunction, disjunction and existential quantification starting from atoms of the from $A_k(t)$ and $P_k(t_1, t_2)$, where $A_k$ is a concept name, $P_k$ a role name, and $t, t_1, t_2$ are *terms* taken from the list of variables $y_0, y_1, \ldots$ and the list of object names $a_0, a_1, \ldots$ (i.e., $\varphi$ is a positive existential formula). More precisely,

$$
\begin{aligned}
t &::= y_i \quad | \quad a_i, \\
\varphi &::= A_k(t) \quad | \quad P_k(t_1, t_2) \quad | \quad \varphi_1 \wedge \varphi_2 \quad | \quad \varphi_1 \vee \varphi_2 \quad | \quad \exists y_i \, \varphi.
\end{aligned}
$$

The free variables of $\varphi$ are called *distinguished variables* of $\mathsf{q}$ and the bound ones are *non-distinguished variables* of $\mathsf{q}$. We write $\mathsf{q}(x_1, \ldots, x_n)$ for a query with distinguished variables $x_1, \ldots, x_n$. A *conjunctive query* is a positive existential query that contains no disjunction (it is constructed from atoms by means of conjunction and existential quantification only).

Given a query $\mathsf{q}(\vec{x}) = \varphi(\vec{x})$ with $\vec{x} = x_1, \ldots, x_n$ and an $n$-tuple $\vec{a}$ of object names, we write $\mathsf{q}(\vec{a})$ for the result of replacing every occurrence of $x_i$ in $\varphi(\vec{x})$ with the $i$th member of $\vec{a}$. Queries containing no distinguished variables will be called *ground* (they are also known as Boolean).

Let $\mathcal{I} = (\Delta^{\mathcal{I}}, \cdot^{\mathcal{I}})$ be an interpretation. An *assignment* $\mathfrak{a}$ in $\Delta^{\mathcal{I}}$ is a function associating with every variable $y$ an element $\mathfrak{a}(y)$ of $\Delta^{\mathcal{I}}$. We will use the following notation: $a_i^{\mathcal{I}, \mathfrak{a}} = a_i^{\mathcal{I}}$ and $y^{\mathcal{I}, \mathfrak{a}} = \mathfrak{a}(y)$. The *satisfaction relation* for positive existential formulas with respect to a given assignment $\mathfrak{a}$ is defined inductively by taking:

$$
\begin{aligned}
\mathcal{I} &\models^{\mathfrak{a}} A_k(t) &\text{iff}\quad& t^{\mathcal{I}, \mathfrak{a}} \in A_k^{\mathcal{I}}, \\
\mathcal{I} &\models^{\mathfrak{a}} P_k(t_1, t_2) &\text{iff}\quad& (t_1^{\mathcal{I}, \mathfrak{a}}, t_2^{\mathcal{I}, \mathfrak{a}}) \in P_k^{\mathcal{I}}, \\
\mathcal{I} &\models^{\mathfrak{a}} \varphi_1 \wedge \varphi_2 &\text{iff}\quad& \mathcal{I} \models^{\mathfrak{a}} \varphi_1 \text{ and } \mathcal{I} \models^{\mathfrak{a}} \varphi_2, \\
\mathcal{I} &\models^{\mathfrak{a}} \varphi_1 \vee \varphi_2 &\text{iff}\quad& \mathcal{I} \models^{\mathfrak{a}} \varphi_1 \text{ or } \mathcal{I} \models^{\mathfrak{a}} \varphi_2, \\
\mathcal{I} &\models^{\mathfrak{a}} \exists y_i \, \varphi &\text{iff}\quad& \mathcal{I} \models^{\mathfrak{b}} \varphi, \text{ for some assignment } \mathfrak{b} \text{ in } \Delta^{\mathcal{I}} \text{ that may differ from } \mathfrak{a} \text{ on } y_i.
\end{aligned}
$$

For a ground query $\mathsf{q}(\vec{a})$, the satisfaction relation does not depend on the assignment $\mathfrak{a}$, and so we write $\mathcal{I} \models \mathsf{q}(\vec{a})$ instead of $\mathcal{I} \models^{\mathfrak{a}} \mathsf{q}(\vec{a})$. The answer to such a query is either 'yes' or 'no.'

For a KB $\mathcal{K} = (\mathcal{T}, \mathcal{A})$, we say that a tuple $\vec{a}$ of object names from $\mathcal{A}$ is a *certain answer* to $\mathsf{q}(\vec{x})$ with respect to $\mathcal{K}$, and write $\mathcal{K} \models \mathsf{q}(\vec{a})$, if $\mathcal{I} \models \mathsf{q}(\vec{a})$ whenever $\mathcal{I} \models \mathcal{K}$. The *query answering problem* can be formulated as follows: given an $\mathcal{L}$-KB $\mathcal{K} = (\mathcal{T}, \mathcal{A})$, a query $\mathsf{q}(\vec{x})$, and a tuple $\vec{a}$ of object names from $\mathcal{A}$, decide whether $\mathcal{K} \models \mathsf{q}(\vec{a})$.

Note that the instance checking problem is a special case of query answering: an object $a$ is an instance of an $\mathcal{L}$-concept $C$ with respect to a KB $\mathcal{K}$ iff the answer to the query $A(a)$ with respect to $\mathcal{K}'$ is 'yes,' where $\mathcal{K}' = (\mathcal{T}', \mathcal{A})$ and $\mathcal{T}' = \mathcal{T} \cup \{C \sqsubseteq A\}$, with $A$ a fresh concept name. For Horn-concepts $B_1 \sqcap \cdots \sqcap B_k$, we consider the query $A_1(a) \wedge \cdots \wedge A_k(a)$ with respect to $\mathcal{K}'$, where $\mathcal{K}' = (\mathcal{T}', \mathcal{A})$ and $\mathcal{T}' = \mathcal{T} \cup \{B_1 \sqsubseteq A_1, \ldots, B_k \sqsubseteq A_k\}$, with the $A_i$ fresh concept names. Similarly, we deal with Krom-concepts $D_1 \sqcap \cdots \sqcap D_k$, where each $D_i$ is a possibly negated basic concept. For core-concepts, the reduction holds just for conjunctions of basic concepts.

## 3.2 Complexity Measures: Data and Combined Complexity

The computational complexity of the reasoning problems discussed above can be analyzed with respect to different complexity measures, which depend on those parameters of the problem that are regarded to be the input (i.e., can vary) and those that are regarded to be fixed. For satisfiability and instance checking, the parameters to consider are the size of the TBox $\mathcal{T}$ and the size of the ABox $\mathcal{A}$, that is the number of symbols in $\mathcal{T}$ and $\mathcal{A}$, denoted $|\mathcal{T}|$ and $|\mathcal{A}|$, respectively. The size $|\mathcal{K}|$ of the knowledge $\mathcal{K} = (\mathcal{T}, \mathcal{A})$ is simply given by $|\mathcal{T}| + |\mathcal{A}|$. For query answering, one more parameter to consider would be the size of the query. However, in our analysis we adopt the standard database assumption that the size of queries is always bounded by some reasonable constant and, in any case, negligible with respect to both the size of the TBox and the size of the ABox. Thus we do not count the query as part of the input.

Hence, we consider our reasoning problems under two complexity measures. If the whole KB $\mathcal{K}$ is regarded as an input, then we deal with *combined complexity*. If, however, only the ABox $\mathcal{A}$ is counted as an input, while the TBox $\mathcal{T}$ (and the query) is regarded to be fixed, then our concern is *data complexity* (Vardi, 1982). Combined complexity is of interest when we are still designing and testing the ontology. On the other hand, data complexity is preferable in all those cases where the TBox is fixed or its size (and the size of the query) is negligible compared to the size of the ABox, which is the case, for instance, in the context of ontology-based data access (Calvanese, De Giacomo, Lembo, Lenzerini, Poggi, & Rosati, 2007) and other data intensive applications (Decker, Erdmann, Fensel, & Studer, 1999; Noy, 2004; Lenzerini, 2002; Calvanese et al., 2008). Since the logics of the *DL-Lite* family were tailored to deal with large data sets stored in relational databases, data complexity of both instance checking and query answering is of particular interest to us.

## 3.3 Remarks on the Complexity Classes LogSpace and AC$^0$

In this paper, we deal with the following complexity classes:

$$\mathrm{AC}^0 \subsetneq \mathrm{LogSpace} \subseteq \mathrm{NLogSpace} \subseteq \mathrm{P} \subseteq \mathrm{NP} \subseteq \mathrm{ExpTime}.$$

Their definitions can be found in the standard textbooks (e.g., Garey & Johnson, 1979; Papadimitriou, 1994; Vollmer, 1999; Kozen, 2006). Here we only remind the reader of the two smallest classes LogSpace and AC$^0$.

A problem belongs to LogSpace if there is a two-tape Turing machine $\mathcal{M}$ such that, starting with an input of length $n$ written on the *read-only input tape*, $\mathcal{M}$ stops in an accepting or rejecting state having used at most $\log n$ cells of the (initially blank) *read/write work tape*. A LogSpace *transducer* is a three-tape Turing machine that, having started with an input of length $n$ written on the read-only input tape, writes the result (of polynomial size) on the *write-only output tape* using at most $\log n$ cells of the (initially blank) read/write work tape. A LogSpace-*reduction* is a reduction computable by a LogSpace transducer; the composition of two LogSpace transducers is also a LogSpace transducer (Kozen, 2006, Lemma 5.1).

The formal definition of the complexity class AC$^0$ (see, e.g., Boppana & Sipser, 1990; Vollmer, 1999 and references therein) is based on the circuit model, where functions are represented as directed acyclic graphs built from unbounded fan-in AND, OR and NOT gates (i.e., AND and OR gates may have an unbounded number of incoming edges). For this definition we assume that decision problems are encoded in the alphabet $\{0, 1\}$ and so can be regarded as Boolean functions. AC$^0$ is the class of problems definable using a family of circuits of constant depth and polynomial size, which can be generated by a deterministic Turing machine (in the size of the input); the latter condition is called LOGTIME-*uniformity*. Intuitively, AC$^0$ allows us to use polynomially many processors but the run-time must be constant. A typical example of an AC$^0$ problem is evaluation of first-order queries over databases (or model checking of first-order sentences over finite models), where only the database (first-order model) is regarded as the input and the query (first-order sentence) is assumed to be fixed (Abiteboul, Hull, & Vianu, 1995; Vollmer, 1999). On the other hand, the undirected graph reachability problem is known to be in LOGSPACE (Reingold, 2008) but not in AC$^0$. A Boolean function $f\colon \{0,1\}^n \to \{0,1\}$ is called AC$^0$-*reducible* (or *constant-depth reducible*) to a function $g\colon \{0,1\}^n \to \{0,1\}$ if there is a (LOGTIME-uniform) family of constant-depth circuits built from AND, OR, NOT and $g$ gates that computes $f$. In this case we say that there is an AC$^0$-reduction. Note that all the reductions considered in Section 3.1 are AC$^0$-reductions. Unless otherwise indicated, in what follows we write 'reduction' for 'AC$^0$-reduction.'

### 3.4 Summary of Complexity Results

In this article, our aim is to investigate (i) the *combined* and *data complexity* of the satisfiability and instance checking problems and (ii) the *data complexity* of the query answering problem for the logics of the extended *DL-Lite* family, both with and without the UNA. The obtained and known results for the first 32 logics from Table 1 (the logics $DL\text{-}Lite_\alpha^{(\mathcal{HF})^+}$ and $DL\text{-}Lite_\alpha^{(\mathcal{HN})^+}$ are not included) are summarized in Table 2 (we remind the reader that satisfiability and instance checking are reducible to the complements of each other and that instance checking is a special case of query answering). In fact, all of the results in the table follow from the lower and upper bounds marked with [≥] and [≤], respectively (by taking into account the hierarchy of languages of the *DL-Lite* family): for example, the NLOGSPACE membership of satisfiability in $DL\text{-}Lite_{krom}^\mathcal{N}$ in Theorem 5.7 implies the same upper bound for $DL\text{-}Lite_{krom}^\mathcal{F}$, $DL\text{-}Lite_{krom}$, $DL\text{-}Lite_{core}^\mathcal{N}$, $DL\text{-}Lite_{core}^\mathcal{F}$ and $DL\text{-}Lite_{core}$ because all of them are sub-languages of $DL\text{-}Lite_{krom}^\mathcal{N}$.

**Remark 3.1** Two further complexity results are to be noted (they are not included in Table 2):

(i) If equality between object names is allowed in the language of *DL-Lite*, which only makes sense if the UNA is dropped, then the AC$^0$ memberships in Table 2 are replaced by LOGSPACE-completeness (see Section 8, Theorem 8.3 and 8.9); inequality constraints do not affect the complexity.

(ii) If we extend any of our languages with role transitivity constraints then the combined complexity of satisfiability remains the same, while for data complexity, instance checking and query answering become NLOGSPACE-hard (see Lemma 6.3), i.e., the membership in AC$^0$ for data complexity is replaced by NLOGSPACE-completeness, while all other complexity results remain the same.

In either case, the property of first-order rewritability—that is, the possibility of rewriting a given query q and a given TBox $\mathcal{T}$ into a single first-order query q′ returning the certain answers to q over $(\mathcal{T}, \mathcal{A})$ for every ABox $\mathcal{A}$, which ensures that the query answering problem is in AC$^0$ for data complexity—is lost.

| Languages | UNA | Complexity | | |
|---|---|---|---|---|
| | | Combined complexity | Data complexity | |
| | | Satisfiability | Instance checking | Query answering |
| $DL\text{-}Lite_{core}^{[\,\mathcal{H}]}$ | | NLogSpace $\geq$ [A] | in AC$^0$ | in AC$^0$ |
| $DL\text{-}Lite_{horn}^{[\,\mathcal{H}]}$ | yes/no | P $\leq$ [Th.8.2] $\geq$ [A] | in AC$^0$ | in AC$^0$ $\leq$ [C] |
| $DL\text{-}Lite_{krom}^{[\,\mathcal{H}]}$ | | NLogSpace $\leq$ [Th.8.2] | in AC$^0$ | coNP $\geq$ [B] |
| $DL\text{-}Lite_{bool}^{[\,\mathcal{H}]}$ | | NP $\leq$ [Th.8.2] $\geq$ [A] | in AC$^0$ $\leq$ [Th.8.3] | coNP |
| $DL\text{-}Lite_{core}^{[\mathcal{FN}](\mathcal{HF})(\mathcal{HN})}$ | | NLogSpace | in AC$^0$ | in AC$^0$ |
| $DL\text{-}Lite_{horn}^{[\mathcal{FN}](\mathcal{HF})(\mathcal{HN})}$ | yes | P $\leq$ [Th.5.8, 5.13] | in AC$^0$ | in AC$^0$ $\leq$ [Th.7.1] |
| $DL\text{-}Lite_{krom}^{[\mathcal{FN}](\mathcal{HF})(\mathcal{HN})}$ | | NLogSpace $\leq$ [Th.5.7,5.13] | in AC$^0$ | coNP |
| $DL\text{-}Lite_{bool}^{[\mathcal{FN}](\mathcal{HF})(\mathcal{HN})}$ | | NP $\leq$ [Th.5.6, 5.13] | in AC$^0$ $\leq$ [Cor.6.2] | coNP |
| $DL\text{-}Lite_{core/horn}^{[\mathcal{F}(\mathcal{HF})]}$ | | P $\leq$ [Cor.8.8] $\geq$ [Th.8.7] | P $\geq$ [Th.8.7] | P |
| $DL\text{-}Lite_{krom}^{[\mathcal{F}(\mathcal{HF})]}$ | | P $\leq$ [Cor.8.8] | P | coNP |
| $DL\text{-}Lite_{bool}^{[\mathcal{F}(\mathcal{HF})]}$ | no | NP | P $\leq$ [Cor.8.8] | coNP |
| $DL\text{-}Lite_{core/horn}^{[\mathcal{N}(\mathcal{HN})]}$ | | NP $\geq$ [Th.8.4] | coNP $\geq$ [Th.8.4] | coNP |
| $DL\text{-}Lite_{krom/bool}^{[\mathcal{N}(\mathcal{HN})]}$ | | NP $\leq$ [Th.8.5] | coNP | coNP |
| $DL\text{-}Lite_{core/horn}^{\mathcal{HF}}$ | | ExpTime $\geq$ [Th.5.10] | P $\geq$ [Th.6.7] | P $\leq$ [D] |
| $DL\text{-}Lite_{krom/bool}^{\mathcal{HF}}$ | yes/no | ExpTime | coNP $\geq$ [Th.6.5] | coNP |
| $DL\text{-}Lite_{core/horn}^{\mathcal{HN}}$ | | ExpTime | coNP $\geq$ [Th.6.6] | coNP |
| $DL\text{-}Lite_{krom/bool}^{\mathcal{HN}}$ | | ExpTime $\leq$ [F] | coNP | coNP $\leq$ [E] |

[A] complexity of the respective fragment of propositional Boolean logic
[B] follows from the proof of the data complexity result for instance checking in $\mathcal{ALE}$ (Schaerf, 1993)
[C] (Calvanese et al., 2006)
[D] follows from Horn-$\mathcal{SHIQ}$ (Hustadt, Motik, & Sattler, 2005; Eiter, Gottlob, Ortiz, & Šimkus, 2008)
[E] follows from $\mathcal{SHIQ}$ (Ortiz, Calvanese, & Eiter, 2006, 2008; Glimm, Horrocks, Lutz, & Sattler, 2007)
[F] follows from $\mathcal{SHIQ}$ (Tobies, 2001)

Table 2: Complexity of *DL-Lite* logics (all the complexity bounds save 'in AC$^0$' are tight).
$DL\text{-}Lite_{\alpha}^{[\beta_1|\cdots|\beta_n]}$ means any of $DL\text{-}Lite_{\alpha}^{\beta_1}$, ..., $DL\text{-}Lite_{\alpha}^{\beta_n}$
   (in particular, $DL\text{-}Lite_{\alpha}^{[\,\mathcal{H}]}$ is either $DL\text{-}Lite_{\alpha}$ or $DL\text{-}Lite_{\alpha}^{\mathcal{H}}$).
$DL\text{-}Lite_{core/horn}^{\beta}$ means $DL\text{-}Lite_{core}^{\beta}$ or $DL\text{-}Lite_{horn}^{\beta}$ (likewise for $DL\text{-}Lite_{krom/bool}^{\beta}$).
'$\leq$ [X]' ('$\geq$ [X]') means that the upper (respectively, lower) bound follows from [X].

Detailed proofs of our results will be given in Sections 5–8. For the variants of logics involving number restrictions, all upper bounds hold also under the assumption that the numbers $q$ in concepts of the form $\geq q\,R$ are given in *binary*. (Intuitively, this follows from the fact that in our proofs we only use those numbers that explicitly occur in the KB.) All lower bounds remain the same for the unary coding, since in the corresponding proofs we only use numbers not exceeding 4.

In the next section we consider the extended *DL-Lite* family in a more general context by identifying its place among other *DL-Lite*-related logics, in particular the OWL 2 profiles.

## 4. The Landscape of *DL-Lite* Logics

The original family of *DL-Lite* logics was created with two goals in mind: to identify description logics that, on the one hand, are capable of representing some basic features of conceptual modeling formalisms (such as UML class diagrams and ER diagrams) and, on the other hand, are computationally tractable, in particular, matching the $AC^0$ data complexity of database query answering.

As we saw in Section 2.2, to represent UML class diagrams one does not need the typical quantification constructs of the basic description logic $\mathcal{ALC}$ (Schmidt-Schauß & Smolka, 1991), namely, universal restriction $\forall R.C$ and *qualified* existential quantification $\exists R.C$: one can always take the role filler $C$ to be $\top$. Indeed, domain and range restrictions for a relationship $P$ can be expressed by the concepts inclusions $\exists P \sqsubseteq B_1$ and $\exists P^- \sqsubseteq B_2$, respectively. Thus, almost all concept inclusions required for capturing UML class diagrams are of the form $B_1 \sqsubseteq B_2$ or $B_1 \sqsubseteq \neg B_2$. These observations motivated the introduction by Calvanese et al. (2005) of the first *DL-Lite* logic, which in our new nomenclature corresponds to $DL\text{-}Lite_{core}^{\mathcal{F}}$. Their main results were a polynomial-time upper bound for the combined complexity of KB satisfiability and a LOGSPACE upper bound for the data complexity of conjunctive query answering (under the UNA). These results were extended by Calvanese et al. (2006) to two larger languages: $DL\text{-}Lite_{horn}^{\mathcal{F}}$ and $DL\text{-}Lite_{horn}^{\mathcal{H}}$, which were originally called $DL\text{-}Lite_{\sqcap,\mathcal{F}}$ and $DL\text{-}Lite_{\sqcap,\mathcal{R}}$, respectively. Calvanese et al. (2007b) introduced another member of the *DL-Lite* family (named $DL\text{-}Lite_{\mathcal{R}}$), which extended $DL\text{-}Lite_{core}^{\mathcal{H}}$ with *role disjointness* axioms of the form $\mathsf{Dis}(R_1, R_2)$. The computational behavior of the new logic turned out to be the same as that of $DL\text{-}Lite_{core}^{\mathcal{H}}$. It may be worth mentioning that $DL\text{-}Lite_{core}^{\mathcal{H}}$ covers the DL fragment of RDFS (Klyne & Carroll, 2004; Hayes, 2004). Note also that Calvanese et al. (2006) considered the variants of both $DL\text{-}Lite_{\sqcap,\mathcal{F}}$ and $DL\text{-}Lite_{\sqcap,\mathcal{R}}$ with arbitrary $n$-ary relations (not only the usual binary roles) and showed that query answering in them is still in LOGSPACE for data complexity. We conjecture that similar results can be obtained for the other *DL-Lite* logics introduced in this paper. Artale et al. (2007b) demonstrated how $n$-ary relations can be represented in $DL\text{-}Lite_{core}^{\mathcal{F}}$ by means of reification.

A further variant of *DL-Lite*, called $DL\text{-}Lite_{\mathcal{A}}$ ('$\mathcal{A}$' for attributes), was introduced by Poggi et al. (2008a) with the aim of capturing as many features of conceptual modeling formalisms as possible, while still maintaining the computational properties of the basic variants of *DL-Lite*. One of the features in $DL\text{-}Lite_{\mathcal{A}}$, borrowed from conceptual modeling formalisms and adopted also in OWL, is the distinction between (abstract) objects and data values, and consequently, between concepts (sets of objects) and datatypes (sets of data values), and between roles (i.e., object properties in OWL, relating objects with objects) and attributes (i.e., data properties in OWL, relating objects with data values). However, as far as the results in this paper are concerned, the distinction between concepts and datatypes, and between roles and attributes has no impact on reasoning whatsoever, since datatypes can simply be treated as special concepts that are mutually disjoint and are also disjoint from the proper concepts. Instead, more relevant for reasoning is the possibility to express in $DL\text{-}Lite_{\mathcal{A}}$ both role inclusions and functionality, i.e., $DL\text{-}Lite_{\mathcal{A}}$ includes both $DL\text{-}Lite_{core}^{\mathcal{H}}$ and $DL\text{-}Lite_{core}^{\mathcal{F}}$, but not $DL\text{-}Lite_{core}^{\mathcal{HF}}$.

As we have already mentioned, role inclusions and functionality constraints cannot be combined in an unrestricted way without losing the good computational properties: in

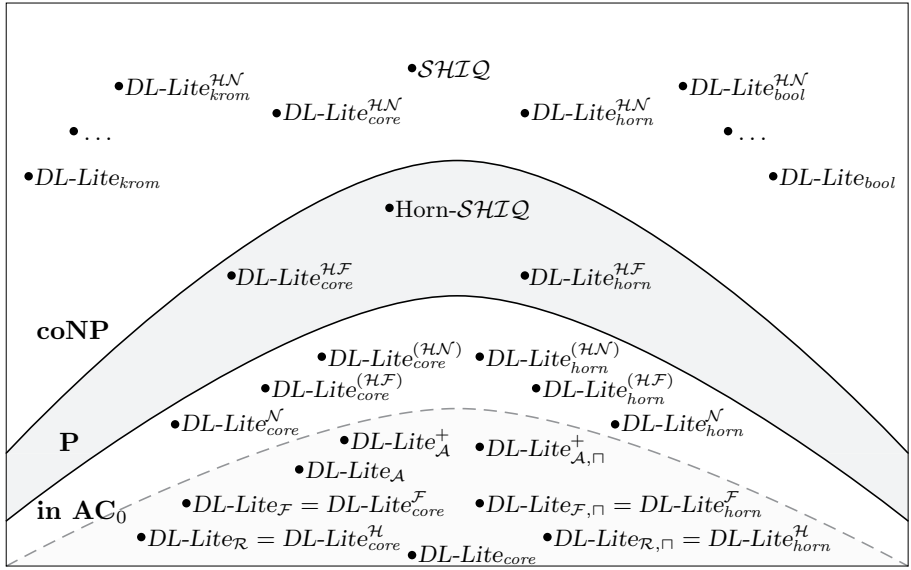

Figure 3: The *DL-Lite* family and relations.

Theorems 5.10 and 6.7, we prove that satisfiability of $DL\text{-}Lite^{\mathcal{HF}}_{core}$ KBs is ExpTime-hard for combined complexity, while instance checking is data-hard for P (NLogSpace-hardness was shown by Calvanese et al., 2006). In $DL\text{-}Lite_{\mathcal{A}}$, to keep query answering in $AC^0$ for data complexity and satisfiability in NLogSpace for combined complexity, functional roles (and attributes) are not allowed to be specialized, i.e., used positively on the right-hand side of role (and attribute) inclusion axioms. So, condition (**A₃**) is a slight generalization of this restriction. $DL\text{-}Lite_{\mathcal{A}}$ also allows axioms of the form $B \sqsubseteq \exists R.C$ for *non-functional* roles $R$, which is covered by conditions (**A₁**) and (**A₂**). Thus, $DL\text{-}Lite_{\mathcal{A}}$ can be regarded as a proper fragment of both $DL\text{-}Lite^{(\mathcal{HF})}_{core}$ and $DL\text{-}Lite^{(\mathcal{HN})}_{horn}$. We show in Sections 5.3 and 7 that these three languages enjoy very similar computational properties under the UNA: tractable satisfiability and query answering in $AC^0$.

We conclude this section with a picture in Figure 3 illustrating the landscape of *DL-Lite*-related logics by grouping them according to the data complexity of positive existential query answering *under the UNA*. The original eight *DL-Lite* logics, called by Calvanese et al. (2007b) '*the DL-Lite family*,' are shown in the bottom sector of the picture (the logics $DL\text{-}Lite^+_{\mathcal{A}}$ and $DL\text{-}Lite^+_{\mathcal{A},\sqcap}$ extend $DL\text{-}Lite_{\mathcal{A}}$ and $DL\text{-}Lite_{\mathcal{A},\sqcap}$ with identification constraints, which are out of the scope of this article). Their nearest relatives are the logic $DL\text{-}Lite^{(\mathcal{HN})}_{horn}$ and its fragments, which are all in $AC^0$ as well. The next layer contains the logics $DL\text{-}Lite^{\mathcal{HF}}_{core}$ and $DL\text{-}Lite^{\mathcal{HF}}_{horn}$, in which query answering is data-complete for P (no matter whether the UNA is adopted or not). In fact, these logics are fragments of the much more expressive DL Horn-$\mathcal{SHIQ}$, which was shown to enjoy the same data complexity of query answering by Eiter et al. (2008). It remains to be seen whether polynomial query answering is practically feasible; recent experiments with the DL $\mathcal{EL}$ (Lutz, Toman, & Wolter, 2008) indicate that this may indeed be the case. Finally, very distant relatives of the *DL-Lite* family comprise the upper layer of the picture, where query answering is data-complete for coNP, that is, the same as for the very expressive DL $\mathcal{SHIQ}$.

## 4.1 The *DL-Lite* Family and OWL 2

The upcoming version 2 of the Web Ontology Language OWL[7] defines three *profiles*,[8] that is, restricted versions of the language that suit specific needs. The *DL-Lite* family, notably $DL\text{-}Lite_{core}^{\mathcal{H}}$ (or the original $DL\text{-}Lite_{\mathcal{R}}$), is at the basis of one of these OWL 2 profiles, called OWL 2 QL. According to `http://www.w3.org/TR/owl2-profiles/`, 'OWL 2 QL is aimed at applications that use very large volumes of instance data, and where query answering is the most important reasoning task. In OWL 2 QL, [. . . ] sound and complete conjunctive query answering can be performed in LogSpace with respect to the size of the data (assertions) [and] polynomial time algorithms can be used to implement the ontology consistency and class expression subsumption reasoning problems. The expressive power of the profile is necessarily quite limited, although it does include most of the main features of conceptual models such as UML class diagrams and ER diagrams.' In this section, we briefly discuss the results obtained in this article in the context of additional constructs that are present in OWL 2.

A very important difference between the *DL-Lite* family and OWL is the status of the unique name assumption (UNA): this assumption is quite common in data management, and hence adopted in the *DL-Lite* family, but not adopted in OWL. Instead, the OWL syntax provides explicit means for stating that object names, say $a$ and $b$, are supposed to denote the same individual, $a \approx b$, or that they should be interpreted differently, $a \not\approx b$ (in OWL, these constructs are called `sameAs` and `differentFrom`).

The complexity results we obtain for logics of the form $DL\text{-}Lite_{\alpha}^{\mathcal{H}}$ do not depend on whether the UNA is adopted or not (because every model of a $DL\text{-}Lite_{\alpha}^{\mathcal{H}}$ KB without UNA can be 'untangled' into a model of the same KB respecting the UNA; see Lemma 8.10). However, this is not the case for the logics $DL\text{-}Lite_{\alpha}^{\mathcal{F}}$ and $DL\text{-}Lite_{\alpha}^{\mathcal{N}}$, where there is an obvious interaction between the UNA and number restrictions (cf. Table 2). For example, under the UNA, instance checking for $DL\text{-}Lite_{core}^{\mathcal{F}}$ is in AC[0] for data complexity, whereas dropping this assumption results in a much higher complexity: in Section 8, we prove that it is P-complete. The addition of the equality construct $\approx$ to $DL\text{-}Lite_{core}^{\mathcal{H}}$ and $DL\text{-}Lite_{horn}^{\mathcal{H}}$ slightly changes data complexity of query answering and instance checking, as it rises from membership in AC[0] to LogSpace-completeness; see Section 8. What is more important, however, is that in this case we loose first-order rewritability of query answering and instance checking, and as a result cannot use the standard database query engines in a straightforward manner.

Since the OWL 2 profiles are defined as syntactic restrictions of the language without changing the basic semantic assumptions, it was chosen not to include in the OWL 2 QL profile any construct that interferes with the UNA and which, in the absence of the UNA, would cause higher complexity. That is why OWL 2 QL does not include number restrictions, not even functionality constraints. Also, *keys* (the mechanism of identifying objects by means of the values of their properties) are not supported, although they are an impor-tant notion in conceptual modeling. Indeed, keys can be considered as a generalization of functionality constraints (Toman & Weddell, 2005, 2008; Calvanese, De Giacomo, Lembo, Lenzerini, & Rosati, 2007a, 2008b), since asserting a unary key, i.e., one involving only a single role $R$, is equivalent to asserting the functionality of the inverse of $R$. Hence, in the absence of the UNA, allowing keys would change the computational properties.

---

7. `http://www.w3.org/2007/OWL/`

8. In logic, *profiles* would be called *fragments* as they are defined by placing restrictions on the OWL 2 syntax only.

As we have already mentioned, some other standard OWL constructs, such as role disjointness, (a)symmetry and (ir)reflexivity constraints, can be added to the *DL-Lite* logics without changing their computational behavior. Role transitivity constraints, $\mathsf{Tra}(R)$, asserting that $R$ must be interpreted as a transitive role, can also be added to *DL-Lite*$_{horn}^{(\mathcal{HN})}$ but this leads to the increase of the data complexity for all reasoning problems to NLogSpace, although satisfiability remains in P for combined complexity. These results can be found in Section 5.3.

Of other constructs of OWL 2 that so far are not supported by the *DL-Lite* logics we mention *nominals* (i.e., singleton concepts), *Boolean operators* on roles, and *role chains*.

## 5. Satisfiability: Combined Complexity

*DL-Lite*$_{bool}^{\mathcal{HN}}$ is clearly a sub-logic of the description logic $\mathcal{SHIQ}$, the satisfiability problem for which is known to be ExpTime-complete (Tobies, 2001).

In Section 5.1 we show, however, that the satisfiability problem for *DL-Lite*$_{bool}^{\mathcal{N}}$ KBs is reducible to the satisfiability problem for the *one-variable fragment*, $\mathcal{QL}^1$, of first-order logic without equality and function symbols. As satisfiability of $\mathcal{QL}^1$-formulas is NP-complete (see, e.g., Börger et al., 1997) and the logics under consideration contain full Booleans on concepts, satisfiability of *DL-Lite*$_{bool}^{\mathcal{N}}$ KBs is NP-complete as well. We shall also see that the translations of Horn and Krom KBs into $\mathcal{QL}^1$ belong to the Horn and Krom fragments of $\mathcal{QL}^1$, respectively, which are known to be P- and NLogSpace-complete (see, e.g., Papadimitriou, 1994; Börger et al., 1997). In Section 5.2, we will show how to simulate the behavior of polynomial-space-bounded alternating Turing machines by means of *DL-Lite*$_{core}^{\mathcal{HF}}$ KBs. This will give the (optimal) ExpTime lower bound for satisfiability of KBs in all the languages of our family containing unrestricted occurrences of both functionality constraints and role inclusions. In Section 5.3, we extend the embedding into $\mathcal{QL}^1$, defined in Section 5.1, to the logic *DL-Lite*$_{bool}^{(\mathcal{HN})}$, thereby establishing the same upper bounds as for *DL-Lite*$_{bool}^{\mathcal{N}}$ and its fragments. Finally, in Section 5.4 we investigate the impact of role transitivity constraints.

### 5.1 *DL-Lite*$_{bool}^{\mathcal{N}}$ and its Fragments: First-Order Perspective

Our aim in this section is to construct a reduction of the satisfiability problem for *DL-Lite*$_{bool}^{\mathcal{N}}$ KBs to satisfiability of $\mathcal{QL}^1$-formulas. We will do this in two steps: first we present a lengthy yet quite 'natural' and transparent (yet exponential) reduction $\cdot^\dagger$, and then we shall see from the proof that this reduction can be substantially optimized to a linear reduction $\cdot^\ddagger$.

Let $\mathcal{K} = (\mathcal{T}, \mathcal{A})$ be a *DL-Lite*$_{bool}^{\mathcal{N}}$ KB. Recall that $role^\pm(\mathcal{K})$ denotes the set of direct and inverse role names occurring in $\mathcal{K}$ and $ob(\mathcal{A})$ the set of object names occurring in $\mathcal{A}$. For $R \in role^\pm(\mathcal{K})$, let $Q_{\mathcal{T}}^R$ be the set of natural numbers containing 1 and all the numbers $q$ for which the concept $\geq q\,R$ occurs in $\mathcal{T}$ (recall that the ABox does not contain number restrictions). Note that $|Q_{\mathcal{T}}^R| \geq 2$ if $\mathcal{T}$ contains a functionality constraint for $R$.

With every object name $a_i \in ob(\mathcal{A})$ we associate the individual constant $a_i$ of $\mathcal{QL}^1$ and with every concept name $A_i$ the unary predicate $A_i(x)$ from the signature of $\mathcal{QL}^1$. For each role $R \in role^{\pm}(\mathcal{K})$, we introduce $|Q_{\mathcal{T}}^R|$-many fresh unary predicates

$$E_q R(x), \qquad \text{for } q \in Q_{\mathcal{T}}^R.$$

The intended meaning of these predicates is as follows: for a role name $P_k$,

- $E_q P_k(x)$ and $E_q P_k^-(x)$ represent the sets of points with *at least $q$ distinct $P_k$-successors* and *at least $q$ distinct $P_k$-predecessors*, respectively. In particular, $E_1 P_k(x)$ and $E_1 P_k^-(x)$ represent the domain and range of $P_k$, respectively.

Additionally, for every pair of roles $P_k, P_k^- \in role^{\pm}(\mathcal{K})$, we take two fresh individual constants

$$dp_k \qquad \text{and} \qquad dp_k^-$$

of $\mathcal{QL}^1$, which will serve as 'representatives' of the points from the domains of $P_k$ and $P_k^-$, respectively (provided that they are not empty). Let $dr(\mathcal{K}) = \{dr \mid R \in role^{\pm}(\mathcal{K})\}$. Furthermore, for each pair of object names $a_i, a_j \in ob(\mathcal{A})$ and each $R \in role^{\pm}(\mathcal{K})$, we take a fresh *propositional variable* $Ra_i a_j$ of $\mathcal{QL}^1$ to encode the ABox assertion $R(a_i, a_j)$.[9]

By induction on the construction of a $DL\text{-}Lite_{bool}^{\mathcal{N}}$ concept $C$ we define the $\mathcal{QL}^1$-formula $C^*$:

$$\perp^* = \perp, \qquad (A_i)^* = A_i(x), \qquad (\geq q\, R)^* = E_q R(x),$$
$$(\neg C)^* = \neg C^*(x), \qquad (C_1 \sqcap C_2)^* = C_1^*(x) \wedge C_2^*(x).$$

The $DL\text{-}Lite_{bool}^{\mathcal{N}}$ TBox $\mathcal{T}$ corresponds then to the $\mathcal{QL}^1$-sentence $\forall x\, \mathcal{T}^*(x)$, where

$$\mathcal{T}^*(x) = \bigwedge_{C_1 \sqsubseteq C_2 \in \mathcal{T}} \big( C_1^*(x) \to C_2^*(x) \big). \tag{1}$$

The ABox $\mathcal{A}$ is translated into the following pair of $\mathcal{QL}^1$-sentences

$$\mathcal{A}^{\dagger 1} = \bigwedge_{A_k(a_i) \in \mathcal{A}} A_k(a_i) \quad \wedge \bigwedge_{\neg A_k(a_i) \in \mathcal{A}} \neg A_k(a_i), \tag{2}$$

$$\mathcal{A}^{\dagger 2} = \bigwedge_{P_k(a_i, a_j) \in \mathcal{A}} P_k a_i a_j \quad \wedge \bigwedge_{\neg P_k(a_i, a_j) \in \mathcal{A}} \neg P_k a_i a_j. \tag{3}$$

For every role $R \in role^{\pm}(\mathcal{K})$, we need two $\mathcal{QL}^1$-formulas:

$$\varepsilon_R(x) = E_1 R(x) \to inv(E_1 R)(inv(dr)), \tag{4}$$

$$\delta_R(x) = \bigwedge_{\substack{q, q' \in Q_{\mathcal{T}}^R, \ q' > q \\ q' > q'' > q \text{ for no } q'' \in Q_{\mathcal{T}}^R}} \big( E_{q'} R(x) \to E_q R(x) \big), \tag{5}$$

---

9. In what follows, we slightly abuse notation and write $R(a_i, a_j) \in \mathcal{A}$ to indicate that $P_k(a_i, a_j) \in \mathcal{A}$ if $R = P_k$, or $P_k(a_j, a_i) \in \mathcal{A}$ if $R = P_k^-$.

where (by overloading the *inv* operator),

$$inv(E_q R) \;=\; \begin{cases} E_q P_k^-, & \text{if } R = P_k, \\ E_q P_k, & \text{if } R = P_k^-, \end{cases} \qquad \text{and} \qquad inv(dr) = \begin{cases} dp_k^-, & \text{if } R = P_k, \\ dp_k, & \text{if } R = P_k^-. \end{cases}$$

Formula (4) says that if the domain of $R$ is not empty then its range is not empty either: it contains the constant $inv(dr)$, the 'representative' of the domain of $inv(R)$.

We also need formulas representing the relationship of the propositional variables $Ra_i a_j$ with the unary predicates for the role domain and range: for a role $R \in role^{\pm}(\mathcal{K})$, let $R^{\dagger}$ be the following $\mathcal{QL}^1$-sentence

$$\bigwedge_{a_i \in ob(\mathcal{A})} \bigwedge_{q \in Q_{\mathcal{T}}^R} \bigwedge_{\substack{a_{j_1},\dots,a_{j_q} \in ob(\mathcal{A}) \\ j_k \neq j_{k'} \text{ for } k \neq k'}} \Big( \bigwedge_{k=1}^q Ra_i a_{j_k} \to E_q R(a_i) \Big) \quad \wedge \bigwedge_{a_i, a_j \in ob(\mathcal{A})} (Ra_i a_j \to inv(R)a_j a_i), \quad (6)$$

where $inv(R)a_j a_i$ is the propositional variable $P_k^- a_j a_i$ if $R = P_k$ and $P_k a_j a_i$ if $R = P_k^-$. Note that the first conjunct of (6) is the only part of the translation that relies on the UNA.

Finally, for the *DL-Lite$_{bool}^{\mathcal{N}}$* knowledge base $\mathcal{K} = (\mathcal{T}, \mathcal{A})$, we set

$$\mathcal{K}^{\dagger} \;=\; \forall x \left[ \mathcal{T}^*(x) \quad \wedge \bigwedge_{R \in role^{\pm}(\mathcal{K})} (\varepsilon_R(x) \wedge \delta_R(x)) \right] \quad \wedge \quad \left[ \mathcal{A}^{\dagger 1} \wedge \mathcal{A}^{\dagger 2} \quad \wedge \bigwedge_{R \in role^{\pm}(\mathcal{K})} R^{\dagger} \right].$$

Thus, $\mathcal{K}^{\dagger}$ is a universal sentence of $\mathcal{QL}^1$.

**Example 5.1** Consider, for example, the KB $\mathcal{K} = (\mathcal{T}, \mathcal{A})$ with

$$\mathcal{T} \;=\; \big\{ A \sqsubseteq \exists P^-, \; \exists P^- \sqsubseteq A, \; A \sqsubseteq \geq 2\, P, \; \top \sqsubseteq \, \leq 1\, P^-, \; \exists P \sqsubseteq A \big\}$$

and $\mathcal{A} = \{A(a), \; P(a, a')\}$. Then we obtain the following first-order translation:

$$\begin{aligned}
\mathcal{K}^{\dagger} \;=\; & \forall x\, \chi(x) \;\wedge\; A(a) \;\wedge\; Paa' \;\wedge \\
& \big(Paa' \to E_1 P(a)\big) \;\wedge\; \big(Paa \to E_1 P(a)\big) \;\wedge \\
& \quad \big(Pa'a \to E_1 P(a')\big) \;\wedge\; \big(Pa'a' \to E_1 P(a')\big) \;\wedge \\
& \big(P^- aa' \to E_1 P^-(a)\big) \;\wedge\; \big(P^- aa \to E_1 P^-(a)\big) \;\wedge \\
& \quad \big(P^- a'a \to E_1 P^-(a')\big) \;\wedge\; \big(P^- a'a' \to E_1 P^-(a')\big) \;\wedge \\
& \big(Paa' \wedge Paa \to E_2 P(a)\big) \;\wedge\; \big(Pa'a \wedge Pa'a' \to E_2 P(a')\big) \;\wedge \\
& \big(P^- aa' \wedge P^- aa \to E_2 P^-(a)\big) \;\wedge\; \big(P^- a'a \wedge P^- a'a' \to E_2 P^-(a')\big) \;\wedge \\
& \big(Paa' \leftrightarrow P^- a'a\big) \wedge \big(Pa'a \leftrightarrow P^- aa'\big) \wedge \big(Paa \leftrightarrow P^- aa\big) \wedge \big(Pa'a' \leftrightarrow P^- a'a'\big).
\end{aligned}$$

where

$$\begin{aligned}
\chi(x) \;=\; & \big(A(x) \to E_1 P^-(x)\big) \;\wedge\; \big(E_1 P^-(x) \to A(x)\big) \;\wedge\; \big(A(x) \to E_2 P(x)\big) \;\wedge \\
& \quad \big(\top \to \neg E_2 P^-(x)\big) \;\wedge\; \big(E_1 P(x) \to A(x)\big) \;\wedge \\
& \big(E_1 P(x) \to E_1 P^-(dp^-)\big) \;\wedge\; \big(E_1 P^-(x) \to E_1 P(dp)\big) \;\wedge \\
& \quad \big(E_2 P(x) \to E_1 P(x)\big) \;\wedge\; \big(E_2 P^-(x) \to E_1 P^-(x)\big). \quad (7)
\end{aligned}$$

**Theorem 5.2** *A DL-Lite$_{bool}^{\mathcal{N}}$ knowledge base $\mathcal{K} = (\mathcal{T}, \mathcal{A})$ is satisfiable iff the $\mathcal{QL}^1$-sentence $\mathcal{K}^\dagger$ is satisfiable.*

**Proof** ($\Leftarrow$) If $\mathcal{K}^\dagger$ is satisfiable then there is a model $\mathfrak{M}$ of $\mathcal{K}^\dagger$ whose domain consists of all the constants occurring in $\mathcal{K}^\dagger$—i.e., $ob(\mathcal{A}) \cup dr(\mathcal{K})$ (say, an Herbrand model of $\mathcal{K}^\dagger$). We denote this domain by $D$ and the interpretations of the (unary) predicates $P$, propositional variables $p$ and constants $a$ of $\mathcal{QL}^1$ in $\mathfrak{M}$ by $P^{\mathfrak{M}}$, $p^{\mathfrak{M}}$ and $a^{\mathfrak{M}}$, respectively. Thus, for every constant $a$, we have $a^{\mathfrak{M}} = a$. Let $D_0$ be the set of all constants $a$, $a \in ob(\mathcal{A})$. Without loss of generality we may assume that $D_0 \neq \emptyset$.

We construct an interpretation $\mathcal{I}$ for *DL-Lite$_{bool}^{\mathcal{N}}$* based on some domain $\Delta^{\mathcal{I}} \supseteq D_0$ that will be inductively defined as the union

$$\Delta^{\mathcal{I}} = \bigcup_{m=0}^{\infty} W_m, \qquad \text{where} \quad W_0 = D_0.$$

The interpretations of the object names $a_i$ in $\mathcal{I}$ are given by their interpretations in $\mathfrak{M}$, namely, $a_i^{\mathcal{I}} = a_i^{\mathfrak{M}} \in W_0$. Each set $W_{m+1}$, for $m \geq 0$, is constructed by adding to $W_m$ some new elements that are fresh *copies* of certain elements from $D \setminus D_0$. If such a new element $w'$ is a copy of $w \in D \setminus D_0$ then we write $cp(w') = w$, while for $w \in D_0$ we let $cp(w) = w$. The set $W_m \setminus W_{m-1}$, for $m \geq 0$, will be denoted by $V_m$ (for convenience, let $W_{-1} = \emptyset$, so that $V_0 = D_0$).

The interpretations $A_k^{\mathcal{I}}$ of concept names $A_k$ in $\mathcal{I}$ are defined by taking

$$A_k^{\mathcal{I}} = \big\{ w \in \Delta^{\mathcal{I}} \mid \mathfrak{M} \models A_k^*[cp(w)] \big\}. \tag{8}$$

The interpretation $P_k^{\mathcal{I}}$ of a role name $P_k$ in $\mathcal{I}$ will be defined inductively as the union

$$P_k^{\mathcal{I}} = \bigcup_{m=0}^{\infty} P_k^m, \qquad \text{where} \quad P_k^m \subseteq W_m \times W_m,$$

along with the construction of $\Delta^{\mathcal{I}}$. First, for a role $R \in role^{\pm}(\mathcal{K})$, we define the *required R-rank $r(R, d)$ of a point* $d \in D$ by taking

$$r(R, d) = \max\big(\{0\} \cup \{q \in Q_{\mathcal{T}}^R \mid \mathfrak{M} \models E_q R[d]\}\big).$$

It follows from (5) that if $r(R, d) = q$ then, for every $q' \in Q_{\mathcal{T}}^R$, we have $\mathfrak{M} \models E_{q'} R[d]$ whenever $q' \leq q$, and $\mathfrak{M} \models \neg E_{q'} R[d]$ whenever $q < q'$. We also define the *actual R-rank $r_m(R, w)$ of a point* $w \in \Delta^{\mathcal{I}}$ at step $m$ by taking

$$r_m(R, w) = \begin{cases} \sharp\{w' \in W_m \mid (w, w') \in P_k^m\}, & \text{if } R = P_k, \\ \sharp\{w' \in W_m \mid (w', w) \in P_k^m\}, & \text{if } R = P_k^-. \end{cases}$$

For the basis of induction we set, for each role name $P_k \in role(\mathcal{K})$,

$$P_k^0 = \big\{ (a_i^{\mathfrak{M}}, a_j^{\mathfrak{M}}) \in W_0 \times W_0 \mid \mathfrak{M} \models P_k a_i a_j \big\}. \tag{9}$$

Observe that, by (6), for all $R \in role^{\pm}(\mathcal{K})$ and $w \in W_0$,

$$r_0(R, w) \leq r(R, cp(w)). \tag{10}$$

Suppose now that $W_m$ and the $P_k^m$, for $m \geq 0$, have already been defined. If we had $r_m(R, w) = r(R, cp(w))$, for all roles $R \in role^{\pm}(\mathcal{K})$ and points $w \in W_m$, then the interpretation $\mathcal{I}$ we need would be constructed. However, in general this is not the case because there may be some 'defects' in the sense that the actual rank of some points is smaller than the required rank.

For a role name $P_k \in role(\mathcal{K})$, consider the following two sets of defects in $P_k^m$:

$$\Lambda_k^m = \big\{ w \in V_m \mid r_m(P_k, w) < r(P_k, cp(w)) \big\},$$
$$\Lambda_k^{m-} = \big\{ w \in V_m \mid r_m(P_k^-, w) < r(P_k^-, cp(w)) \big\}.$$

The purpose of, say, $\Lambda_k^m$ is to identify those 'defective' points $w \in V_m$ from which precisely $r(P_k, cp(w))$ distinct $P_k$-arrows should start (according to $\mathfrak{M}$), but some arrows are still missing (only $r_m(P_k, w)$ many arrows exist). To 'cure' these defects, we extend $W_m$ and $P_k^m$ respectively to $W_{m+1}$ and $P_k^{m+1}$ according to the following rules:

$(\Lambda_k^m)$ Let $w \in \Lambda_k^m$, $q = r(P_k, cp(w)) - r_m(P_k, w)$ and $d = cp(w)$. We have $\mathfrak{M} \models E_{q'} P_k[d]$ for some $q' \in Q_{\mathcal{T}}^R$ with $q' \geq q > 0$. Then, by (5), $\mathfrak{M} \models E_1 P_k[d]$ and, by (4), $\mathfrak{M} \models E_1 P_k^-[dp_k^-]$. In this case we take $q$ *fresh* copies $w_1', \dots, w_q'$ of $dp_k^-$ (and set $cp(w_i') = dp_k^-$, for $1 \leq i \leq q$), add them to $W_{m+1}$ and add the pairs $(w, w_i')$, $1 \leq i \leq q$, to $P_k^{m+1}$.

$(\Lambda_k^{m-})$ Let $w \in \Lambda_k^{m-}$, $q = r(P_k^-, cp(w)) - r_m(P_k^-, w)$ and $d = cp(w)$. Then $\mathfrak{M} \models E_{q'} P_k^-[d]$ for some $q' \in Q_{\mathcal{T}}^R$ with $q' \geq q > 0$. So, by (5), we have $\mathfrak{M} \models E_1 P_k^-[d]$ and, by (4), $\mathfrak{M} \models E_1 P_k[dp_k]$. Take $q$ *fresh* copies $w_1', \dots, w_q'$ of $dp_k$ (and set $cp(w_i') = dp_k$, for $1 \leq i \leq q$), add them to $W_{m+1}$ and add the pairs $(w_i', w)$, $1 \leq i \leq q$, to $P_k^{m+1}$.

**Example 5.3** Consider again the KB $\mathcal{K}$ and its first-order translation $\mathcal{K}^\dagger$ from Example 5.1. Consider also a model $\mathfrak{M}$ of $\mathcal{K}^\dagger$ with the domain $D = \{a, a', dp, dp^-\}$, where

$$A^{\mathfrak{M}} = (E_1 P)^{\mathfrak{M}} = (E_1 P^-)^{\mathfrak{M}} = (E_2 P)^{\mathfrak{M}} = D, \quad (E_2 P^-)^{\mathfrak{M}} = \emptyset,$$
$$(Paa')^{\mathfrak{M}} = (P^- a' a)^{\mathfrak{M}} = \mathbf{t}.$$

We begin the construction of the interpretation $\mathcal{I}$ of $\mathcal{K}$ by setting $W_0 = V_0 = D_0 = \{a, a'\}$ and $P^0 = \{(a, a')\}$. Then we compute the required and actual ranks $r(R, w)$ and $r_0(R, w)$, for $R \in \{P, P^-\}$ and $w \in V_0$:

(i) $r(P, a) = 2$ and $r_0(P, a) = 1$,      (ii) $r(P, a') = 2$ and $r_0(P, a') = 0$,
(iii) $r(P^-, a) = 1$ and $r_0(P^-, a) = 0$,     (iv) $r(P^-, a') = 1$ and $r_0(P^-, a') = 1$.

At the next step, we draw a $P$-arrow from $a$ to a fresh copy of $dp^-$ to cure defect (i), draw two $P$-arrows from $a'$ to two more fresh copies of $dp^-$ in order to cure defects (ii), and finally we take a fresh copy of $dp$ and connect it to $a$ by a $P$-arrow, thereby curing defect (iii).

One more step of this 'unraveling' construction is shown in Figure 4.

Observe the following important property of the construction: for $m, m_0 \geq 0$, $w \in V_{m_0}$ and $R \in role^{\pm}(\mathcal{K})$,

$$r_m(R, w) = \begin{cases} 0, & \text{if } m < m_0, \\ q, & \text{if } m = m_0, \text{ for some } q \leq r(R, cp(w)), \\ r(R, cp(w)), & \text{if } m > m_0. \end{cases} \tag{11}$$

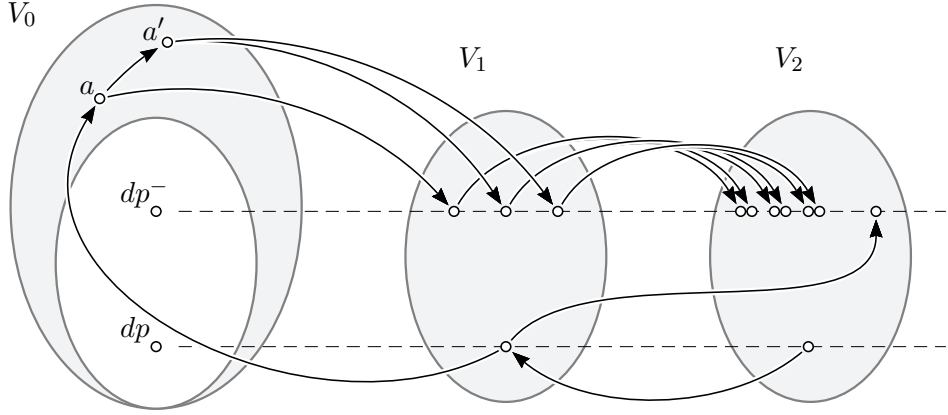

Figure 4: Unraveling model $\mathfrak{M}$ (first three steps).

To prove this property, consider all possible cases:

- If $m < m_0$ then the point $w$ has not been added to $W_m$ yet, i.e., $w \notin W_m$, and so we have $r_m(R, w) = 0$.

- If $m = m_0$ and $m_0 = 0$ then $r_m(R, w) \leq r(R, cp(w))$ follows from (10).

- If $m = m_0$ and $m_0 > 0$ then $w$ was added at step $m_0$ to cure a defect of some point $w' \in W_{m_0-1}$. This means that there is $P_k \in role(\mathcal{K})$ such that either $(w', w) \in P_k^{m_0}$ and $w' \in \Lambda_k^{m_0-1}$ or $(w, w') \in P_k^{m_0}$ and $w' \in \Lambda_k^{(m_0-1)-}$. Consider the former case. We have $cp(w) = dp_k^-$. Since *fresh* witnesses are picked up every time the rule $(\Lambda_k^{m_0-1})$ is applied, $r_{m_0}(P_k^-, w) = 1$, $r_{m_0}(P_k, w) = 0$ and $r_{m_0}(R, w) = 0$, for every $R \neq P_k, P_k^-$. So it suffices to show that $r(P_k^-, dp_k^-) \geq 1$. Indeed, as $\mathfrak{M} \models E_q P_k[cp(w')]$ for some $q \in Q_{\mathcal{T}}^R$, we have, by (5), $\mathfrak{M} \models E_1 P_k[cp(w')]$, and so, by (4), $\mathfrak{M} \models E_1 P_k^-[dp_k^-]$. By the definition of $r$, we have $r(P_k^-, dp_k^-) \geq 1$. The latter case is considered analogously.

- If $m = m_0 + 1$ then, for each role name $P_k$, all defects of $w$ are cured at step $m_0 + 1$ by applying the rules $(\Lambda_k^{m_0})$ and $(\Lambda_k^{m_0-})$. Therefore, $r_{m_0+1}(R, w) = r(R, cp(w))$.

- If $m > m_0 + 1$ then (11) follows from the observation that new arrows involving $w$ can only be added at step $m_0 + 1$, that is, for all $m \geq 0$ and each *role name* $P_k \in role(\mathcal{K})$,

$$P_k^{m+1} \setminus P_k^m \quad \subseteq \quad V_m \times V_{m+1} \quad \cup \quad V_{m+1} \times V_m. \tag{12}$$

It follows that, for all $R \in role^{\pm}(\mathcal{K})$, $q \in Q_{\mathcal{T}}^R$ and $w \in \Delta^{\mathcal{I}}$, we have:

$$\mathfrak{M} \models E_q R[cp(w)] \qquad \text{iff} \qquad w \in (\geq q\, R)^{\mathcal{I}}. \tag{13}$$

Indeed, if $\mathfrak{M} \models E_q R[cp(w)]$ then, by definition, $r(R, cp(w)) \geq q$. Let $w \in V_{m_0}$. Then, by (11), $r_m(R, w) = r(R, cp(w)) \geq q$, for all $m > m_0$. It follows from the definition of $r_m(R, w)$ and $R^{\mathcal{I}}$ that $w \in (\geq q\, R)^{\mathcal{I}}$. Conversely, let $w \in (\geq q\, R)^{\mathcal{I}}$ and $w \in V_{m_0}$. Then, by (11), $q \leq r_m(R, w) = r(R, cp(w))$, for all $m > m_0$. So, by the definition of $r(R, cp(w))$ and (5), $\mathfrak{M} \models E_q R[cp(w)]$.

By induction on the construction of concepts $C$ in $\mathcal{K}$ one can readily see that, for every $w \in \Delta^{\mathcal{I}}$, we have

$$\mathfrak{M} \models C^*[cp(w)] \qquad \text{iff} \qquad w \in C^{\mathcal{I}}. \tag{14}$$

Indeed, the basis is trivial for $B = \bot$ and follows from (8) for $B = A_k$ and from (13) for $B = \geq q\, R$, while the induction step for the Booleans ($C = \neg C_1$ and $C = C_1 \sqcap C_2$) immediately follows from the induction hypothesis.

Finally, we show that for each $\psi \in \mathcal{T} \cup \mathcal{A}$,

$$\mathfrak{M} \models \psi^\dagger \qquad \text{iff} \qquad \mathcal{I} \models \psi.$$

The case $\psi = C_1 \sqsubseteq C_2$ follows from (14); for $\psi = A_k(a_i)$ and $\psi = \neg A_k(a_i)$ from the definition of $A_k^{\mathcal{I}}$. For $\psi = P_k(a_i, a_j)$ and $\psi = \neg P_k(a_i, a_j)$, we have $(a_i^{\mathcal{I}}, a_j^{\mathcal{I}}) \in P_k^{\mathcal{I}}$ iff, by (12), $(a_i^{\mathcal{I}}, a_j^{\mathcal{I}}) \in P_k^0$ iff, by (9), $\mathfrak{M} \models P_k a_i a_j$.

Thus, we have established that $\mathcal{I} \models \mathcal{K}$.

($\Rightarrow$) Conversely, suppose that $\mathcal{I} \models \mathcal{K}$ is an interpretation with domain $\Delta^{\mathcal{I}}$. We construct a model $\mathfrak{M}$ of $\mathcal{K}^\dagger$ based on the same $\Delta^{\mathcal{I}}$. For every $a_i \in ob(\mathcal{A})$, we let $a_i^{\mathfrak{M}} = a_i^{\mathcal{I}}$ and, for every $R \in role^\pm(\mathcal{K})$, we take some $d \in (\geq 1\, R)^{\mathcal{I}}$ if $(\geq 1\, R)^{\mathcal{I}} \neq \emptyset$ and an arbitrary element $d \in \Delta^{\mathcal{I}}$ otherwise, and let $dr^{\mathfrak{M}} = d$. Next, for every concept name $A_k$, we let $A_k^{\mathfrak{M}} = A_k^{\mathcal{I}}$ and, for every role $R \in role^\pm(\mathcal{K})$ and $q \in Q_{\mathcal{T}}^R$, we set $E_q R^{\mathfrak{M}} = (\geq q\, R)^{\mathcal{I}}$. Finally, for every role $R \in role^\pm(\mathcal{K})$ and every pair of objects $a_i, a_j \in ob(\mathcal{A})$, we define $(Ra_i a_j)^{\mathfrak{M}}$ to be true iff $\mathcal{I} \models R(a_i, a_j)$. One can readily check that $\mathfrak{M} \models \mathcal{K}^\dagger$. Details are left to the reader. $\qquad\square$

The first-order translation $\mathcal{K}^\dagger$ of $\mathcal{K}$ is obviously too lengthy to provide us with reasonably low complexity results: $|\mathcal{K}^\dagger| \leq |\mathcal{K}| + (2 + q_{\mathcal{T}}^2) \cdot |role(\mathcal{K})| + 2 \cdot |role(\mathcal{K})| \cdot |ob(\mathcal{A})|^{q_{\mathcal{T}}}$. However, it follows from the proof above that a lot of information in this translation is redundant and can be safely omitted.

Now we define a more concise translation $\mathcal{K}^\ddagger$ of $\mathcal{K} = (\mathcal{T}, \mathcal{A})$ into $\mathcal{QL}^1$ by taking:

$$\mathcal{K}^\ddagger = \forall x \left[ \mathcal{T}^*(x) \quad \wedge \bigwedge_{R \in role^\pm(\mathcal{K})} \big( \varepsilon_R(x) \wedge \delta_R(x) \big) \right] \quad \wedge \quad \mathcal{A}^{\dagger^1} \quad \wedge \quad \mathcal{A}^{\ddagger^2},$$

where $\mathcal{T}^*(x)$, $\varepsilon_R(x)$, $\delta_R(x)$ and $\mathcal{A}^{\dagger^1}$ are defined as before by means of (1), (4), (5) and (2), respectively, and

$$\mathcal{A}^{\ddagger^2} = \bigwedge_{\substack{a \in ob(\mathcal{A})}} \bigwedge_{\substack{R \in role^\pm(\mathcal{K}) \\ \exists a' \in ob(\mathcal{A})\ R(a, a') \in \mathcal{A}}} E_{q_{R,a}} R(a) \quad \wedge \bigwedge_{\neg P_k(a_i, a_j) \in \mathcal{A}} (\neg P_k(a_i, a_j))^\perp, \tag{15}$$

where $q_{R,a}$ is the maximum number in $Q_{\mathcal{T}}^R$ such that there are $q_{R,a}$ many distinct $a_i$ with $R(a, a_i) \in \mathcal{A}$ (here we use the UNA) and $(\neg P_k(a_i, a_j))^\perp = \bot$ if $P_k(a_i, a_j) \in \mathcal{A}$ and $\top$ otherwise. Now both the size of $\mathcal{A}^{\ddagger^2}$ and the size of $\mathcal{K}^\ddagger$ are linear in the size of $\mathcal{A}$ and $\mathcal{K}$, respectively, *no matter whether the numbers are coded in unary or in binary.*

More importantly, the translation $\cdot^{\ddagger}$ can actually be done in LOGSPACE. Indeed, this is trivially the case for $\mathcal{T}^*(x)$, $\varepsilon_R(x)$, $\delta_R(x)$, $\mathcal{A}^{\dagger^1}$ and the last conjunct of $\mathcal{A}^{\ddagger^2}$. As for the first conjunct of $\mathcal{A}^{\ddagger^2}$ then, for $R \in role^{\pm}(\mathcal{K})$ and $a \in ob(\mathcal{A})$, the maximum $q_{R,a}$ in $Q_{\mathcal{T}}^R$ such that there are $q_{R,a}$ many distinct $a_i$ with $R(a, a_i) \in \mathcal{A}$, can be computed using $\log \min(\max Q_{\mathcal{T}}^R, |ob(\mathcal{A})|) + \log |ob(\mathcal{A})|$ cells. Initially we set $q = 0$, and then enumerate all object names $a_i$ in $\mathcal{A}$ incrementing the current $q$ each time we find $R(a, a_i) \in \mathcal{A}$. We stop if $q = \max Q_{\mathcal{T}}^R$ or we reach the end of the object name list. The resulting $q_{R,a}$ is the maximum number in $Q_{\mathcal{T}}^R$ not exceeding $q$.

**Example 5.4** The translation $\mathcal{K}^{\ddagger}$ of the KB $\mathcal{K}$ from Example 5.1 looks as follows:

$$\mathcal{K}^{\ddagger} = \forall x \, \chi(x) \quad \wedge \quad A(a) \quad \wedge \quad E_1 P(a) \quad \wedge \quad E_1 P^-(a'),$$

where $\chi(x)$ is defined by (7).

**Corollary 5.5** *A DL-Lite$_{bool}^{\mathcal{N}}$ KB $\mathcal{K}$ is satisfiable iff the $\mathcal{QL}^1$-sentence $\mathcal{K}^{\ddagger}$ is satisfiable.*

**Proof** The claim follows from the fact that $\mathcal{K}^{\dagger}$ is satisfiable iff $\mathcal{K}^{\ddagger}$ is satisfiable. Indeed, if $\mathfrak{M} \models \mathcal{K}^{\dagger}$ then clearly $\mathfrak{M} \models \mathcal{K}^{\ddagger}$. Conversely, if $\mathfrak{M} \models \mathcal{K}^{\ddagger}$ then one can construct a new model $\mathfrak{M}'$ based on the same domain $D$ as $\mathfrak{M}$ by taking:

- $A_k^{\mathfrak{M}'} = A_k^{\mathfrak{M}}$, for all concept names $A_k$;

- $E_q R^{\mathfrak{M}'} = E_q R^{\mathfrak{M}}$, for all $R \in role^{\pm}(\mathcal{K})$ and $q \in Q_{\mathcal{T}}^R$;

- $(R a_i a_j)^{\mathfrak{M}'}$ is true iff $R(a_i, a_j) \in \mathcal{A}$;

- $a_i^{\mathfrak{M}'} = a_i^{\mathfrak{M}}$, for all $a_i \in ob(\mathcal{A})$;

- $dr^{\mathfrak{M}'} = dr^{\mathfrak{M}}$, for all $R \in role^{\pm}(\mathcal{K})$.

We claim that $\mathfrak{M}' \models \mathcal{K}^{\dagger}$. Indeed, $E_q R^{\mathfrak{M}'} = E_q R^{\mathfrak{M}}$, for every $R \in role^{\pm}(\mathcal{K})$ and $q \in Q_{\mathcal{T}}^R$. It follows then that $\mathfrak{M}' \models \forall x \, \mathcal{T}^*(x)$ and $\mathfrak{M}' \models \forall x \, \varepsilon_R(x)$. By definition, $\mathfrak{M} \models \mathcal{A}^{\dagger^1}$, $\mathfrak{M}' \models \mathcal{A}^{\dagger^2}$ and $\mathfrak{M}' \models \forall x \, \delta_R(x)$. It remains to show that $\mathfrak{M}' \models R^{\dagger}$. Suppose $\mathfrak{M}' \models \bigwedge_{i=1}^{q} R a a_{j_i}$, that is $R(a, a_{j_i}) \in \mathcal{A}$, for distinct $a_{j_1}, \ldots, a_{j_q}$, and $q \in Q_{\mathcal{T}}^R$. Clearly, we have $q \leq q_{R,a}$ and $\mathfrak{M} \models E_q R(a)$ and thus $\mathfrak{M}' \models E_q R(a)$. □

As an immediate consequence of Corollary 5.5, the facts that the translation $\cdot^{\ddagger}$ can be done in LOGSPACE, that the satisfiability problem for $\mathcal{QL}^1$-formulas is NP-complete and that *DL-Lite$_{bool}$* contains all the Booleans—and so can encode full propositional logic—we obtain the following result:

**Theorem 5.6** *Satisfiability of DL-Lite$_{bool}^{\mathcal{N}}$, DL-Lite$_{bool}^{\mathcal{F}}$ and DL-Lite$_{bool}$ knowledge bases is NP-complete for combined complexity.*

Observe now that if $\mathcal{K}$ is a *DL-Lite$_{krom}^{\mathcal{N}}$* KB then $\mathcal{K}^{\ddagger}$ is in the Krom fragment of $\mathcal{QL}^1$.

**Theorem 5.7** *Satisfiability of DL-Lite$_{\alpha}^{\mathcal{N}}$, DL-Lite$_{\alpha}^{\mathcal{F}}$ and DL-Lite$_{\alpha}$ knowledge bases, where $\alpha \in \{core, krom\}$, is NLOGSPACE-complete for combined complexity.*

**Proof** As the satisfiability problem for Krom formulas with the prefix of the form $\forall x$ (as in $\mathcal{K}^{\ddagger}$) is NLogSpace-complete (see, e.g., Börger et al., 1997, Exercise 8.3.7) and $\cdot^{\ddagger}$ is a LogSpace reduction, satisfiability is in NLogSpace for all the logics mentioned in the theorem. As for the lower bound, it suffices to recall that the NLogSpace-hardness for satisfiability of propositional Krom formulas is proved by reduction of the directed graph reachability problem using only 'core' propositional formulas (Börger et al., 1997), and so satisfiability in all of the above logics is NLogSpace-hard. ❑

If $\mathcal{K}$ is a $DL\text{-}Lite_{horn}^{\mathcal{N}}$ KB then $\mathcal{K}^{\ddagger}$ belongs to the universal Horn fragment of $\mathcal{QL}^1$.

**Theorem 5.8** *Satisfiability of $DL\text{-}Lite_{horn}^{\mathcal{N}}$, $DL\text{-}Lite_{horn}^{\mathcal{F}}$ and $DL\text{-}Lite_{horn}$ KBs is P-complete for combined complexity.*

**Proof** As $\mathcal{QL}^1$ contains no function symbols and $\mathcal{K}^{\ddagger}$ is universal, satisfiability of $\mathcal{K}^{\ddagger}$ is LogSpace-reducible to satisfiability of a set of propositional Horn formulas, namely, the formulas that are obtained from $\mathcal{K}^{\ddagger}$ by replacing $x$ with each of the constants occurring in $\mathcal{K}^{\ddagger}$. It remains to recall that the satisfiability problem for propositional Horn formulas is P-complete (see, e.g., Papadimitriou, 1994), which gives the required upper bound for $DL\text{-}Lite_{horn}^{\mathcal{N}}$ and lower bound for $DL\text{-}Lite_{horn}$. ❑

## 5.2 $DL\text{-}Lite_{core}^{\mathcal{HF}}$ is ExpTime-hard

Unfortunately, the translation $\cdot^{\ddagger}$ constructed in the previous section cannot be extended to logics of the form $DL\text{-}Lite_{\alpha}^{\mathcal{HN}}$ with *both* number restrictions *and* role inclusions. In this section we show that the satisfiability problem for $DL\text{-}Lite_{core}^{\mathcal{HF}}$ KBs is ExpTime-hard, which matches the upper bound for satisfiability of $DL\text{-}Lite_{bool}^{\mathcal{HN}}$ KBs even under binary coding of natural numbers (Tobies, 2001).

Note first that, although intersection is not allowed on the left-hand side of $DL\text{-}Lite_{core}^{\mathcal{HF}}$ concept inclusions, in certain cases (when the right-hand side is consistent) we can 'simulate' it by using role inclusions and functionality constraints. Suppose that a knowledge base $\mathcal{K}$ contains a concept inclusion of the form $C_1 \sqcap C_2 \sqsubseteq C$. Define a new KB $\mathcal{K}'$ by replacing this axiom in $\mathcal{K}$ with the following set of new axioms, where $R_1, R_2, R_3, R_{12}, R_{23}$ are fresh role names:

$$C_1 \ \sqsubseteq \ \exists R_1 \qquad\qquad C_2 \ \sqsubseteq \ \exists R_2, \qquad (16)$$
$$R_1 \ \sqsubseteq \ R_{12}, \qquad\qquad R_2 \ \sqsubseteq \ R_{12}, \qquad (17)$$
$$\geq 2\,R_{12} \ \sqsubseteq \ \bot, \qquad\qquad\qquad\qquad\qquad\qquad (18)$$
$$\exists R_1^- \ \sqsubseteq \ \exists R_3^-, \qquad\qquad\qquad\qquad\qquad\qquad (19)$$
$$\exists R_3 \ \sqsubseteq \ C, \qquad\qquad\qquad\qquad\qquad\qquad (20)$$
$$R_3 \ \sqsubseteq \ R_{23}, \qquad\qquad R_2 \ \sqsubseteq \ R_{23}, \qquad (21)$$
$$\geq 2\,R_{23}^- \ \sqsubseteq \ \bot. \qquad\qquad\qquad\qquad\qquad\qquad (22)$$

**Lemma 5.9** (i) *If $\mathcal{I} \models \mathcal{K}'$ then $\mathcal{I} \models \mathcal{K}$, for every interpretation $\mathcal{I}$.*

(ii) *If $\mathcal{I} \models \mathcal{K}$ and $C^{\mathcal{I}} \neq \emptyset$ then there is a model $\mathcal{I}'$ of $\mathcal{K}'$ which has the same domain as $\mathcal{I}$ and agrees with it on every symbol from $\mathcal{K}$.*

**Proof** (i) Suppose that $\mathcal{I} \models \mathcal{K}'$ and $x \in C_1^{\mathcal{I}} \cap C_2^{\mathcal{I}}$. By (16), there is $y$ with $(x,y) \in R_1^{\mathcal{I}}$, and so $y \in (\exists R_1^-)^{\mathcal{I}}$, and there is $z$ with $(x,z) \in R_2^{\mathcal{I}}$. By (17), $\{(x,y),(x,z)\} \subseteq R_{12}^{\mathcal{I}}$, whence $y = z$ in view of (18). By (19), $y \in (\exists R_3^-)^{\mathcal{I}}$ and hence there is $u$ with $(u,y) \in R_3^{\mathcal{I}}$ and $u \in (\exists R_3)^{\mathcal{I}}$. By (20), $u \in C^{\mathcal{I}}$. By (21), $(u,y) \in R_{23}^{\mathcal{I}}$ and $(x,y) \in R_{23}^{\mathcal{I}}$. Finally, it follows from (22) that $u = x$, and so $x \in C^{\mathcal{I}}$. Thus, $\mathcal{I} \models \mathcal{K}$.

(ii) Take some point $c \in C^{\mathcal{I}}$ and define an extension $\mathcal{I}'$ of $\mathcal{I}$ to the new role names by setting:

- $R_1^{\mathcal{I}'} = \{(x,x) \mid x \in C_1^{\mathcal{I}}\}$,

- $R_2^{\mathcal{I}'} = \{(x,x) \mid x \in C_2^{\mathcal{I}}\}$,

- $R_3^{\mathcal{I}'} = \{(x,x) \mid x \in (C_1 \sqcap C_2)^{\mathcal{I}}\} \cup \{(c,x) \mid x \in (C_1 \sqcap \neg C_2)^{\mathcal{I}}\}$,

- $R_{12}^{\mathcal{I}'} = R_1^{\mathcal{I}'} \cup R_2^{\mathcal{I}'}$ $\qquad$ and $\qquad$ $R_{23}^{\mathcal{I}'} = R_2^{\mathcal{I}'} \cup R_3^{\mathcal{I}'}$.

It is readily seen that $\mathcal{I}'$ satisfies all the axioms (16)–(22), and so $\mathcal{I}' \models \mathcal{K}'$. $\qquad\square$

We are now in a position to prove the following:

**Theorem 5.10** *Satisfiability of DL-Lite$_{core}^{\mathcal{HF}}$ KBs is* EXPTIME*-hard for combined complexity (with or without the UNA).*

**Proof** We will prove this theorem in two steps. First we consider the logic *DL-Lite$_{horn}^{\mathcal{HF}}$* and show how to encode the behavior of polynomial-space-bounded *alternating Turing machines* (ATMs, for short) by means of *DL-Lite$_{horn}^{\mathcal{HF}}$* KBs. As APSPACE = EXPTIME, where APSPACE is the class of problems recognized by polynomial-space-bounded ATMs (see, e.g., Kozen, 2006), this will establish EXPTIME-hardness of satisfiability for *DL-Lite$_{horn}^{\mathcal{HF}}$*. Then, using Lemma 5.9, we will show how to get rid of conjunctions on the left-hand side of the concept inclusions involved in this encoding of ATMs and thus establish EXPTIME-hardness of *DL-Lite$_{core}^{\mathcal{HF}}$*.

Without loss of generality, we can consider only ATMs $\mathcal{M}$ with *binary* computational trees. This means that, for every non-halting state $q$ and every symbol $a$ from the tape alphabet, $\mathcal{M}$ has precisely two instructions of the form

$$(q,a) \rightsquigarrow_{\mathcal{M}}^0 (q',a',d') \qquad \text{and} \qquad (q,a) \rightsquigarrow_{\mathcal{M}}^1 (q'',a'',d''), \tag{23}$$

where $d',d'' \in \{\rightarrow, \leftarrow\}$ and $\rightarrow$ (resp., $\leftarrow$) means 'move the head right (resp., left) one cell'. We remind the reader that each non-halting state of $\mathcal{M}$ is either an *and-state* or an *or-state*.

Given such an ATM $\mathcal{M}$, a polynomial function $p(n)$ such that every run of $\mathcal{M}$ on every input of length $n$ does not use more than $p(n)$ tape cells, and an input word $\vec{a} = a_1, \ldots, a_n$, we construct a *DL-Lite$_{horn}^{\mathcal{HF}}$* knowledge base $\mathcal{K}_{\mathcal{M},\vec{a}}$ with the following properties: (i) the size of $\mathcal{K}_{\mathcal{M},\vec{a}}$ is polynomial in the size of $\mathcal{M}$, $\vec{a}$, and (ii) $\mathcal{M}$ accepts $\vec{a}$ iff $\mathcal{K}_{\mathcal{M},\vec{a}}$ is not satisfiable. Denote by $Q$ the set of states and by $\Sigma$ the tape alphabet of $\mathcal{M}$.

To encode the instructions of $\mathcal{M}$, we need the following roles:

- $S_q, S_q^0, S_q^1$, for each $q \in Q$: informally, $x \in (\exists S_q^-)^{\mathcal{I}}$, for some interpretation $\mathcal{I}$, means that $x$ represents a configuration of $\mathcal{M}$ with the state $q$, and $x \in (\exists S_q^k)^{\mathcal{I}}$ means that the next state, according to the transition $\rightsquigarrow_{\mathcal{M}}^k$, is $q$, where $k \in \{0,1\}$;

- $H_i, H_i^0, H_i^1$, for each $i \leq p(n)$: $x \in (\exists H_i^-)^{\mathcal{I}}$ means that $x$ represents a configuration of $\mathcal{M}$ where the head scans the $i$th cell, and $x \in (\exists H_i^k)^{\mathcal{I}}$ that, according to the transition $\rightsquigarrow_{\mathcal{M}}^k$, $k \in \{0,1\}$, in the next configuration the head scans the $i$th cell;

- $C_{ia}, C_{ia}^0, C_{ia}^1$, for each $i \leq p(n)$ and each $a \in \Sigma$: $x \in (\exists C_{ia}^-)^{\mathcal{I}}$ means that $x$ represents a configuration of $\mathcal{M}$ where the $i$th cell contains $a$, and $x \in (\exists C_{ia}^k)^{\mathcal{I}}$ that, according to $\rightsquigarrow_{\mathcal{M}}^k$, $k \in \{0,1\}$, in the next configuration the $i$th cell contains $a$.

This intended meaning can be encoded using the following concept inclusions: for every instruction $(q,a) \rightsquigarrow_{\mathcal{M}}^k (q',a',\rightarrow)$ of $\mathcal{M}$ and every $i < p(n)$,

$$\exists S_q^- \sqcap \exists H_i^- \sqcap \exists C_{ia}^- \ \sqsubseteq \ \exists H_{i+1}^k \sqcap \exists S_{q'}^k \sqcap \exists C_{ia'}^k, \tag{24}$$

and for every instruction $(q,a) \rightsquigarrow_{\mathcal{M}}^k (q',a',\leftarrow)$ of $\mathcal{M}$ and every $i$, $1 < i \leq p(n)$,

$$\exists S_q^- \sqcap \exists H_i^- \sqcap \exists C_{ia}^- \ \sqsubseteq \ \exists H_{i-1}^k \sqcap \exists S_{q'}^k \sqcap \exists C_{ia'}^k. \tag{25}$$

To preserve the symbols on the tape that are not in the active cell, we use the following concept inclusions, for $k \in \{0,1\}$, $i,j \leq p(n)$ with $j \neq i$, and $a \in \Sigma$:

$$\exists H_j^- \sqcap \exists C_{ia}^- \ \sqsubseteq \ \exists C_{ia}^k. \tag{26}$$

To 'synchronize' our roles, we need two more (functional) roles $T_k$ and a number of role inclusions to be added to the TBox: for all $k \in \{0,1\}$, $i \leq p(n)$, $q \in Q$, and $a \in \Sigma$,

$$C_{ia}^k \ \sqsubseteq \ C_{ia}, \qquad\qquad H_i^k \ \sqsubseteq \ H_i, \qquad\qquad S_q^k \ \sqsubseteq \ S_q, \tag{27}$$

$$C_{ia}^k \ \sqsubseteq \ T_k, \qquad\qquad H_i^k \ \sqsubseteq \ T_k, \qquad\qquad S_q^k \ \sqsubseteq \ T_k, \tag{28}$$

$$\geq 2\, T_k \ \sqsubseteq \ \bot. \tag{29}$$

It remains to encode the acceptance conditions for $\mathcal{M}$ on $\vec{a}$. This can be done with the help of the role names $Y_k$, for $k \in \{0,1\}$, and the concept name $A$:

$$\exists S_q^- \ \sqsubseteq \ A, \quad q \text{ an accepting state}, \tag{30}$$

$$Y_k \ \sqsubseteq \ T_k, \tag{31}$$

$$\geq 2\, T_k^- \ \sqsubseteq \ \bot, \tag{32}$$

$$\exists T_k^- \sqcap A \ \sqsubseteq \ \exists Y_k^-, \tag{33}$$

$$\exists S_q^- \sqcap \exists Y_k \ \sqsubseteq \ A, \quad q \text{ an or-state}, \tag{34}$$

$$\exists S_q^- \sqcap \exists Y_0 \sqcap \exists Y_1 \ \sqsubseteq \ A, \quad q \text{ an and-state}. \tag{35}$$

The TBox $\mathcal{T}$ of the $DL\text{-}Lite_{horn}^{\mathcal{HF}}$ knowledge base $\mathcal{K}_{\mathcal{M},\vec{a}}$ we are constructing consists of the axioms (24)–(35) together with the auxiliary axiom

$$A \sqcap D \ \sqsubseteq \ \bot, \tag{36}$$

where $D$ is a fresh concept name. The ABox $\mathcal{A}$ of $\mathcal{K}_{\mathcal{M},\vec{a}}$ is comprised of the following assertions, for some object names $s$ and $u$:

$$S_{q_0}(u,s), \quad q_0 \text{ the initial state}, \tag{37}$$

$$H_1(u,s), \tag{38}$$

$$C_{ia_i}(u,s), \quad i \leq p(n),\, a_i \text{ the } i\text{th symbol on the input tape}, \tag{39}$$

$$D(s). \tag{40}$$

Clearly, $\mathcal{K}_{\mathcal{M},\vec{a}} = (\mathcal{T}, \mathcal{A})$ is a $DL\text{-}Lite_{horn}^{\mathcal{HF}}$ KB and its size is polynomial in the size of $\mathcal{M}$, $\vec{a}$.

**Lemma 5.11** *The ATM $\mathcal{M}$ accepts $\vec{a}$ iff the KB $\mathcal{K}_{\mathcal{M},\vec{a}}$ is not satisfiable.*

**Proof** ($\Rightarrow$) Suppose that $\mathcal{M}$ accepts $\vec{a}$ but $\mathcal{I} \models \mathcal{K}_{\mathcal{M},\vec{a}}$ for some interpretation $\mathcal{I}$. Then we can reconstruct the full computation tree for $\mathcal{M}$ on $\vec{a}$ by induction in the following way.

Let the root of the tree be the point $s^{\mathcal{I}}$. By (37)–(39), $s$ represents the initial configuration of $\mathcal{M}$ on $\vec{a}$ in accordance with the intended meaning of the roles $S_{q_0}$, $H_1$ and $C_{ia_i}$ explained above (it does not matter if, for instance, we also have $s^{\mathcal{I}} \in (\exists H_5^-)^{\mathcal{I}}$).

Assume now that we have already found a point $x \in \Delta^{\mathcal{I}}$ representing some configuration

$$\mathfrak{c} = b_1, \ldots, b_{i-1}, (q, b_i), b_{i+1}, \ldots, b_{p(n)}, \tag{41}$$

where $q$ is the current non-halting state and the head scans the $i$th cell containing $b_i$. This means that we have

$$x \in (\exists S_q^-)^{\mathcal{I}} \ \cap \ (\exists H_i^-)^{\mathcal{I}} \quad \text{and} \quad x \in (\exists C_{jb_j}^-)^{\mathcal{I}}, \quad \text{for all } j \leq p(n).$$

Assume also that $\mathcal{M}$ contains two instructions of the form (23) for $(q, b_i)$, that is $q$ is non-halting. If we have $(q, b_i) \leadsto_{\mathcal{M}}^k (q', b', \rightarrow)$, for $k = 0$ or 1, then, by (24) and (26), there are points $y_s$, $y_h$ and $y_j$, for $j \leq p(n)$, in $\Delta^{\mathcal{I}}$ such that

$$(x, y_s) \in (S_{q'}^k)^{\mathcal{I}}, \quad (x, y_h) \in (H_{i+1}^k)^{\mathcal{I}}, \quad (x, y_i) \in (C_{ib'}^k)^{\mathcal{I}}, \quad (x, y_j) \in (C_{jb_j}^k)^{\mathcal{I}}, \quad \text{for } j \neq i.$$

By (28)–(29), $S_{q'}^0$, $H_{i+1}^0$, $C_{ib'_i}^0$ and the $C_{jb_j}^0$, $j \neq i$, are all sub-roles of the functional role $T_k$, and so all the points $y_s$, $y_h$ and $y_j$ coincide; we denote this point by $x_k$. By (27), we then have:

$$(x, x_k) \in T_k^{\mathcal{I}}, \quad x_k \in (\exists S_{q'}^-)^{\mathcal{I}} \ \cap \ (\exists H_{i+1}^-)^{\mathcal{I}} \ \cap \ (\exists C_{ib'}^-)^{\mathcal{I}} \quad \text{and} \quad x_k \in (\exists C_{jb_j}^-)^{\mathcal{I}}, \quad \text{for } j \neq i.$$

Similarly, if we have $(q, b_i) \leadsto_{\mathcal{M}}^k (q'', b'', \leftarrow)$, for $k = 0$ or 1, then, by (25) and (26), there is a point $x_k \in \Delta^{\mathcal{I}}$ such that

$$(x, x_k) \in T_k^{\mathcal{I}}, \quad x_k \in (\exists S_{q''}^-)^{\mathcal{I}} \ \cap \ (\exists H_{i-1}^-)^{\mathcal{I}} \ \cap \ (\exists C_{ib''}^-)^{\mathcal{I}} \quad \text{and} \quad x_k \in (\exists C_{jb_j}^-)^{\mathcal{I}}, \quad \text{for } j \neq i.$$

Thus, for $k = 0, 1$, $x_k$ is a $T_k$-successor of $x$ representing the configuration $\mathfrak{c}_k$ of $\mathcal{M}$ after it has executed $(q, b_i) \leadsto_{\mathcal{M}}^k (q'', b'', d)$; in this case $\mathfrak{c}_k$ is called the $k$-successor of $\mathfrak{c}$.

According to (30), every point in the constructed computation tree for $\mathcal{M}$ on $\vec{a}$ representing a configuration with an accepting state is in $A^{\mathcal{I}}$. Suppose now, inductively, that $x$ represents some configuration $\mathfrak{c}$ of the form (41), $q$ is an or-state, $x_k$ represents the $k$-successor of $\mathfrak{c}$ and $(x, x_k) \in T_k^{\mathcal{I}}$, for $k = 0, 1$, and one of the $x_k$, say $x_0$, is in $A^{\mathcal{I}}$. In view of (33), we have $x_0 \in (\exists Y_0^-)^{\mathcal{I}}$. As $T_0^-$ is functional by (32) and $Y_0$ is a sub-role of $T_0$ by (31), $(x, x_0) \in Y_0^{\mathcal{I}}$, and so, by (34), $x \in A^{\mathcal{I}}$. The case of $x$ being an and-state is considered analogously with the help of (35).

Since $\mathcal{M}$ accepts $\vec{a}$, we then conclude that $s^{\mathcal{I}} \in A^{\mathcal{I}}$, contrary to (36) and (40).

($\Leftarrow$) Conversely, suppose now that $\mathcal{M}$ does not accept $\vec{a}$. Consider the full computation tree $(\Delta, <_0 \cup <_1)$ with nodes labeled with configurations of $\mathcal{M}$ in such a way that the root is labeled with the initial configuration

$$(q_0, a_1), a_2, \ldots, a_n, a_{n+1}, \ldots, a_{p(n)},$$

(where the $a_i$, for $n+1 \leq i \leq p(n)$, are all 'blank'), and if some node $x$ in the tree is labeled with a non-halting $\mathfrak{c}$ of the form (41) and $\mathcal{M}$ contains two instructions of the form (23), then $x$ has one $<_0$-successor labeled with the 0-successor of $\mathfrak{c}$ and one $<_1$-successor labeled with the 1-successor of $\mathfrak{c}$. (It should be emphasized that $(\Delta, <_0 \cup <_1)$ is a *tree*, where different nodes may be labeled with the same configuration.)

We use this tree to construct an interpretation $\mathcal{I} = (\Delta^{\mathcal{I}}, \cdot^{\mathcal{I}})$ as follows:

- $\Delta^{\mathcal{I}} = \Delta \cup \{u\}$, for some $u \notin \Delta$;

- $s^{\mathcal{I}}$ is the root of $\Delta$ and $u^{\mathcal{I}} = u$;

- $D^{\mathcal{I}} = \{s^{\mathcal{I}}\}$;

- $(x, x_k) \in (S_{q'}^k)^{\mathcal{I}}$, $(x, x_k) \in (H_{i+1}^k)^{\mathcal{I}}$, $(x, x_k) \in (C_{ib'}^k)^{\mathcal{I}}$, and $(x, x_k) \in (C_{jb_j}^k)^{\mathcal{I}}$, for $j \neq i$, iff $x$ is labeled with $\mathfrak{c}$ of the form (41), $(q, b_i) \leadsto_{\mathcal{M}}^k (q', b', \rightarrow)$ and $x <_k x_k$, for $k = 0, 1$;

- $(x, x_k) \in (S_{q'}^k)^{\mathcal{I}}$, $(x, x_k) \in (H_{i-1}^k)^{\mathcal{I}}$, $(x, x_k) \in (C_{ib'}^k)^{\mathcal{I}}$, and $(x, x_k) \in (C_{jb_j}^k)^{\mathcal{I}}$, for $j \neq i$, iff $x$ is labeled with $\mathfrak{c}$ of the form (41), $(q, b_i) \leadsto_{\mathcal{M}}^k (q', b', \leftarrow)$ and $x <_k x_k$, for $k = 0, 1$;

- $(u, s^{\mathcal{I}}) \in (S_{q_0})^{\mathcal{I}}$, $(u, s^{\mathcal{I}}) \in (H_1)^{\mathcal{I}}$, $(u, s^{\mathcal{I}}) \in (C_{ia_i})^{\mathcal{I}}$, $i \leq p(n)$ and over $\Delta$ the extensions for the roles $S_q$, $H_i$ and $C_{ia}$ are defined according to (27);

- $T_k^{\mathcal{I}} = <_k$, for $k = 0, 1$;

- $Y_0^{\mathcal{I}}$, $Y_1^{\mathcal{I}}$ and $A^{\mathcal{I}}$ are defined inductively:

  - *Induction basis*: if $x \in \Delta$ is labeled with an accepting configuration, then $x \in A^{\mathcal{I}}$.
  - *Induction step*: (i) if $x <_k x_k$, for $k = 0, 1$, and $x_k \in A^{\mathcal{I}}$, then $(x, x_k) \in Y_k^{\mathcal{I}}$; (ii) if $x$ is an or-state (respectively, and-state) and $(x, x_k) \in Y_k^{\mathcal{I}}$ for some (respectively, all) $k \in \{0, 1\}$, then $x \in A^{\mathcal{I}}$.

It follows from the given definition that $\mathcal{I} \models \mathcal{K}_{\mathcal{M}, \vec{a}}$. Details are left to the reader. □

The lemma we have just proved establishes that satisfiability of $DL\text{-}Lite_{horn}^{\mathcal{HF}}$ KBs is ExpTime-hard. Our next aim is to show how one can eliminate the conjunctions in the left-hand side of the TBox axioms (24)–(26), (33)–(35). We will do this with the help of Lemma 5.9. Before applying it, we check first that if $\mathcal{K}_{\mathcal{M}, \vec{a}}$ is satisfiable then it is satisfiable in an interpretation $\mathcal{I}$ such that $\mathcal{I} \models \mathcal{K}_{\mathcal{M}, \vec{a}}$ and $C^{\mathcal{I}} \neq \emptyset$, for every $C$ occurring in an axiom of the form $C_1 \sqcap C_2 \sqsubseteq C$ in $\mathcal{K}$. Consider, for instance, axiom (24) and assume that $\mathcal{I} \models \mathcal{K}_{\mathcal{M}, \vec{a}}$, but $(\exists S_{q'}^k)^{\mathcal{I}} = \emptyset$. Then, we can construct a new interpretation $\mathcal{I}'$ by adding two new points, say $x$ and $y$, to the domain of $\mathcal{I}$, and setting $(x, y) \in (S_{q'}^k)^{\mathcal{I}'}$, $(x, y) \in (S_{q'})^{\mathcal{I}'}$, $(x, y) \in (T_k)^{\mathcal{I}'}$. Furthermore, if $q'$ is an accepting state, we also set $y \in A^{\mathcal{I}'}$ and $(x, y) \in Y_k^{\mathcal{I}'}$. One can readily check that $\mathcal{I}'$ is still a model for $\mathcal{K}_{\mathcal{M}, \vec{a}}$. The other conjuncts of (24) and the remaining axioms are considered analogously.

After an application of Lemma 5.9 to an axiom of the form $C_1 \sqcap C_2 \sqsubseteq C$ with $C_2 = C_2' \sqcap C_2''$ we obtain, by (16)–(22), a new KB $\mathcal{K}'$ with the concept inclusion of the form $C_2' \sqcap C_2'' \sqsubseteq \exists R_1$, which also requires treatment by means of the same lemma. To be able to do this, we again have to check that $\mathcal{K}'$ is satisfiable in some interpretation $\mathcal{I}''$ with $(\exists R_1)^{\mathcal{I}''} \neq \emptyset$. Suppose that $\mathcal{I}' \models \mathcal{K}'$ and $(\exists R_1)^{\mathcal{I}'} = \emptyset$. Then we can construct $\mathcal{I}''$ by adding two new points, say $x$ and $y$, to the domain of $\mathcal{I}'$, adding $x$ to $C^{\mathcal{I}'}$ and $(x, y)$ to each of $R_1^{\mathcal{I}'}$, $R_{12}^{\mathcal{I}'}$, $R_{23}^{\mathcal{I}'}$ and $R_3^{\mathcal{I}'}$. It is readily seen that $\mathcal{I}'' \models \mathcal{K}'$.

It is to be noted that the proof above does not depend on whether the UNA is adopted or not. ❑

As an immediate consequence we obtain:

**Corollary 5.12** *Satisfiability of DL-Lite$_\alpha^{\mathcal{HF}}$ and DL-Lite$_\alpha^{\mathcal{HN}}$ KBs with or without the UNA is* ExpTime-*complete for combined complexity, where $\alpha \in \{ core, krom, horn, bool \}$.*

### 5.3 Reconciling Number Restrictions and Role Inclusions

As we have seen in the previous section, the unrestricted interaction between number restrictions and role inclusions allowed in the logics of the form *DL-Lite$_\alpha^{\mathcal{HN}}$* results in high combined complexity of satisfiability. In Section 6.2, we shall see that the data complexity of instance checking and query answering also becomes unacceptably high for these logics. A quick look at the proof of Theorem 5.10 reveals the 'culprit:' the interplay between role inclusions $R_1 \sqsubseteq R$, $R_2 \sqsubseteq R$ and functionality constraints $\geq 2\,R \sqsubseteq \bot$, which effectively mean that if $R_1(x, y)$ and $R_2(x, z)$ then $y = z$. In this section we study the case when such an interplay is not allowed.

Recall from Section 2.1 that *DL-Lite$_\alpha^{(\mathcal{HN})}$* TBoxes $\mathcal{T}$, for $\alpha \in \{ core, krom, horn, bool \}$, satisfy the following conditions:

**(A$_1$)** $\mathcal{T}$ may contain only *positive occurrences* of qualified number restrictions $\geq q\,R.C$, where $C$ is a conjunction of concepts allowed on the right-hand side of $\alpha$-concept inclusions;

**(A$_2$)** if $\geq q\,R.C$ occurs in $\mathcal{T}$, then $\mathcal{T}$ does not contain *negative occurrences* of number restrictions $\geq q'\,R$ or $\geq q'\,inv(R)$ with $q' \geq 2$;

**(A$_3$)** if $R$ has a proper sub-role in $\mathcal{T}$, then $\mathcal{T}$ does not contain *negative occurrences* of $\geq q\,R$ or $\geq q\,inv(R)$ with $q \geq 2$.

*DL-Lite$_\alpha^{(\mathcal{HN})}$* TBoxes can contain role constraints such as $\mathsf{Dis}(R_1, R_2)$, $\mathsf{Asym}(P_k)$, $\mathsf{Sym}(P_k)$, $\mathsf{Irr}(P_k)$, and $\mathsf{Ref}(P_k)$.

Our main aim in this section is to prove the following theorem and develop the technical tools we need to investigate the data complexity of reasoning with *DL-Lite$_{bool}^{(\mathcal{HN})}$* and its sublogics later on in the paper.

**Theorem 5.13** *For combined complexity,* (i) *satisfiability of DL-Lite$_{bool}^{(\mathcal{HN})}$ KBs is* NP-*complete;* (ii) *satisfiability of DL-Lite$_{horn}^{(\mathcal{HN})}$ KBs is* P-*complete; and* (iii) *satisfiability of DL-Lite$_{krom}^{(\mathcal{HN})}$ and DL-Lite$_{core}^{(\mathcal{HN})}$ KBs is* NLogSpace-*complete.*

Let us consider first the sub-language of $DL\text{-}Lite_{bool}^{(\mathcal{HN})}$ without qualified number restrictions and the role constraints mentioned above; we denote it by $DL\text{-}Lite_{bool}^{(\mathcal{HN})^-}$. This sub-language is required for purely technical reasons. In Section 7, we will also use $DL\text{-}Lite_{horn}^{(\mathcal{HN})^-}$, but we do *not* need the *core* or *Krom* fragments.

Suppose we are given a $DL\text{-}Lite_{bool}^{(\mathcal{HN})^-}$ KB $\mathcal{K} = (\mathcal{T}, \mathcal{A})$. Let $Id$ be a distinguished role name. We will use it to simulate the *identity relation* required for encoding the role constraints. We assume that either $\mathcal{K}$ does not contain $Id$ at all or satisfies the following conditions:

**(Id$_1$)** $Id(a_i, a_j) \in \mathcal{A}$ iff $i = j$, for all $a_i, a_j \in ob(\mathcal{A})$,

**(Id$_2$)** $\{\top \sqsubseteq \exists Id, \;\; Id^- \sqsubseteq Id\} \subseteq \mathcal{T}$, and $Q_{\mathcal{T}}^{Id} = Q_{\mathcal{T}}^{Id^-} = \{1\}$,

**(Id$_3$)** $Id$ is only allowed in role inclusions of the form $Id^- \sqsubseteq Id$ and $Id \sqsubseteq R$.

In what follows, without loss of generality, we will assume that

**(Q)** $Q_{\mathcal{T}}^R \subseteq Q_{\mathcal{T}}^{R'}$ whenever $R \sqsubseteq_{\mathcal{T}}^* R'$

(for if this is not the case we can always add the missing numbers to $Q_{\mathcal{T}}^{R'}$, e.g., by introducing fictitious concept inclusions of the form $\bot \sqsubseteq \geq q\, R'$).

Now, in the same way as in Section 5.1, we define two translations $\cdot^{\dagger\mathbf{e}}$ and $\cdot^{\ddagger\mathbf{e}}$ of $\mathcal{K}$ into the one-variable fragment $\mathcal{QL}^1$ of first-order logic. The former translation, $\cdot^{\dagger\mathbf{e}}$, retains the information about the relationships between ABox objects, and we show that every model of $\mathcal{K}^{\dagger\mathbf{e}}$ can again be 'unraveled' into a model of $\mathcal{K}$. We define $\cdot^{\dagger\mathbf{e}}$ by taking:

$$\mathcal{K}^{\dagger\mathbf{e}} = \forall x \left[ \mathcal{T}^*(x) \;\wedge\; \mathcal{T}^{\mathcal{R}}(x) \;\wedge\; \bigwedge_{R \in role^\pm(\mathcal{K})} \big(\varepsilon_R(x) \wedge \delta_R(x)\big) \right] \wedge$$
$$\left[ \mathcal{A}^{\dagger 1} \;\wedge\; \mathcal{A}^{\dagger 2} \;\wedge\; \bigwedge_{R \in role^\pm(\mathcal{K})} R^\dagger \;\wedge\; \bigwedge_{\substack{R \sqsubseteq R' \in \mathcal{T} \\ a_i, a_j \in ob(\mathcal{A})}} \big(R a_i a_j \to R' a_i a_j\big) \right],$$

where $\mathcal{T}^*(x)$, $\mathcal{A}^{\dagger 1}$, $\mathcal{A}^{\dagger 2}$, $\varepsilon_R(x)$, $\delta_R(x)$ and $R^\dagger$ are as in (1)–(6) and

$$\mathcal{T}^{\mathcal{R}}(x) \;=\; \bigwedge_{\substack{R \sqsubseteq R' \in \mathcal{T} \text{ or} \\ inv(R) \sqsubseteq inv(R') \in \mathcal{T}}} \; \bigwedge_{q \in Q_{\mathcal{T}}^R} \big(E_q R(x) \to E_q R'(x)\big). \tag{42}$$

The following lemma is an analogue of Theorem 5.2:

**Lemma 5.14** *A $DL\text{-}Lite_{bool}^{(\mathcal{HN})^-}$ KB $\mathcal{K}$ is satisfiable iff the $\mathcal{QL}^1$-sentence $\mathcal{K}^{\dagger\mathbf{e}}$ is satisfiable.*

**Proof** The proof basically follows the lines of the proof of Theorem 5.2 with some modifications. We present a modified unraveling construction here; the converse direction is exactly the same as in Theorem 5.2.

In each equivalence class $[R_i] = \{R_j \mid R_i \equiv_{\mathcal{T}}^* R_j\}$ we select a single role (a representative of that class) and denote it by $rep_{\mathcal{T}}^*(R_i)$. When extending $P_k^m$ to $P_k^{m+1}$, we use the following modified 'curing' rules:

$(\Lambda_k^m)$ If $P_k \neq rep_{\mathcal{T}}^*(P_k)$ do nothing: the defects are cured for $rep_{\mathcal{T}}^*(P_k)$. Otherwise, let $w \in \Lambda_k^m$, $q = r(P_k, cp(w)) - r_m(P_k, w)$ and $d = cp(w)$. We have $\mathfrak{M} \models E_{q'} P_k[d]$ for some $q' \geq q > 0$. Then, by (5), $\mathfrak{M} \models E_1 P_k[d]$ and, by (4), $\mathfrak{M} \models E_1 P_k^-[dp_k^-]$. In this case we take $q$ *fresh* copies $w_1', \ldots, w_q'$ of $dp_k^-$ (and set $cp(w_i') = dp_k^-$, for $1 \leq i \leq q$), add them to $W_{m+1}$ and

- add the pairs $(w, w_i')$, $1 \leq i \leq q$, to each $P_j^{m+1}$ with $P_k \sqsubseteq_{\mathcal{T}}^* P_j$ (including $P_j = P_k$);
- add the pairs $(w_i', w)$, $1 \leq i \leq q$, to each $P_j^{m+1}$ with $P_k^- \sqsubseteq_{\mathcal{T}}^* P_j$;
- if $Id$ occurs in $\mathcal{K}$, add the pairs $(w_i', w_i')$, $1 \leq i \leq q$, to each $P_j^{m+1}$ with $Id \sqsubseteq_{\mathcal{T}}^* P_j$.

$(\Lambda_k^{m-})$ This rule is the mirror image of $(\Lambda_k^m)$: $P_k$ and $dp_k^-$ are replaced everywhere with $P_k^-$ and $dp_k$, respectively; see the proof of Theorem 5.2.

It follows from this definition that $Id$ never has any defects and is interpreted in the resulting interpretation $\mathcal{I}$ by the identity relation $Id^{\mathcal{I}} = \{(w, w) \mid w \in \Delta^{\mathcal{I}}\}$; the interpretations of roles respect all the role inclusions, i.e., $R_1^{\mathcal{I}} \subseteq R_2^{\mathcal{I}}$ whenever $R_1 \sqsubseteq_{\mathcal{T}}^* R_2$.

It remains to show that the constructed interpretation $\mathcal{I}$ is indeed a model of $\mathcal{K}$. First, (11) trivially holds for $Id$ as both the required and actual ranks are equal to 1. Second, (11) holds for $R$ such that $R \neq Id$ and $R$ has no proper sub-roles: the proof is exactly the same as in Theorem 5.2, taking into account that we cure defects only for a single role in each equivalence class and that, by (42), for all $R' \in [R]$, we have $r(R', cp(w)) = r(R, cp(w))$ and $r(inv(R), cp(w)) = r(inv(R'), cp(w))$. It follows that (13) holds for $Id$ and any role $R$ without proper sub-roles. However, (13) does not necessarily hold for roles $R$ with proper sub-roles: as follows from the construction, the actual rank may be greater than the required rank, in which case we only have the following:

$$\text{if} \quad \mathfrak{M} \models E_q R[cp(w)] \qquad \text{then} \qquad w \in (\geq q\,R)^{\mathcal{I}}.$$

However, this is enough for our purposes. By induction on the structure of concepts and using $(\mathbf{A_3})$, one can show that $\mathcal{I} \models C_1 \sqsubseteq C_2$ whenever $\mathfrak{M} \models \forall x\,(C_1^*(x) \to C_2^*(x))$, for each concept inclusion $C_1 \sqsubseteq C_2 \in \mathcal{T}$, and therefore, $\mathcal{I} \models \mathcal{T}$. We also have $\mathcal{I} \models \mathcal{A}$ (see the proof of Theorem 5.2) and thus $\mathcal{I} \models \mathcal{K}$. ❑

**Remark 5.15** It follows from the proofs of Theorem 5.2 and Lemma 5.14 that, for the $DL\text{-}Lite_{bool}^{(\mathcal{HN})^-}$ KB $\mathcal{K} = (\mathcal{T}, \mathcal{A})$, every model $\mathfrak{M}$ of $\mathcal{K}^{\ddagger e}$ induces a model $\mathcal{I}_{\mathfrak{M}}$ of $\mathcal{K}$ with the following properties:

**(ABox)** For all $a_i, a_j \in ob(\mathcal{A})$, we have $(a_i^{\mathcal{I}_{\mathfrak{M}}}, a_j^{\mathcal{I}_{\mathfrak{M}}}) \in R^{\mathcal{I}_{\mathfrak{M}}}$ iff $R(a_i, a_j) \in \mathsf{Cl}_{\mathcal{T}}^{\mathbf{e}}(\mathcal{A})$, where

$$\mathsf{Cl}_{\mathcal{T}}^{\mathbf{e}}(\mathcal{A}) = \{R_2(a_i, a_j) \mid R_1(a_i, a_j) \in \mathcal{A},\ R_1 \sqsubseteq_{\mathcal{T}}^* R_2\}.$$

**(forest)** The object names $a \in ob(\mathcal{A})$ induce a partitioning of $\Delta^{\mathcal{I}_{\mathfrak{M}}}$ into disjoint labeled trees $\mathfrak{T}_a = (T_a, E_a, \ell_a)$ with nodes $T_a$, edges $E_a$, root $a^{\mathcal{I}_{\mathfrak{M}}}$, and a labeling function $\ell_a \colon E_a \to role^{\pm}(\mathcal{K}) \setminus \{Id, Id^-\}$.

**(copy)** There is a function $cp \colon \Delta^{\mathcal{I}_{\mathfrak{M}}} \to ob(\mathcal{A}) \cup dr(\mathcal{K})$ such that

- $cp(a^{\mathcal{I}_{\mathfrak{M}}}) = a$ for $a \in ob(\mathcal{A})$, and
- $cp(w) = dr$ if, for some $a$ and $w' \in T_a$, $(w', w) \in E_a$ and $\ell_a(w', w) = inv(R)$.

**(iso)** For each $R \in role^{\pm}(\mathcal{K})$, all labeled subtrees generated by elements $w \in \Delta^{\mathcal{I}_{\mathfrak{M}}}$ with $cp(w) = dr$ are isomorphic.

**(concept)** $w \in B^{\mathcal{I}_{\mathfrak{M}}}$ iff $\mathfrak{M} \models B^*[cp(w)]$, for each basic concept $B$ in $\mathcal{K}$ and each $w \in \Delta^{\mathcal{I}_{\mathfrak{M}}}$.

**(role)** $Id^{\mathcal{I}_{\mathfrak{M}}} = \big\{ (w, w) \mid w \in \Delta^{\mathcal{I}_{\mathfrak{M}}} \big\}$ and, for every other role name $P_k$,

$$P_k^{\mathcal{I}_{\mathfrak{M}}} \;=\; \big\{ (a_i^{\mathcal{I}_{\mathfrak{M}}}, a_j^{\mathcal{I}_{\mathfrak{M}}}) \mid R(a_i, a_j) \in \mathcal{A}, \ R \sqsubseteq_{\mathcal{T}}^* P_k \big\} \quad \cup \quad \big\{ (w, w) \mid Id \sqsubseteq_{\mathcal{T}}^* P_k \big\} \quad \cup$$
$$\bigcup_{a \in ob(\mathcal{A})} \big\{ (w, w') \in E_a \mid \ell_a(w, w') = R, \ R \sqsubseteq_{\mathcal{T}}^* P_k \big\}.$$

Such a model will be called an *untangled model of $\mathcal{K}$* (*the untangled model of $\mathcal{K}$ induced by $\mathfrak{M}$*, to be more precise).

The translation $\cdot^{\ddagger e}$ generalizes $\cdot^{\dagger}$ and thus suffers from the same exponential blowup. So we define an optimized translation, $\cdot^{\ddagger e}$, which is linear in the size of $\mathcal{K}$, by taking:

$$\mathcal{K}^{\ddagger e} \;=\; \forall x \left[ \mathcal{T}^*(x) \ \wedge \ \mathcal{T}^{\mathcal{R}}(x) \ \wedge \ \bigwedge_{R \in role^{\pm}(\mathcal{K})} \big( \varepsilon_R(x) \wedge \delta_R(x) \big) \right] \ \wedge \ \mathcal{A}^{\dagger 1} \ \wedge \ \mathcal{A}^{\ddagger 2}_{e},$$

where $\mathcal{T}^*(x)$, $\mathcal{T}^{\mathcal{R}}(x)$, $\varepsilon_R(x)$, $\delta_R(x)$ and $\mathcal{A}^{\dagger 1}$ are defined by (1), (42), (4), (5) and (2), respectively, and

$$\mathcal{A}^{\ddagger 2}_{e} \;=\; \bigwedge_{a \in ob(\mathcal{A})} \ \bigwedge_{\substack{R \in role^{\pm}(\mathcal{K}) \\ \exists a' \in ob(\mathcal{A}) \ R(a, a') \in \mathsf{Cl}_{\mathcal{T}}^{e}(\mathcal{A})}} E_{q^{e}_{R,a}} R(a) \quad \wedge \quad \bigwedge_{\neg P_k(a_i, a_j) \in \mathcal{A}} (\neg P_k(a_i, a_j))^{\perp e}, \qquad (43)$$

where $q^{e}_{R,a}$ is the maximum number in $Q^R_{\mathcal{T}}$ such that there are $q^{e}_{R,a}$ many distinct $a_i$ with $R(a, a_i) \in \mathsf{Cl}_{\mathcal{T}}^{e}(\mathcal{A})$ (here we use the UNA) and $(\neg P_k(a_i, a_j))^{\perp e} = \perp$ if $P_k(a_i, a_j) \in \mathsf{Cl}_{\mathcal{T}}^{e}(\mathcal{A})$ and $\top$ otherwise; cf. (15). We note again that if $Q^R_{\mathcal{T}} = \{1\}$, for all roles $R \in role^{\pm}(\mathcal{K})$, then the translation does not depend on whether the UNA is adopted or not.

The following corollary is proved similarly to Corollary 5.5:

**Corollary 5.16** *A DL-Lite$^{(\mathcal{HN})^-}_{bool}$ KB $\mathcal{K}$ is satisfiable iff the $\mathcal{QL}^1$-sentence $\mathcal{K}^{\ddagger e}$ is satisfiable.*

It should be clear that the translation $\cdot^{\ddagger e}$ can be computed in NLOGSPACE (for combined complexity). Indeed, this is readily seen for $\mathcal{T}^*(x)$, $\mathcal{T}^{\mathcal{R}}(x)$, $\varepsilon_R(x)$, $\delta_R(x)$, and $\mathcal{A}^{\dagger 1}$. In order to compute $\mathcal{A}^{\ddagger 2}_{e}$, we need to be able to check whether $R(a_i, a_j) \in \mathsf{Cl}_{\mathcal{T}}^{e}(\mathcal{A})$: this test can be performed by a *non-deterministic* algorithm using *logarithmic* space in $|role^{\pm}(\mathcal{K})|$ (it is basically the same as the standard directed graph reachability problem, which is NLOGSPACE-complete; see, e.g., Kozen, 2006); it can be done using $N \cdot \log |role^{\pm}(\mathcal{K})| + 2 \log |ob(\mathcal{A})|$ cells on the work tape, where $N$ is a constant (in fact, $N = 3$ is enough: one has to store $R$, the current role $R'$ and the path length for the graph reachability subroutine, which is also bounded by $\log |role^{\pm}(\mathcal{K})|$. Therefore, the translation $\cdot^{\dagger e}$ can be computed by an NLOGSPACE transducer.

Now we show how satisfiability of $DL\text{-}Lite_{bool}^{(\mathcal{HN})}$ KBs can be easily reduced to satisfiability of $DL\text{-}Lite_{bool}^{(\mathcal{HN})^-}$ KBs. First, we assume that $DL\text{-}Lite_{bool}^{(\mathcal{HN})}$ KBs contain no role symmetry and asymmetry constraints because $\mathsf{Asym}(P_k)$ can be equivalently replaced with $\mathsf{Dis}(P_k, P_k^-)$ and $\mathsf{Sym}(P_k)$ with $P_k^- \sqsubseteq P_k$ (it should be noted that the introduction of $P_k^- \sqsubseteq P_k$ in the TBox does not violate $(\mathbf{A_3})$). The following lemma allows us to get rid of qualified number restrictions as well as role disjointness, reflexivity and irreflexivity constraints:

**Lemma 5.17** *For every $DL\text{-}Lite_{bool}^{(\mathcal{HN})}$ KB $\mathcal{K}' = (\mathcal{T}', \mathcal{A}')$, one can construct a $DL\text{-}Lite_{bool}^{(\mathcal{HN})^-}$ KB $\mathcal{K} = (\mathcal{T}, \mathcal{A})$ such that*

- *every untangled model $\mathcal{I}_\mathfrak{M}$ of $\mathcal{K}$ is a model of $\mathcal{K}'$, provided that*

$$
\begin{aligned}
&\text{there are no } R_1(a_i, a_j), R_2(a_i, a_j) \in \mathsf{Cl}_{\mathcal{T}}^{\mathsf{e}}(\mathcal{A}) \text{ with } \mathsf{Dis}(R_1, R_2) \in \mathcal{T}', \\
&\text{there is no } R(a_i, a_i) \in \mathsf{Cl}_{\mathcal{T}}^{\mathsf{e}}(\mathcal{A}) \text{ with } \mathsf{Irr}(R) \in \mathcal{T}';
\end{aligned}
\tag{44}
$$

- *every model $\mathcal{I}'$ of $\mathcal{K}'$ gives rise to a model $\mathcal{I}$ of $\mathcal{K}$ based on the same domain as $\mathcal{I}'$ and such that $\mathcal{I}$ agrees with $\mathcal{I}'$ on all symbols from $\mathcal{K}'$.*

*If $\mathcal{K}'$ is a $DL\text{-}Lite_{horn}^{(\mathcal{HN})}$ KB then $\mathcal{K}$ is a $DL\text{-}Lite_{horn}^{(\mathcal{HN})^-}$ KB.*

**Proof** First, for every pair $R$, $C$ such that $\geq q\, R.C$ occurs in $\mathcal{T}'$, we introduce a fresh role name $R_C$. Then we replace each (positive) occurrence of $\geq q\, R.C$ in $\mathcal{T}'$ with $\geq q\, R_C$ and add the following concept and role inclusions to the TBox:

$$
\exists R_C^- \sqsubseteq C \qquad \text{and} \qquad R_C \sqsubseteq R.
$$

We repeat this procedure until all the occurrences of qualified number restrictions are eliminated. Denote by $\mathcal{T}''$ the resulting TBox. Observe that $(\mathbf{A_1})$ and $(\mathbf{A_2})$ ensure that $\mathcal{T}''$ satisfies $(\mathbf{A_3})$. We also notice that $C$ occurs only on the right-hand side of those extra axioms and thus $\mathcal{T}''$ belongs to the same fragment as $\mathcal{T}'$. It should be clear that, since the $\geq q\, R.C$ occur only positively, every model of $\mathcal{T}''$ is a model of $\mathcal{T}'$. Conversely, for every model $\mathcal{I}'$ of $\mathcal{T}'$, there is a model $\mathcal{I}''$ of $\mathcal{T}''$ based on the same domain such that $\mathcal{I}''$ coincides with $\mathcal{I}'$ on all symbols in $\mathcal{T}'$ and $R_C^{\mathcal{I}''} = \{(w, u) \in R^{\mathcal{I}'} \mid u \in C^{\mathcal{I}'}\}$, for each new role $R_C$. So, without loss of generality we may assume that $\mathcal{T}' = \mathcal{T}''$.

Let

$$
\mathcal{T}' \;=\; \mathcal{T}_0' \;\cup\; \mathcal{T}_{ref}' \;\cup\; \mathcal{T}_{irref}' \;\cup\; \mathcal{T}_{disj}',
$$

where $\mathcal{T}_{ref}'$, $\mathcal{T}_{irref}'$ and $\mathcal{T}_{disj}'$ are the sets of role reflexivity, irreflexivity and disjointness constraints in $\mathcal{T}'$ and $\mathcal{T}_0'$ is the remaining $DL\text{-}Lite_{bool}^{(\mathcal{HN})^-}$ TBox. Let

$$
\begin{aligned}
\mathcal{T}_1' &= \big\{\, \top \sqsubseteq \exists Id,\; Id^- \sqsubseteq Id \,\big\} \;\cup\; \big\{\, Id \sqsubseteq P \mid \mathsf{Ref}(P) \in \mathcal{T}_{ref}' \,\big\}, \\
\mathcal{A}_1' &= \big\{\, Id(a_i, a_i) \mid a_i \in ob(\mathcal{A}') \,\big\}.
\end{aligned}
$$

We construct $\mathcal{K}$ by modifying the $DL\text{-}Lite_{bool}^{(\mathcal{HN})^-}$ KB $\mathcal{K}_0 = (\mathcal{T}_0' \cup \mathcal{T}_1', \mathcal{A}' \cup \mathcal{A}_1')$ in two steps:

*Step 1.* For every reflexivity constraint $\mathsf{Ref}(P) \in \mathcal{T}_{ref}'$ take a fresh role name $S_P$ and

- add a new role inclusion $S_P \sqsubseteq P$ to the TBox;

- replace every basic concept $B$ in $\mathcal{T}_0'$ with $B^{S_P}$, which is defined inductively as follows:

    - $A^{S_P} = A$, for each concept name $A$,
    - $(\geq q\,R)^{S_P} = \,\geq q\,R$, for each role $R \notin \{P, P^-\}$,
    - $(\geq q\,P)^{S_P} = \,\geq (q-1)\,S_P$ and $(\geq q\,P^-)^{S_P} = \,\geq (q-1)\,S_P^-$, for $q \geq 2$,
    - $(\exists P)^{S_P} = \top$ and $(\exists P^-)^{S_P} = \top$;

- replace $R(a_i, a_j) \in \mathcal{A}'$ such that $R \equiv_{\mathcal{T}'}^* P$ with $S_P(a_i, a_j)$ whenever $i \neq j$.

Intuitively, we split the role $P$ into its irreflexive part $S_P$ and $Id$. Note that if $P$ has a reflexive proper sub-role then, by $(\mathbf{A_3})$, there are no restrictions on the maximal number of $P$-successors and $P$-predecessors, and therefore on $S_P$ if $\mathsf{Ref}(P) \in \mathcal{T}'$. Let $(\mathcal{T}_1, \mathcal{A})$ be the resulting $DL\text{-}Lite_{bool}^{(\mathcal{HN})^-}$ KB. Clearly, $(\mathcal{T}_1, \mathcal{A})$ satisfies $(\mathbf{Id_1})$–$(\mathbf{Id_3})$. Observe that

$$\mathsf{Cl}_{\mathcal{T}_1}^\mathsf{e}(\mathcal{A}) \upharpoonright_{role(\mathcal{K}')} \;=\; \mathsf{Cl}_{\mathcal{T}_0' \cup \mathcal{T}_1'}^\mathsf{e}(\mathcal{A}'), \tag{45}$$

where $\upharpoonright_{role(\mathcal{K}')}$ means the restriction to the role names in $\mathcal{K}'$.

Let $\mathcal{I}_\mathfrak{M}$ be an untangled model of $(\mathcal{T}_1, \mathcal{A})$. We show that $\mathcal{I}_\mathfrak{M} \models \mathcal{T}_0'$. Consider a role $P$ with $\mathsf{Ref}(P) \in \mathcal{T}'$. Notice that $S_P$ has no proper sub-roles in $\mathcal{T}_1$ and $Id^{\mathcal{I}_\mathfrak{M}}$ is disjoint with $S_P^{\mathcal{I}_\mathfrak{M}}$. Thus, $S_P^{\mathcal{I}_\mathfrak{M}} \cup Id^{\mathcal{I}_\mathfrak{M}} \subseteq P^{\mathcal{I}_\mathfrak{M}}$ and

**(\*)** $(B^{S_P})^{\mathcal{I}_\mathfrak{M}} \subseteq B^{\mathcal{I}_\mathfrak{M}}$, for $B = \,\geq q\,R$ with $q \geq 2$, whenever $\mathsf{Ref}(P) \in \mathcal{T}'$, $R \in \{P, P^-\}$ and $P$ has a proper sub-role in $\mathcal{T}'$.

If $P$ has no proper sub-roles in $\mathcal{T}'$ (i.e., no proper sub-roles in $\mathcal{T}_1$ different from $S_P$ and $Id$) then we have $S_P^{\mathcal{I}_\mathfrak{M}} \cup Id^{\mathcal{I}_\mathfrak{M}} = P^{\mathcal{I}_\mathfrak{M}}$. So, for all basic concepts $B$ in $\mathcal{T}_0'$ not covered by **(\*)**, we have $B^{\mathcal{I}_\mathfrak{M}} = (B^{S_P})^{\mathcal{I}_\mathfrak{M}}$. It follows from $(\mathbf{A_3})$ that $\mathcal{I}_\mathfrak{M} \models \mathcal{T}_0'$.

*Step 2.* Next we take into account the set $\mathcal{D} = \mathcal{T}_{disj}' \cup \{\mathsf{Dis}(P_k, Id) \mid \mathsf{Irr}(P_k) \in \mathcal{T}_{irref}'\}$ of disjointness constraints by modifying the KB $(\mathcal{T}_1, \mathcal{A})$ constructed at the previous step. Observe that $\exists R_1 \sqsubseteq \bot$ is a logical consequence of any $\mathcal{T} \cup \{\mathsf{Dis}(R_1, R_2)\}$ whenever $R_1 \sqsubseteq_\mathcal{T}^* R_2$. Let $\mathcal{T} = \mathcal{T}_1 \cup \mathcal{T}_2$, where $\mathcal{T}_2$ is defined by taking

$$\mathcal{T}_2 \;=\; \{\exists R_1 \sqsubseteq \bot \mid R_1 \sqsubseteq_\mathcal{T_1}^* R_2 \text{ and either } \mathsf{Dis}(R_1, R_2) \in \mathcal{D} \text{ or } \mathsf{Dis}(R_2, R_1) \in \mathcal{D}\}.$$

By $(\mathbf{role})$, for any untangled model $\mathcal{I}_\mathfrak{M}$ of $(\mathcal{T}, \mathcal{A})$ and $R_1, R_2 \in role^\pm(\mathcal{K})$, $\mathcal{I}_\mathfrak{M} \models \mathsf{Dis}(R_1, R_2)$ if there are no $R_1(a_i, a_j), R_2(a_i, a_j) \in \mathsf{Cl}_{\mathcal{T}_1}^\mathsf{e}(\mathcal{A})$, which, by (45), means that there are no $R_1(a_i, a_j), R_2(a_i, a_j) \in \mathsf{Cl}_{\mathcal{T}_0' \cup \mathcal{T}_1'}^\mathsf{e}(\mathcal{A}')$. So, if (44) holds then every untangled model $\mathcal{I}_\mathfrak{M}$ of $(\mathcal{T}, \mathcal{A})$ is also a model of $\mathcal{T}_1 \cup \mathcal{D}$ and thus, $\mathcal{I}_\mathfrak{M} \models \mathcal{T}_{disj}'$. As $Id^{\mathcal{I}_\mathfrak{M}}$ is the identity relation, we have $\mathcal{I}_\mathfrak{M} \models \mathcal{T}_{ref}' \cup \mathcal{T}_{irref}'$. By (45), $\mathcal{I}_\mathfrak{M} \models \mathcal{A}'$ and as we have shown above, $\mathcal{I}_\mathfrak{M} \models \mathcal{T}_0'$. Therefore, $\mathcal{I}_\mathfrak{M} \models \mathcal{K}'$.

Conversely, suppose $\mathcal{I}'$ is a model of $\mathcal{K}'$. Let $\mathcal{I}$ be an interpretation such that $Id^\mathcal{I}$ is the identity relation, $S_P^\mathcal{I} = P^{\mathcal{I}'} \setminus Id^{\mathcal{I}'}$, for all $P$ with $\mathsf{Ref}(P) \in \mathcal{T}'$, and $A^\mathcal{I} = A^{\mathcal{I}'}$, $P^\mathcal{I} = P^{\mathcal{I}'}$ and $a^\mathcal{I} = a^{\mathcal{I}'}$, for all concept, role and object names $A$, $P$ and $a$ in $\mathcal{K}'$. Clearly, $\mathcal{I} \models (\mathcal{T}_0' \cup \mathcal{T}_1', \mathcal{A}' \cup \mathcal{A}_1')$. By the definition of the $S_P$, $\mathcal{I} \models \mathcal{T}_1$ and, since $\mathcal{I} \models \mathcal{D}$, we obtain $\mathcal{I} \models \mathcal{T}_2$ and thus $\mathcal{I} \models \mathcal{T}$. By (45), $\mathcal{I} \models \mathcal{A}$, whence $\mathcal{I} \models \mathcal{K}$. $\qquad\qquad\square$

Now, as follows from Lemma 5.17, given a $DL\text{-}Lite_\alpha^{(\mathcal{HN})}$ KB $\mathcal{K}'$, for $\alpha \in \{krom, horn, bool\}$, we can compute the $DL\text{-}Lite_{bool}^{(\mathcal{HN})^-}$ KB $\mathcal{K}$ using a LOGSPACE transducer (which is essentially required for checking whether $R \equiv_{\mathcal{T}'}^* P$). We immediately obtain Theorem 5.13 from Lemma 5.14 by observing that, for each $\alpha \in \{krom, horn, bool\}$, $\mathcal{K}^{\ddagger e}$ belongs to the respective first-order fragment and that condition (44) can be checked in NLOGSPACE (computing $\mathsf{Cl}_{\mathcal{T}}^e(\mathcal{A})$ requires directed graph accessibility checks). The result for $DL\text{-}Lite_{core}^{(\mathcal{HN})}$ follows from the corresponding result for $DL\text{-}Lite_{krom}^{(\mathcal{HN})}$.

## 5.4 Role Transitivity Constraints

We now consider the languages $DL\text{-}Lite_\alpha^{(\mathcal{HN})^+}$, $\alpha \in \{core, krom, horn, bool\}$, which extend $DL\text{-}Lite_\alpha^{(\mathcal{HN})}$ with *role transitivity constraints* of the form $\mathsf{Tra}(P_k)$. We remind the reader that a role is called *simple* (see, e.g., Horrocks et al., 2000) if it has no transitive sub-roles (including itself) and that only simple roles $R$ are allowed in concepts of the form $\geq q\,R$, for $q \geq 2$. In particular, if $\mathcal{T}$ contains $\mathsf{Tra}(P)$ then $P$ and $P^-$ are not simple, and so $\mathcal{T}$ cannot contain occurrences of concepts of the form $\geq q\,P$ and $\geq q\,P^-$, for $q \geq 2$.

For a $DL\text{-}Lite_\alpha^{(\mathcal{HN})^+}$ KB $\mathcal{K} = (\mathcal{T}, \mathcal{A})$, define the *transitive closure* $\mathsf{Tra}_{\mathcal{T}}(\mathcal{A})$ of $\mathcal{A}$ by taking

$$\mathsf{Tra}_{\mathcal{T}}(\mathcal{A}) \;=\; \mathcal{A} \cup \big\{ P(a_{i_1}, a_{i_n}) \mid \exists a_{i_2} \ldots a_{i_{n-1}}\; P(a_{i_1}, a_{i_{j+1}}) \in \mathcal{A},\; 1 \leq j < n,\; \mathsf{Tra}(P) \in \mathcal{T} \big\}.$$

Clearly, $\mathsf{Tra}_{\mathcal{T}}(\mathcal{A})$ can be computed in NLOGSPACE: for each pair $(a_i, a_j)$ of objects in $ob(\mathcal{A})$, we add $P(a_i, a_j)$ to $\mathsf{Tra}_{\mathcal{T}}(\mathcal{A})$ iff there is a $P$-path of length $< |ob(\mathcal{A})|$ between $a_i$ and $a_j$ in $\mathcal{A}$ (recall that the directed graph reachability problem is NLOGSPACE-complete).

**Lemma 5.18** *A $DL\text{-}Lite_\alpha^{(\mathcal{HN})^+}$ KB $(\mathcal{T}, \mathcal{A})$ is satisfiable iff the $DL\text{-}Lite_\alpha^{(\mathcal{HN})}$ KB $(\mathcal{T}', \mathcal{A}')$ is satisfiable, where $\mathcal{T}'$ results from $\mathcal{T}$ by removing all the transitivity axioms and*

$$\mathcal{A}' \;=\; \mathsf{Cl}_{\mathcal{T}}^e(\mathsf{Tra}_{\mathcal{T}}(\mathsf{Cl}_{\mathcal{T}}^e(\mathcal{A}))).$$

**Proof** Indeed, if the KB $(\mathcal{T}', \mathcal{A}')$ is satisfiable then we construct a model $\mathcal{I}$ for it as described in the proofs of Lemmas 5.14 and 5.17 and then take the transitive closure of $P^{\mathcal{I}}$ for every $P$ with $\mathsf{Tra}(P) \in \mathcal{T}$ (and update each $R^{\mathcal{I}}$ with $P \sqsubseteq_{\mathcal{T}}^* R$). As $P$ and $P^-$ are simple, $\mathcal{T}$ contains no axioms imposing upper bounds on the number of $P$-successors and predecessors, and so the resulting interpretation must be a model of $(\mathcal{T}, \mathcal{A})$. The converse direction is trivial. ❏

We note that an analogue of Remark 5.15 also holds in this case: just replace $\mathsf{Cl}_{\mathcal{T}}^e(\mathcal{A})$ with $\mathsf{Cl}_{\mathcal{T}}^e(\mathsf{Tra}_{\mathcal{T}}(\mathsf{Cl}_{\mathcal{T}}^e(\mathcal{A})))$ in **(ABox)** and take the transitive closure for each transitive sub-role in **(role)**.

**Remark 5.19** It should be noted that there are two different reasons for the reduction in Lemma 5.18 to be in NLOGSPACE rather than in LOGSPACE (as the reduction $\cdot^{\ddagger}$ is). First, in order to compute $\mathsf{Cl}_{\mathcal{T}}^e(\mathcal{A})$, for each pair of $a_i, a_j$, one has to find a path in the directed graph induced by the role inclusion axioms. Second, in order to compute $\mathsf{Tra}_{\mathcal{T}}(\mathsf{Cl}_{\mathcal{T}}^e(\mathcal{A}))$, one has to find a path in the graph induced by the ABox $\mathcal{A}$ itself. So, if we are concerned with the data complexity, $\mathsf{Cl}_{\mathcal{T}}^e(\mathcal{A})$ can be computed in LOGSPACE (in fact, in $AC^0$, as we shall see in Section 6.1) because the role inclusion graph (and hence its size) does not depend on $\mathcal{A}$. The second reason, however, is more 'dangerous' for data complexity as we shall see in Section 6.1.

As a consequence of Lemma 5.18 and Theorem 5.13 we obtain the following:

**Corollary 5.20** *For combined complexity,* (i) *satisfiability of DL-Lite$_{bool}^{(\mathcal{HN})^+}$ KBs is* NP-*complete;* (ii) *satisfiability of DL-Lite$_{horn}^{(\mathcal{HN})^+}$ KBs is* P-*complete; and* (iii) *satisfiability of DL-Lite$_{krom}^{(\mathcal{HN})^+}$ and DL-Lite$_{core}^{(\mathcal{HN})^+}$ KBs is* NLogSpace-*complete.*

Note again that if the KBs do not contain number restrictions of the form $\geq q\,R$, for $q \geq 2$, (as in the extensions of the *DL-Lite$_\alpha^{\mathcal{H}}$* languages) then the result does not depend on the UNA.

**Remark 5.21** It should be noted that role disjointness, symmetry, asymmetry and transitivity constraints can be added to any of the logics *DL-Lite$_\alpha^{\mathcal{HF}}$* and *DL-Lite$_\alpha^{\mathcal{HN}}$*, for $\alpha \in \{core, krom, horn, bool\}$, without changing the combined complexity of their satisfiability problems (which, by Corollary 5.12, are all ExpTime-complete). Indeed, as follows from Theorem 10 of Glimm et al. (2007), KB satisfiability in the extension of $\mathcal{SHIQ}$ with role conjunction is in ExpTime if the length of role conjunctions is bounded by some constant (in our case, this constant is 2 because $\mathsf{Dis}(R_1, R_2)$ can be encoded by $\exists(R_1 \sqcap R_2).\top \sqsubseteq \bot$; $\mathsf{Asym}(R)$ is dealt with similarly). We conjecture that role reflexivity and irreflexivity constraints do not change complexity either.

## 6. Instance Checking: Data Complexity

So far we have assumed the whole KB $\mathcal{K} = (\mathcal{T}, \mathcal{A})$ to be the input for the satisfiability problem. According to the classification suggested by Vardi (1982), we have been considering its *combined complexity*. Two other types of complexity for knowledge bases are:

- the *schema* (or *TBox*) *complexity*, where only the TBox $\mathcal{T}$ is regarded to be the input, while the ABox $\mathcal{A}$ is assumed to be fixed; and

- the *data* (or *ABox*) *complexity*, where only the ABox $\mathcal{A}$ is regarded to be the input.

It is easy to see that the schema complexity of the satisfiability problem for all our logics considered above coincides with the corresponding combined complexity. In this section, we analyze the data complexity of satisfiability and instance checking.

### 6.1 *DL-Lite$_{bool}^{\mathcal{N}}$, DL-Lite$_{bool}^{\mathcal{H}}$* and *DL-Lite$_{bool}^{(\mathcal{HN})}$* are in AC$^0$

In what follows, without loss of generality we assume that all role and concept names of a given knowledge base $\mathcal{K} = (\mathcal{T}, \mathcal{A})$ occur in its TBox and write $role(\mathcal{T})$, $role^\pm(\mathcal{T})$ and $dr(\mathcal{T})$ instead of $role(\mathcal{K})$, $role^\pm(\mathcal{K})$ and $dr(\mathcal{K})$, respectively; the set of concept names in $\mathcal{T}$ is denoted by $con(\mathcal{T})$. In this section we reduce satisfiability of *DL-Lite$_{bool}^{(\mathcal{HN})}$* KBs to model checking in first-order logic. To this end, we fix a signature containing two unary predicates $A_k$ and $\overline{A_k}$, for each concept name $A_k$, and two binary predicates $P_k$ and $\overline{P_k}$, for each role name $P_k$.

Consider first the case of a $DL\text{-}Lite_{bool}^{(\mathcal{HN})^-}$ KB $\mathcal{K}$. We represent the ABox $\mathcal{A}$ of $\mathcal{K}$ as a first-order model $\mathfrak{A}_\mathcal{A}$ of the above signature. The domain of $\mathfrak{A}_\mathcal{A}$ is $ob(\mathcal{A})$ and, for all $a_i, a_j \in ob(\mathcal{A})$ and all predicates $A_k$, $\overline{A_k}$, $P_k$ and $\overline{P_k}$ in the signature,

$$\mathfrak{A}_\mathcal{A} \models A_k[a_i] \quad \text{iff} \quad A_k(a_i) \in \mathcal{A}, \qquad \mathfrak{A}_\mathcal{A} \models \overline{A_k}[a_i] \quad \text{iff} \quad \neg A_k(a_i) \in \mathcal{A},$$

$$\mathfrak{A}_\mathcal{A} \models P_k[a_i, a_j] \quad \text{iff} \quad P_k(a_i, a_j) \in \mathcal{A}, \qquad \mathfrak{A}_\mathcal{A} \models \overline{P_k}[a_i, a_j] \quad \text{iff} \quad \neg P_k(a_i, a_j) \in \mathcal{A}.$$

Now we construct a first-order sentence $\varphi_\mathcal{T}$ in the same signature such that (i) $\varphi_\mathcal{T}$ depends on $\mathcal{T}$ but does not depend on $\mathcal{A}$, and (ii) $\mathfrak{A}_\mathcal{A} \models \varphi_\mathcal{T}$ iff $\mathcal{K}^{\ddagger e}$ is satisfiable.

To simplify presentation, we denote by $\mathsf{ext}(\mathcal{T})$ the extension of $\mathcal{T}$ with the following concept inclusions:

- $\geq q' R \sqsubseteq \, \geq q R$, for all $R \in role^\pm(\mathcal{T})$ and $q, q' \in Q_\mathcal{T}^R$ such that $q' > q$ and $q' > q'' > q$ for no $q'' \in Q_\mathcal{T}^R$, and

- $\geq q R \sqsubseteq \, \geq q R'$, for all $q \in Q_\mathcal{T}^R$ and $R \sqsubseteq R' \in \mathcal{T}$ or $inv(R) \sqsubseteq inv(R') \in \mathcal{T}$.

Clearly, $(\mathsf{ext}(\mathcal{T}))^*(x)$ is equivalent (in first-order logic) to $\mathcal{T}^*(x) \wedge \mathcal{T}^\mathcal{R}(x) \wedge \bigwedge_{R \in role^\pm(\mathcal{T})} \delta_R(x)$; see (1), (5) and (42).

Let $Bcon(\mathcal{T})$ be the set of basic concepts occurring in $\mathcal{T}$ (i.e., concepts of the form $A$ and $\geq q R$, for $A \in con(\mathcal{T})$, $R \in role^\pm(\mathcal{T})$ and $q \in Q_\mathcal{T}^R$). To indicate which basic concepts hold or do not hold on a domain element of a first-order model of $\mathcal{K}^{\ddagger e}$, we use functions $\xi \colon Bcon(\mathcal{T}) \to \{\top, \bot\}$, which will be called *types*. Denote by $\mathbf{Tp}$ the set of all such types (there are $2^{|Bcon(\mathcal{T})|}$ of them). For a complex concept $C$, we define $\xi(C)$ by induction: $\xi(\neg C) = \neg \xi(C)$ and $\xi(C_1 \sqcap C_2) = \xi(C_1) \wedge \xi(C_2)$. The propositional variable-free formula

$$\xi^\mathcal{T} \;=\; \bigwedge_{C_1 \sqsubseteq C_2 \in \mathsf{ext}(\mathcal{T})} \big( \xi(C_1) \to \xi(C_2) \big)$$

ensures that the type $\xi$ is consistent with concept and role inclusions in $\mathcal{T}$. It should be emphasized that $\xi^\mathcal{T}$ is built from $\bot$ and $\top$ using the Boolean connectives and therefore does not depend on a particular domain element of $\mathfrak{A}_\mathcal{A}$. The following formula is true if a given element $x$ of $\mathfrak{A}_\mathcal{A}$ is of type $\xi$ (see $\mathcal{A}^{\dagger 1}$ and $\mathcal{A}^{\ddagger e}$; (2) and (43), respectively):

$$\xi^*(x) \;=\; \bigwedge_{A_k \in con(\mathcal{T})} \big( (A_k(x) \to \xi(A_k)) \wedge (\overline{A_k}(x) \to \neg \xi(A_k)) \big) \quad \wedge$$

$$\bigwedge_{R \in role^\pm(\mathcal{T})} \bigwedge_{q \in Q_\mathcal{T}^R} \big( E_q R^\mathcal{T}(x) \to \xi(\geq q R) \big) \quad \wedge \quad \bigwedge_{P_k \in role(\mathcal{T})} \forall x \forall y \, \big( P_k^\mathcal{T}(x, y) \wedge \overline{P_k}(x, y) \to \bot \big),$$

where $E_q R^\mathcal{T}(x)$ and $R^\mathcal{T}(x, y)$, for $R \in role^\pm(\mathcal{T})$, are abbreviations defined by

$$E_q R^\mathcal{T}(x) \;=\; \exists y_1 \ldots \exists y_q \, \big( \bigwedge_{1 \leq i < j \leq q} (y_i \neq y_j) \wedge \bigwedge_{1 \leq i \leq q} R^\mathcal{T}(x, y_i) \big), \tag{46}$$

$$R^\mathcal{T}(x, y) \;=\; \bigvee_{P_k \sqsubseteq_\mathcal{T}^\star R} P_k(x, y) \quad \vee \quad \bigvee_{P_k^- \sqsubseteq_\mathcal{T}^\star R} P_k(y, x). \tag{47}$$

Clearly, we have $R(a_i, a_j) \in \mathsf{Cl}^\mathsf{e}_\mathcal{T}(\mathcal{A})$ iff $\mathfrak{A}_\mathcal{A} \models R^\mathcal{T}[a_i, a_j]$ and $\mathfrak{A}_\mathcal{A} \models E_q R^\mathcal{T}[a]$ iff $a$ has at least $q$ distinct $R$-successors in $\mathsf{Cl}^\mathsf{e}_\mathcal{T}(\mathcal{A})$ (and thus in every model of $\mathcal{K}$).

Without loss of generality we may assume that $role^\pm(\mathcal{T}) = \{R_1, \ldots, R_k\} \neq \emptyset$. Denote by $\mathbf{Tp}^k$ the set of $k$-tuples $\vec{\xi}$ containing a type $\xi_{dr_i} \in \mathbf{Tp}$ for each role $R_i \in role^\pm(\mathcal{T})$. We then set

$$\varphi_\mathcal{T} \;=\; \bigvee_{\vec{\xi} \in \mathbf{Tp}^k} \forall x \, \vartheta^{\vec{\xi}}_\mathcal{T}(x),$$

where

$$\vartheta^{(\xi_{dr_1}, \ldots, \xi_{dr_k})}_\mathcal{T}(x) \;=\; \bigvee_{\xi \in \mathbf{Tp}} \Big( \xi^*(x) \;\wedge\; \xi^\mathcal{T} \;\wedge\; \bigwedge_{R_i \in role^\pm(\mathcal{T})} \xi^\mathcal{T}_{dr_i} \;\wedge$$

$$\bigwedge_{R_i \in role^\pm(\mathcal{T})} \Big( \big( \xi(\exists R_i) \vee \bigvee_{S \in role^\pm(\mathcal{T})} \xi_{ds}(\exists R_i) \big) \to \xi_{inv(dr_i)}(\exists inv(R_i)) \Big) \Big).$$

To explain the meaning of the subformulas of $\varphi_\mathcal{T}$, assume that $(\mathcal{T}, \mathcal{A})$ is satisfiable. In order to construct a model $\mathfrak{M}$ for $\mathcal{K}^{\ddagger \mathsf{e}}$ from the first-order model $\mathfrak{A}_\mathcal{A}$, we have to specify the basic concepts that contain a given constant of $\mathcal{K}^{\ddagger \mathsf{e}}$. In other words, we have to select a type for each $dr_i \in dr(\mathcal{T})$ and each $a \in ob(\mathcal{A})$. The formula $\varphi_\mathcal{T}$ says that one can select a $k$-tuple of types $\vec{\xi} = (\xi_{dr_1}, \ldots, \xi_{dr_k}) \in \mathbf{Tp}^k$ such that one of its disjuncts is true in $\mathfrak{A}_\mathcal{A}$. Such a $k$-tuple fixes the 'witness' part of the model $\mathfrak{M}$, consisting of the $dr_i$, and determines the basic concepts these $dr_i$ belong to. Then each disjunct of $\varphi_\mathcal{T}$ says that (having fixed the 'witness' part of the model), for every $a \in ob(\mathcal{A})$, there is a type $\xi$ (determining the basic concepts $a$ belongs to) such that

- $\xi$ is consistent with the information about $a$ in $\mathcal{A}$ (cf. $\xi^*(x)$);

- $\xi$ is also consistent with the concept and role inclusions of $\mathcal{T}$ (cf. $\xi^\mathcal{T}$);

- each of $\xi_{dr_1}, \ldots, \xi_{dr_k}$ is consistent with the concept and role inclusions of $\mathcal{T}$ (cf. $\xi^\mathcal{T}_{dr_i}$);

- each role $R_i$ with a nonempty domain (i.e., either $\xi$ or any of $\xi_{ds}$ is $\top$ on $\exists R_i$) has a nonempty range, in particular, $\xi_{inv(dr_i)}(\exists inv(R_i)) = \top$; see also $\varepsilon_R(x)$ as defined by (4).

**Lemma 6.1** $\mathfrak{A}_\mathcal{A} \models \varphi_\mathcal{T}$ iff $\mathcal{K}^{\ddagger \mathsf{e}}$ is satisfiable.

**Proof** ($\Rightarrow$) Fix some $\vec{\xi} = (\xi_{dr_1}, \ldots, \xi_{dr_k}) \in \mathbf{Tp}^k$ such that $\mathfrak{A}_\mathcal{A} \models \forall x \, \vartheta^{\vec{\xi}}_\mathcal{T}(x)$. Then, for each $a \in ob(\mathcal{A})$, fix some type such that the respective disjunct of $\vartheta^{\vec{\xi}}_\mathcal{T}(x)$ holds on $a$ in $\mathfrak{A}_\mathcal{A}$ and denote it by $\xi_a$. Define a first-order model $\mathfrak{M}$ over the domain $ob(\mathcal{A}) \cup dr(\mathcal{T})$ by taking:

- $\mathfrak{M} \models B^*[c]$ iff $\xi_c(B) = \top$, for all $c \in ob(\mathcal{A}) \cup dr(\mathcal{T})$ and $B \in Bcon(\mathcal{T})$

($B^*$ is the unary predicate for $B$ as defined on p. 22). It is easy to check that $\mathfrak{M} \models \mathcal{K}^{\ddagger \mathsf{e}}$.

($\Leftarrow$) Suppose now that $\mathcal{K}^{\ddagger \mathsf{e}}$ is satisfiable. Then there is a model $\mathfrak{M}$ of $\mathcal{K}^{\ddagger \mathsf{e}}$ with domain $ob(\mathcal{A}) \cup dr(\mathcal{T})$. To see that $\mathfrak{A}_\mathcal{A} \models \varphi_\mathcal{T}$, it suffices to take the functions $\xi_{dr_i}$ and $\xi_a$ defined by:

- $\xi_{dr_i}(B) = \top$ iff $\mathfrak{M} \models B^*[dr_i]$, for $dr_i \in dr(\mathcal{T})$ and $B \in Bcon(\mathcal{T})$,

- $\xi_a(B) = \top$ iff $\mathfrak{M} \models B^*[a]$, for $a \in ob(\mathcal{A})$ and $B \in Bcon(\mathcal{T})$.

Details are left to the reader. ❑

It follows from Lemmas 6.1 and 5.17 and Corollary 5.16 that we have:

**Corollary 6.2** *The satisfiability and instance checking problems for DL-Lite$_{bool}^{\mathcal{N}}$, DL-Lite$_{bool}^{\mathcal{H}}$ and DL-Lite$_{bool}^{(\mathcal{HN})}$ KBs are in* AC$^0$ *for data complexity.*

**Proof** DL-Lite$_{bool}^{\mathcal{N}}$ and DL-Lite$_{bool}^{\mathcal{H}}$ are sub-languages of DL-Lite$_{bool}^{(\mathcal{HN})^-}$ and for them the result immediately follows from Lemma 6.1 and Corollary 5.16. For a DL-Lite$_{bool}^{(\mathcal{HN})}$ KB $\mathcal{K}' = (\mathcal{T}', \mathcal{A}')$, by Lemma 5.17, we construct a DL-Lite$_{bool}^{(\mathcal{HN})^-}$ KB $\mathcal{K} = (\mathcal{T}, \mathcal{A})$ such that $\mathcal{K}'$ is satisfiable iff $\mathcal{K}$ is satisfiable and (44) holds. The latter condition corresponds to the following first-order sentence

$$\gamma_{\mathcal{T}'} = \bigwedge_{\mathsf{Dis}(R_1,R_2)\in\mathcal{T}'} \forall x \forall y \left( R_1^{\mathcal{T}}(x,y) \wedge R_2^{\mathcal{T}}(x,y) \to \bot \right) \quad \wedge \quad \bigwedge_{\mathsf{Irr}(P_k)\in\mathcal{T}'} \forall x \left( P_k^{\mathcal{T}}(x,x) \to \bot \right),$$

evaluated in $\mathfrak{A}_{\mathcal{A}}$. Therefore, $\mathcal{K}'$ is satisfiable iff $\mathfrak{A}_{\mathcal{A}} \models \varphi_{\mathcal{T}} \wedge \gamma_{\mathcal{T}'}$. Let $\psi = \varphi_{\mathcal{T}} \wedge \gamma_{\mathcal{T}'}$ and $\psi'$ be the result of replacing each $S_P(t_1, t_2)$, for $\mathsf{Ref}(P) \in \mathcal{T}'$, with $P(t_1, t_2) \wedge (t_1 \neq t_2)$; see the proof of Lemma 5.17. It remains to observe that $\mathfrak{A}_{\mathcal{A}} \models \psi$ iff $\mathfrak{A}_{\mathcal{A}'} \models \psi'$. ❑

As before, this result does not depend on the UNA for any member of the *DL-Lite* family that has no number restrictions of the form $\geq q\,R$, for $q \geq 2$ (in particular, for DL-Lite$_{bool}^{\mathcal{H}}$ and its fragments).

We also note that transitive roles cannot be included in our languages for free if we are concerned with data complexity:

**Lemma 6.3** *Satisfiability and instance checking of DL-Lite$_{core}$ KBs extended with role transitivity constraints are* NLOGSPACE*-hard for data complexity.*

**Proof** Suppose we are given a directed graph. Let $P$ be a role name. Define an ABox $\mathcal{A}$ by taking $P(a_i, a_j) \in \mathcal{A}$ iff there is an edge $(a_i, a_j)$ in the graph. Then a node $a_n$ is reachable from a node $a_0$ iff the DL-Lite$_{core}$ ABox $\mathcal{A} \cup \{\neg P(a_0, a_n)\}$ is not satisfiable in models with transitive $P$. This encoding immediately gives the claim of the lemma because the directed graph reachability problem is NLOGSPACE-complete, NLOGSPACE is closed under the complement (see, e.g., Kozen, 2006) and the TBox $\{\mathsf{Tra}(P)\}$ does not depend on the input. ❑

On the other hand, as the reduction of Lemma 5.18 is computable in NLOGSPACE, we obtain the following:

**Corollary 6.4** *Satisfiability and instance checking of DL-Lite$_{bool}^{(\mathcal{HN})^+}$ KBs are* NLOGSPACE*-complete for data complexity.*

**Proof** The upper bound is obtained by applying the NLOGSPACE reduction of Lemma 5.18 and using Corollary 6.2. The lower bound follows from Lemma 6.3. ❑

## 6.2 P- and coNP-hardness for Data Complexity

Let us now turn to the data complexity of instance checking for the *DL-Lite* logics with arbitrary number restrictions and role inclusions. As follows from the results of Ortiz et al. (2006) for $\mathcal{SHIQ}$, instance checking (and in fact query answering) for *DL-Lite*$_{bool}^{\mathcal{HN}}$ is in coNP for data complexity, while the results of Hustadt et al. (2005) and Eiter et al. (2008) for Horn-$\mathcal{SHIQ}$ imply a polynomial-time upper bound for *DL-Lite*$_{horn}^{\mathcal{HF}}$.

Here we show that these upper bounds are optimal in the following sense: on the one hand, instance checking in *DL-Lite*$_{core}^{\mathcal{HF}}$ is P-hard for data complexity; on the other hand, it becomes coNP-hard for both *DL-Lite*$_{krom}^{\mathcal{HF}}$ and *DL-Lite*$_{core}^{\mathcal{HN}}$ (that is, if we allow negated concept names or arbitrary number restrictions—in fact, $\geq 2\,R$ is enough). Note that the results of this section do not depend on whether we adopt the UNA or not.

**Theorem 6.5** *The instance checking (and query answering) problem for DL-Lite*$_{krom}^{\mathcal{HF}}$ *KBs is data-hard for* coNP *(with or without the UNA).*

**Proof** The proof is by reduction of the unsatisfiability problem for 2+2CNF, which is known to be coNP-complete (Schaerf, 1993). Given a 2+2CNF formula

$$\varphi \;=\; \bigwedge_{k=1}^{n}(a_{k,1} \vee a_{k,2} \vee \neg a_{k,3} \vee \neg a_{k,4}),$$

where each $a_{k,j}$ is one of the propositional variables $a_1, \dots, a_m$, we construct a KB $(\mathcal{T}, \mathcal{A}_\varphi)$ whose TBox $\mathcal{T}$ does not depend on $\varphi$. We will use the object names $f$, $c_k$, for $1 \leq k \leq n$, and $a_i$, for $1 \leq i \leq m$, role names $S$, $S_{\mathbf{f}}$ and $P_j$, $P_{j,\mathbf{t}}$, $P_{j,\mathbf{f}}$, for $1 \leq j \leq 4$, and concept names $A$ and $D$.

Define $\mathcal{A}_\varphi$ to be the set of the following assertions, for $1 \leq k \leq n$:

$$S(f, c_k), \quad P_1(c_k, a_{k,1}), \quad P_2(c_k, a_{k,2}), \quad P_3(c_k, a_{k,3}), \quad P_4(c_k, a_{k,4}),$$

and let $\mathcal{T}$ consist of the axioms

$$\begin{aligned}
\geq 2\,P_j &\sqsubseteq \bot, && \text{for}\,1 \leq j \leq 4, && (48)\\
P_{j,\mathbf{f}} \sqsubseteq P_j, \qquad P_{j,\mathbf{t}} &\sqsubseteq P_j, && \text{for } 1 \leq j \leq 4, && (49)\\
\neg \exists P_{j,\mathbf{t}} &\sqsubseteq \exists P_{j,\mathbf{f}}, && \text{for } 1 \leq j \leq 4, && (50)\\
\exists P_{j,\mathbf{f}}^- \sqsubseteq \neg A, \qquad \exists P_{j,\mathbf{t}}^- &\sqsubseteq A, && \text{for } 1 \leq j \leq 4, && (51)\\
\exists P_{1,\mathbf{f}}^- \sqcap \exists P_{2,\mathbf{f}}^- \sqcap \exists P_{3,\mathbf{t}}^- \sqcap \exists P_{4,\mathbf{t}}^- &\sqsubseteq \exists S_{\mathbf{f}}^-, &&&& (52)\\
\geq 2\,S^- &\sqsubseteq \bot, &&&& (53)\\
S_{\mathbf{f}} &\sqsubseteq S, &&&& (54)\\
\exists S_{\mathbf{f}} &\sqsubseteq D. &&&& (55)
\end{aligned}$$

Note that axiom (52) does not belong to *DL-Lite*$_{krom}^{\mathcal{HF}}$ because of the conjunctions in its left-hand side. However, it can be eliminated with the help of Lemma 5.9. So let us prove that $(\mathcal{T}, \mathcal{A}_\varphi) \models D(f)$ iff $\varphi$ is not satisfiable.

($\Leftarrow$) Suppose that $\varphi$ is not satisfiable and $\mathcal{I} \models (\mathcal{T}, \mathcal{A}_\varphi)$. Define an assignment $\mathfrak{a}$ of the truth values $\mathbf{t}$ and $\mathbf{f}$ to propositional variables by taking $\mathfrak{a}(a_i) = \mathbf{t}$ iff $a_i^{\mathcal{I}} \in A^{\mathcal{I}}$. As $\varphi$ is false under $\mathfrak{a}$, there is $k$, $1 \leq k \leq n$, such that $\mathfrak{a}(a_{k,1}) = \mathfrak{a}(a_{k,2}) = \mathbf{f}$ and $\mathfrak{a}(a_{k,3}) = \mathfrak{a}(a_{k,4}) = \mathbf{t}$. In view of (50), for each $j$, $1 \leq j \leq 4$, we have $c_k^{\mathcal{I}} \in (\exists P_{j,\mathbf{t}})^{\mathcal{I}} \cup (\exists P_{j,\mathbf{f}})^{\mathcal{I}}$, and by (49), $c_k^{\mathcal{I}} \in (\exists P_j)^{\mathcal{I}}$. Therefore, by (48) and (51), $c_k^{\mathcal{I}} \in (\exists P_{j,\mathbf{t}})^{\mathcal{I}}$ if $\mathfrak{a}(a_{k,j}) = \mathbf{t}$ and $c_k^{\mathcal{I}} \in (\exists P_{j,\mathbf{f}})^{\mathcal{I}}$ if $\mathfrak{a}(a_{k,j}) = \mathbf{f}$, and hence, by (52), $c_k^{\mathcal{I}} \in (\exists S_{\mathbf{f}}^{-})^{\mathcal{I}}$. Then by (53) and (54), we have $f^{\mathcal{I}} \in (\exists S_{\mathbf{f}})^{\mathcal{I}}$, from which, by (55), $f^{\mathcal{I}} \in D^{\mathcal{I}}$. It follows that $(\mathcal{T}, \mathcal{A}_\varphi) \models D(f)$.

($\Rightarrow$) Conversely, suppose that $\varphi$ is satisfiable. Then there is an assignment $\mathfrak{a}$ such that $\mathfrak{a}(a_{k,1}) = \mathbf{t}$ or $\mathfrak{a}(a_{k,2}) = \mathbf{t}$ or $\mathfrak{a}(a_{k,3}) = \mathbf{f}$ or $\mathfrak{a}(a_{k,4}) = \mathbf{f}$, for all $1 \leq k \leq n$. Define $\mathcal{I}$ by taking

- $\Delta^{\mathcal{I}} = \{x_i \mid 1 \leq i \leq m\} \cup \{y_k \mid 1 \leq k \leq n\} \cup \{z\}$,

- $a_i^{\mathcal{I}} = x_i$, for $1 \leq i \leq m$, $\quad c_k^{\mathcal{I}} = y_k$, for $1 \leq k \leq n$, $\quad f^{\mathcal{I}} = z$,

- $A^{\mathcal{I}} = \{x_i \mid \mathfrak{a}(a_i) = \mathbf{t}\} \cup \{y_k \mid 1 \leq k \leq n\} \cup \{z\}$,

- $P_{j,\mathbf{t}}^{\mathcal{I}} = \{(y_k, a_{k,j}^{\mathcal{I}}) \mid 1 \leq k \leq n,\ \mathfrak{a}(a_{k,j}) = \mathbf{t}\} \cup \{(x_i, x_i) \mid \mathfrak{a}(a_i) = \mathbf{t}\} \cup \{(z, z)\}$,

- $P_{j,\mathbf{f}}^{\mathcal{I}} = \{(y_k, a_{k,j}^{\mathcal{I}}) \mid 1 \leq k \leq n,\ \mathfrak{a}(a_{k,j}) = \mathbf{f}\} \cup \{(x_i, x_i) \mid \mathfrak{a}(a_i) = \mathbf{f}\}$,

- $P_j^{\mathcal{I}} = P_{j,\mathbf{t}}^{\mathcal{I}} \cup P_{j,\mathbf{f}}^{\mathcal{I}}$, for $1 \leq j \leq 4$,

- $S_{\mathbf{f}}^{\mathcal{I}} = \{(z, y_k) \mid \mathfrak{a}(a_{k,1} \vee a_{k,2} \vee \neg a_{k,3} \vee \neg a_{k,4}) = \mathbf{f}\} = \emptyset$,

- $S^{\mathcal{I}} = \{(z, y_k) \mid 1 \leq k \leq n\}$,

- $D^{\mathcal{I}} = \{z \mid \mathfrak{a}(\varphi) = \mathbf{f}\} = \emptyset$.

It is not hard to check that $\mathcal{I} \models (\mathcal{T}, \mathcal{A}_\varphi)$ and $\mathcal{I} \not\models D(f)$. $\qquad \Box$

**Theorem 6.6** *The instance checking (and the query answering) problem for DL-Lite$_{core}^{\mathcal{HN}}$ KBs is data-hard for* CONP *(with or without the UNA).*

**Proof** The proof is again by reduction of the unsatisfiability problem for 2+2CNF. The main difference from the previous one is that *DL-Lite*$_{core}^{\mathcal{HN}}$, unlike *DL-Lite*$_{krom}^{\mathcal{HF}}$, cannot express 'covering conditions' like (50). It turns out, however, that we can use number restrictions to represent constraints of this kind. Given a 2+2CNF formula $\varphi$, we take the same ABox $\mathcal{A}_\varphi$ constructed in the proof of Theorem 6.5. The ($\varphi$ independent) TBox $\mathcal{T}$, describing the meaning of any such representation of $\varphi$ in terms of $\mathcal{A}_\varphi$, is also defined in the same way as in that proof, except that the axiom (50) is now replaced by the following set of axioms:

$$T_{j,1} \sqsubseteq T_j, \qquad\qquad T_{j,2} \sqsubseteq T_j, \qquad\qquad T_{j,3} \sqsubseteq T_j, \qquad (56)$$

$$\geq 2\, T_j^{-} \sqsubseteq \bot, \qquad\qquad\qquad\qquad\qquad\qquad (57)$$

$$\exists P_j \sqsubseteq \exists T_{j,1}, \qquad\qquad \exists P_j \sqsubseteq \exists T_{j,2}, \qquad\qquad (58)$$

$$\exists T_{j,1}^{-} \sqcap \exists T_{j,2}^{-} \sqsubseteq \exists T_{j,3}^{-}, \qquad\qquad\qquad\qquad\qquad (59)$$

$$\geq 2\, T_j \sqsubseteq \exists P_{j,\mathbf{t}} \qquad\qquad \exists T_{j,3} \sqsubseteq \exists P_{j,\mathbf{f}}, \qquad (60)$$

where $T_j, T_{j,1}, T_{j,2}, T_{j,3}$ are fresh role names, for each $j$, $1 \leq j \leq 4$. Note that axioms (52) and (59) do not belong to *DL-Lite*$_{core}^{\mathcal{HN}}$ because of the conjunctions in their left-hand side, but we can easily eliminate them using Lemma 5.9. So it remains to prove that $(\mathcal{T}, \mathcal{A}_\varphi) \models D(f)$ iff $\varphi$ is not satisfiable.

($\Leftarrow$) Suppose that $\varphi$ is not satisfiable and $\mathcal{I} \models (\mathcal{T}, \mathcal{A}_\varphi)$. Define an assignment $\mathfrak{a}$ of the truth values $\mathbf{t}$ and $\mathbf{f}$ to propositional variables by taking $\mathfrak{a}(a_i) = \mathbf{t}$ iff $a_i^{\mathcal{I}} \in A^{\mathcal{I}}$. As $\varphi$ is false under $\mathfrak{a}$, there is $k$, $1 \le k \le n$, such that $\mathfrak{a}(a_{k,1}) = \mathfrak{a}(a_{k,2}) = \mathbf{f}$, $\mathfrak{a}(a_{k,3}) = \mathfrak{a}(a_{k,4}) = \mathbf{t}$. For each $j$, $1 \le j \le 4$, we have $c_k^{\mathcal{I}} \in (\exists P_j)^{\mathcal{I}}$; by (58), $c_k^{\mathcal{I}} \in (\exists T_{j,1})^{\mathcal{I}}, (\exists T_{j,2})^{\mathcal{I}}$. So there are $v_1, v_2$ such that $(c_k^{\mathcal{I}}, v_1) \in T_{j,1}^{\mathcal{I}}$ and $(c_k^{\mathcal{I}}, v_2) \in T_{j,2}^{\mathcal{I}}$. If $v_1 \ne v_2$ then $c_k^{\mathcal{I}} \in (\ge 2\, T_j)^{\mathcal{I}}$ and, by (60), $c_k^{\mathcal{I}} \in (P_{j,\mathbf{t}})^{\mathcal{I}}$. Otherwise, if $v_1 = v_2 = v$, we have $v \in (\exists T_{j,3}^-)^{\mathcal{I}}$ by (59), and so by (56) and (57), $c_k^{\mathcal{I}} \in (\exists T_{j,3})^{\mathcal{I}}$, from which, by (60), $c_k^{\mathcal{I}} \in (P_{j,\mathbf{t}})^{\mathcal{I}}$. Therefore, $c_k^{\mathcal{I}} \in (\exists P_{j,\mathbf{t}})^{\mathcal{I}} \cup (\exists P_{j,\mathbf{f}})^{\mathcal{I}}$, and by (49), $c_k^{\mathcal{I}} \in (\exists P_j)^{\mathcal{I}}$. Thus, by (48) and (51), $c_k^{\mathcal{I}} \in (\exists P_{j,\mathbf{t}})^{\mathcal{I}}$ if $\mathfrak{a}(a_{k,j}) = \mathbf{t}$ and $c_k^{\mathcal{I}} \in (\exists P_{j,\mathbf{f}})^{\mathcal{I}}$ if $\mathfrak{a}(a_{k,j}) = \mathbf{f}$, and hence, by (52), $c_k^{\mathcal{I}} \in (\exists S_{\mathbf{f}}^-)^{\mathcal{I}}$. Then by (53) and (54), we have $f^{\mathcal{I}} \in (\exists S_{\mathbf{f}})^{\mathcal{I}}$, from which, by (55), $f^{\mathcal{I}} \in D^{\mathcal{I}}$. It follows that $(\mathcal{T}, \mathcal{A}_\varphi) \models D(f)$.

($\Rightarrow$) Conversely, suppose that $\varphi$ is satisfiable. Then there is an assignment $\mathfrak{a}$ such that $\mathfrak{a}(a_{k,1}) = \mathbf{t}$ or $\mathfrak{a}(a_{k,2}) = \mathbf{t}$ or $\mathfrak{a}(a_{k,3}) = \mathbf{f}$ or $\mathfrak{a}(a_{k,4}) = \mathbf{f}$, for all $1 \le k \le n$. Define $\mathcal{I}$ by taking

- $\Delta^{\mathcal{I}} = \{x_i \mid 1 \le i \le m\} \cup \{y_k \mid 1 \le k \le n\} \cup \{u_{k,j,1}, u_{k,j,2} \mid 1 \le j \le 4, 1 \le k \le n\} \cup \{z\}$,

- $a_i^{\mathcal{I}} = x_i$, for $1 \le i \le m$, $\qquad c_k^{\mathcal{I}} = y_k$, for $1 \le k \le n$, $\qquad f^{\mathcal{I}} = z$,

- $A^{\mathcal{I}} = \{x_i \mid \mathfrak{a}(a_i) = \mathbf{t}\}$,

- $P_{j,\mathbf{t}}^{\mathcal{I}} = \big\{(y_k, a_{k,j}^{\mathcal{I}}) \mid 1 \le k \le n,\ \mathfrak{a}(a_{k,j}) = \mathbf{t}\big\}$, for $1 \le j \le 4$,

- $P_{j,\mathbf{f}}^{\mathcal{I}} = \big\{(y_k, a_{k,j}^{\mathcal{I}}) \mid 1 \le k \le n,\ \mathfrak{a}(a_{k,j}) = \mathbf{f}\big\}$, for $1 \le j \le 4$,

- $P_j^{\mathcal{I}} = P_{j,\mathbf{t}}^{\mathcal{I}} \cup P_{j,\mathbf{f}}^{\mathcal{I}}$, for $1 \le j \le 4$,

- $T_{j,1}^{\mathcal{I}} = \big\{(y_k, u_{k,j,1}) \mid 1 \le k \le n\big\}$, for $1 \le j \le 4$,

- $T_{j,2}^{\mathcal{I}} = \big\{(y_k, u_{k,j,2}) \mid 1 \le k \le n,\ \mathfrak{a}(a_{k,j}) = \mathbf{t}\big\} \cup$
  $\qquad\qquad\qquad \big\{(y_k, u_{k,j,1}) \mid 1 \le k \le n,\ \mathfrak{a}(a_{k,j}) = \mathbf{f}\big\}$, for $1 \le j \le 4$,

- $T_{j,3}^{\mathcal{I}} = \big\{(y_i, u_{k,j,1}) \mid 1 \le k \le n,\ \mathfrak{a}(a_{k,j}) = \mathbf{f}\big\}$, for $1 \le j \le 4$,

- $T_j^{\mathcal{I}} = T_{j,1}^{\mathcal{I}} \cup T_{j,2}^{\mathcal{I}}$,

- $S_{\mathbf{f}}^{\mathcal{I}} = \big\{(z, y_k) \mid \mathfrak{a}(a_{k,1} \lor a_{k,2} \lor \neg a_{k,3} \lor \neg a_{k,4}) = \mathbf{f}\big\} = \emptyset$,

- $S^{\mathcal{I}} = \big\{(z, y_k) \mid 1 \le k \le n\big\}$,

- $D^{\mathcal{I}} = \{z \mid \mathfrak{a}(\varphi) = \mathbf{f}\} = \emptyset$.

It is not hard to check that $\mathcal{I} \models (\mathcal{T}, \mathcal{A}_\varphi)$ and $\mathcal{I} \not\models D(f)$. $\qquad\qquad \square$

Our next lower bound would follow from Theorem 6, item 2 in the work of Calvanese et al. (2006); unfortunately, the proof there is incorrect and cannot be repaired.

**Theorem 6.7** *The instance checking (and query answering) problem for* DL-Lite$_{core}^{\mathcal{HF}}$ *KBs is data-hard for* P *(with or without the UNA).*

**Proof** The proof is by reduction of the entailment problem for Horn-CNF, which is known to be P-complete (see, e.g., Börger et al., 1997, Exercise 2.2.4). Given a Horn-CNF formula

$$\varphi \;=\; \bigwedge_{k=1}^{n} (\neg a_{k,1} \vee \neg a_{k,2} \vee a_{k,3}) \quad \wedge \quad \bigwedge_{l=1}^{p} a_{l,0},$$

where each $a_{k,j}$ and each $a_{l,0}$ is one of the propositional variables $a_1, \ldots, a_m$, we construct a KB $(\mathcal{T}, \mathcal{A}_\varphi)$ whose TBox $\mathcal{T}$ does not depend on $\varphi$. We will need the object names $c_1, \ldots, c_n$ and $v_{k,j,i}$, for $1 \leq k \leq n$, $1 \leq j \leq 3$, $1 \leq i \leq m$ (for each variable, we take one object name for each possible occurrence of this variable in each non-unit clause), role names $S$, $S_{\mathbf{t}}$ and $P_j$, $P_{j,\mathbf{t}}$, for $1 \leq j \leq 3$, and a concept name $A$.

Define $\mathcal{A}_\varphi$ to be the set containing the assertions:

$$S(v_{1,1,i}, v_{1,2,i}), S(v_{1,2,i}, v_{1,3,i}), S(v_{1,3,i}, v_{2,1,i}), S(v_{2,1,i}, v_{2,2,i}), S(v_{2,2,i}, v_{2,3,i}), \ldots$$
$$\ldots, S(v_{n,2,i}, v_{n,3,i}), S(v_{n,3,i}, v_{1,1,i}), \qquad \text{for } 1 \leq i \leq m,$$
$$P_j(v_{k,j,i}, c_k) \quad \text{iff} \quad a_{k,j} = a_i, \qquad \text{for } 1 \leq i \leq m, \quad 1 \leq k \leq n, \quad 1 \leq j \leq 3,$$
$$A(v_{1,1,i}) \quad \text{iff} \quad a_{l,0} = a_i, \qquad \text{for } 1 \leq i \leq m, \quad 1 \leq l \leq p$$

(all objects for each variable are organized in an $S$-cycle and $P_j(v_{k,j,i}, c_k) \in \mathcal{A}_\varphi$ iff the variable $a_i$ occurs in the $k$th non-unit clause of $\varphi$ in the $j$th position). Let $\mathcal{T}$ consist of the following concept and role inclusions:

$$S_{\mathbf{t}} \sqsubseteq S, \tag{61}$$
$$\geq 2\, S \sqsubseteq \bot, \tag{62}$$
$$A \sqsubseteq \exists S_{\mathbf{t}}, \tag{63}$$
$$\exists S_{\mathbf{t}}^{-} \sqsubseteq A, \tag{64}$$
$$\geq 2\, P_1 \sqsubseteq \bot \qquad\qquad \geq 2\, P_2 \sqsubseteq \bot, \tag{65}$$
$$P_{1,\mathbf{t}} \sqsubseteq P_1, \qquad\qquad P_{2,\mathbf{t}} \sqsubseteq P_2, \tag{66}$$
$$A \sqsubseteq \exists P_{1,\mathbf{t}}, \qquad\qquad A \sqsubseteq \exists P_{2,\mathbf{t}}, \tag{67}$$
$$\geq 2\, P_3^{-} \sqsubseteq \bot, \tag{68}$$
$$P_{3,\mathbf{t}} \sqsubseteq P_3, \tag{69}$$
$$\exists P_{1,\mathbf{t}}^{-} \sqcap \exists P_{2,\mathbf{t}}^{-} \sqsubseteq \exists P_{3,\mathbf{t}}^{-}, \tag{70}$$
$$\exists P_{3,\mathbf{t}} \sqsubseteq A. \tag{71}$$

As before, here we have an axiom, namely (70), that does not belong to $DL\text{-}Lite_{core}^{\mathcal{HF}}$ because of the conjunction in its left-hand side, but again it can be eliminated with the help of Lemma 5.9. Our aim is to show that $(\mathcal{T}, \mathcal{A}_\varphi) \models A(v_{1,1,i_0})$ iff $\varphi \models a_{i_0}$.

($\Leftarrow$) Suppose that $\varphi \models a_{i_0}$. Consider an arbitrary model $\mathcal{I}$ of $(\mathcal{T}, \mathcal{A}_\varphi)$ and define $\mathfrak{a}$ to be the assignment of the truth values $\mathbf{t}$ and $\mathbf{f}$ to propositional variables such that $\mathfrak{a}(a_i) = \mathbf{t}$ iff $v_{1,1,i}^{\mathcal{I}} \in A^{\mathcal{I}}$, for $1 \leq i \leq m$. By (61)–(64), for each $i$, $1 \leq i \leq m$, we have either $v_{k,j,i}^{\mathcal{I}} \in A^{\mathcal{I}}$, for *all* $k, j$ with $1 \leq k \leq n$, $1 \leq j \leq 3$, or $v_{k,j,i}^{\mathcal{I}} \notin A^{\mathcal{I}}$, for *all* $k, j$ with $1 \leq k \leq n$, $1 \leq j \leq 3$. Now, if we have $\mathfrak{a}(a_{k,1}) = \mathbf{t}$ and $\mathfrak{a}(a_{k,2}) = \mathbf{t}$, for $1 \leq k \leq n$ then, by (65)–(67), $c_k^{\mathcal{I}} \in (\exists P_{1,\mathbf{t}}^{-})^{\mathcal{I}}, (\exists P_{2,\mathbf{t}}^{-})^{\mathcal{I}}$. By (70), $c_k^{\mathcal{I}} \in (\exists P_{3,\mathbf{t}}^{-})^{\mathcal{I}}$ and hence, by (68) and (69), $v_{k,3,i}^{\mathcal{I}} \in (\exists P_{3,\mathbf{t}})^{\mathcal{I}},$ where $a_{k,3} = a_i$, which means, by (71), that $v_{k,3,i}^{\mathcal{I}} \in A^{\mathcal{I}}$, and so $v_{1,1,i}^{\mathcal{I}} \in A^{\mathcal{I}}$ and $\mathfrak{a}(a_i) = \mathbf{t}$. It follows that $\mathfrak{a}(\varphi) = \mathbf{t}$, and hence $\mathfrak{a}(a_{i_0}) = \mathbf{t}$, which, by definition, means that $v_{1,1,i_0}^{\mathcal{I}} \in A^{\mathcal{I}}$. So we can conclude that $(\mathcal{T}, \mathcal{A}_\varphi) \models A(v_{1,1,i_0})$.

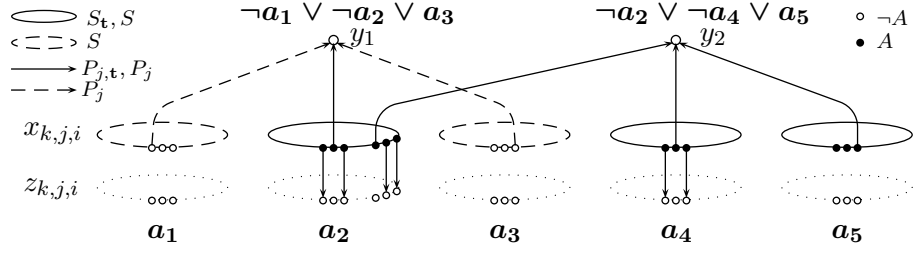

Figure 5: The model $\mathcal{I}$ satisfying $(\mathcal{T}, \mathcal{A}_\varphi)$, for $\varphi = (\neg a_1 \vee \neg a_2 \vee a_3) \wedge (\neg a_2 \vee \neg a_4 \vee a_5)$.

($\Rightarrow$) Conversely, suppose that $\varphi \not\models a_{i_0}$. Then there is an assignment $\mathfrak{a}$ with $\mathfrak{a}(\varphi) = \mathbf{t}$ and $\mathfrak{a}(a_{i_0}) = \mathbf{f}$. We construct a model $\mathcal{I}$ of $(\mathcal{T}, \mathcal{A}_\varphi)$ such that $\mathcal{I} \not\models A(v_{1,1,i_0})$. Define $\mathcal{I}$ by taking

- $\Delta^{\mathcal{I}} = \{x_{k,j,i}, z_{k,j,i} \mid 1 \leq k \leq n, 1 \leq j \leq 3, 1 \leq i \leq m\} \cup \{y_k \mid 1 \leq k \leq n\}$,

- $c_k^{\mathcal{I}} = y_k$, for $1 \leq k \leq n$,

- $v_{k,j,i}^{\mathcal{I}} = x_{k,j,i}$, for $1 \leq k \leq n$, $1 \leq j \leq 3$, $1 \leq i \leq m$,

- $A^{\mathcal{I}} = \{x_{k,j,i} \mid 1 \leq k \leq n, \ 1 \leq j \leq 3, \ \mathfrak{a}(a_i) = \mathbf{t}\}$,

- $S^{\mathcal{I}} = \bigcup\limits_{1 \leq i \leq m} S_i$, where $S_i = \{(x_{k,1,i}, x_{k,2,i}), (x_{k,2,i}, x_{k,3,i}), (x_{k,3,i}, x_{k\oplus 1,1,i}) \mid 1 \leq k \leq n\}$
  and $k \oplus 1 = k+1$ if $k < n$, and $k \oplus 1 = 1$ if $k = n$,

- $S_{\mathbf{t}}^{\mathcal{I}} = \bigcup\limits_{\substack{1 \leq i \leq m \\ \mathfrak{a}(a_i) = \mathbf{t}}} S_i$,

- $P_j^{\mathcal{I}} = \{(x_{k,j,i}, y_k) \mid 1 \leq k \leq n, \ a_i = a_{k,j}\} \cup$
  $\{(x_{k,j,i}, z_{k,j,i}) \mid 1 \leq k \leq n, \ a_i \neq a_{k,j}\}$, for $1 \leq j \leq 2$,

- $P_3^{\mathcal{I}} = \{(x_{k,3,i}, y_k) \mid 1 \leq k \leq n, \ a_i = a_{k,3}\}$,

- $P_{j,\mathbf{t}}^{\mathcal{I}} = \{(x_{k,j,i}, y_k) \mid 1 \leq k \leq n, \ a_i = a_{k,j}, \ \mathfrak{a}(a_i) = \mathbf{t}\} \cup$
  $\{(x_{k,j,i}, z_{k,j,i}) \mid 1 \leq k \leq n, \ a_i \neq a_{k,j}\}$, for $1 \leq j \leq 2$,

- $P_{3,\mathbf{t}}^{\mathcal{I}} = \{(x_{k,3,i}, y_k) \mid 1 \leq k \leq n, \ a_i = a_{k,3}, \ \mathfrak{a}(a_i) = \mathbf{t}\}$.

It is routine to check that we indeed have $\mathcal{I} \models (\mathcal{T}, \mathcal{A}_\varphi)$ and $\mathcal{I} \not\models A(v_{1,1,i_0})$. See Figure 5 for an example. □

# 7. Query Answering: Data Complexity

The positive existential query answering problem is known to be data-complete for coNP in the case of $DL\text{-}Lite_{bool}^{\mathcal{HN}}$: the upper bound follows from the results of Ortiz et al. (2006), while the lower bound was established for $DL\text{-}Lite_{krom}$ by Calvanese et al. (2006), Schaerf (1993). In the case of $DL\text{-}Lite_{horn}^{\mathcal{HF}}$, query answering is data-complete for P, as follows from the results of Hustadt et al. (2005) and Eiter et al. (2008) for Horn-$\mathcal{SHIQ}$, while for $DL\text{-}Lite_{horn}^{\mathcal{H}}$ it is in $\mathrm{AC}^0$ (Calvanese et al., 2006).

In fact, the coNP upper bound holds for the extension of $DL\text{-}Lite_{bool}^{\mathcal{HN}}$ with role disjointness and (a)symmetry constraints (this follows from Glimm et al., 2007, Theorem 10; cf. Remark 5.21). We conjecture that the same result holds for role (ir)reflexivity constraints.

Our main result in this section is the following:

**Theorem 7.1** *The positive existential query answering problem for the logics $DL\text{-}Lite_{horn}^{\mathcal{N}}$, $DL\text{-}Lite_{horn}^{\mathcal{H}}$ and $DL\text{-}Lite_{horn}^{(\mathcal{HN})}$ is in $\mathrm{AC}^0$ for data complexity.*

**Proof** Suppose that we are given a *consistent $DL\text{-}Lite_{horn}^{(\mathcal{HN})}$* KB $\mathcal{K}' = (\mathcal{T}', \mathcal{A}')$ (with all its concept and role names occurring in the TBox $\mathcal{T}'$) and a positive existential query in prenex form $\mathsf{q}(\vec{x}) = \exists \vec{y}\, \varphi(\vec{x}, \vec{y})$ in the signature of $\mathcal{K}'$. Consider the $DL\text{-}Lite_{horn}^{(\mathcal{HN})^-}$ KB $\mathcal{K} = (\mathcal{T}, \mathcal{A})$ provided by Lemma 5.17 (the language $DL\text{-}Lite_{horn}^{(\mathcal{HN})^-}$ is defined in Section 5.3).

**Lemma 7.2** *For every tuple $\vec{a}$ of object names in $\mathcal{K}'$, we have $\mathcal{K}' \models \mathsf{q}(\vec{a})$ iff $\mathcal{I} \models \mathsf{q}(\vec{a})$ for all untangled models $\mathcal{I}$ of $\mathcal{K}$.*

**Proof** ($\Rightarrow$) Suppose that $\mathcal{K}' \models \mathsf{q}(\vec{a})$ and $\mathcal{I}$ is an untangled model $\mathcal{I}$ of $\mathcal{K}$. By Lemma 5.17 and in view of consistency of $\mathcal{K}'$, which ensures that (44) holds, we then have $\mathcal{I} \models \mathcal{K}'$ and therefore, $\mathcal{I} \models \mathsf{q}(\vec{a})$.

($\Leftarrow$) Suppose $\mathcal{I}' \models \mathcal{K}'$. By Lemma 5.17, there is a model $\mathcal{I}$ of $\mathcal{K}$ with the same domain as $\mathcal{I}'$ that coincides with $\mathcal{I}'$ on all symbols in $\mathcal{K}'$. As $\mathcal{I} \models \mathsf{q}(\vec{a})$, we must then have $\mathcal{I}' \models \mathsf{q}(\vec{a})$, and so $\mathcal{K}' \models \mathsf{q}(\vec{a})$ as required. ❑

Next we show that, as $\mathcal{K}^{\ddagger \mathsf{e}}$ is a Horn sentence, it is enough to consider just one special model $\mathcal{I}_0$ of $\mathcal{K}$ in the formulation of Lemma 7.2. Let $\mathfrak{M}_0$ be the *minimal Herbrand model* of (the universal Horn sentence) $\mathcal{K}^{\ddagger \mathsf{e}}$. We remind the reader (for details consult, e.g., Apt, 1990; Rautenberg, 2006) that $\mathfrak{M}_0$ can be constructed by taking the intersection of all Herbrand models for $\mathcal{K}^{\ddagger \mathsf{e}}$, that is, of all models based on the domain that consists of constant symbols from $\mathcal{K}^{\ddagger \mathsf{e}}$—i.e., $\Lambda = ob(\mathcal{A}) \cup dr(\mathcal{T})$; cf. Remark 5.15. We then have the following

$$\mathfrak{M}_0 \models B^*[c] \qquad \text{iff} \qquad \mathcal{K}^{\ddagger \mathsf{e}} \models B^*(c), \qquad \text{for } B \in Bcon(\mathcal{T}) \text{ and } c \in \Lambda.$$

Let $\mathcal{I}_0$ be the untangled model of $\mathcal{K}$ induced by $\mathfrak{M}_0$. Denote the domain of $\mathcal{I}_0$ by $\Delta^{\mathcal{I}_0}$. Property **(copy)** of Remark 5.15 provides us with a function $cp \colon \Delta^{\mathcal{I}_0} \to \Lambda$.

There are two consequences of Lemma 5.14. First, we have

$$a_i^{\mathcal{I}_0} \in B^{\mathcal{I}_0} \quad \text{iff} \quad \mathcal{K} \models B(a_i), \qquad \text{for } B \in Bcon(\mathcal{T}) \text{ and } a_i \in ob(\mathcal{A}). \tag{72}$$

Second, for every $R \in role^{\pm}(\mathcal{T})$, if $R^{\mathcal{I}_0} \neq \emptyset$ then $R^{\mathcal{I}} \neq \emptyset$, for all models $\mathcal{I}$ of $\mathcal{K}$. Indeed, if $R^{\mathcal{I}_0} \neq \emptyset$ then $\mathfrak{M}_0 \models (\exists R)^*[dr]$. Therefore, $(\mathcal{T} \cup \{\exists R \sqsubseteq \bot\}, \mathcal{A})$ is not satisfiable, and thus $R^{\mathcal{I}} \neq \emptyset$, for all $\mathcal{I}$ with $\mathcal{I} \models \mathcal{K}$. Moreover, if $R^{\mathcal{I}_0} \neq \emptyset$ then

$$w \in B^{\mathcal{I}_0} \quad \text{iff} \quad \mathcal{K} \models \exists R \sqsubseteq B, \qquad \text{for } B \in Bcon(\mathcal{T}) \text{ and } w \in \Delta^{\mathcal{I}_0} \text{ with } cp(w) = dr. \quad (73)$$

**Lemma 7.3** *If $\mathcal{I}_0 \models \mathsf{q}(\vec{a})$ then $\mathcal{I} \models \mathsf{q}(\vec{a})$ for all untangled models $\mathcal{I}$ of $\mathcal{K}$.*

**Proof** Suppose $\mathcal{I} \models \mathcal{K}$. As $\mathsf{q}(\vec{a})$ is a positive existential sentence, it is enough to construct a homomorphism $h \colon \mathcal{I}_0 \to \mathcal{I}$. We remind the reader that, by **(forest)**, the domain $\Delta^{\mathcal{I}_0}$ of $\mathcal{I}_0$ is partitioned into disjoint trees $\mathfrak{T}_a$, for $a \in ob(\mathcal{A})$. Define the *depth* of a point $w \in \Delta^{\mathcal{I}_0}$ to be the length of the shortest path in the respective tree to its root. Denote by $W_m$ the set of points of depth $\leq m$; in particular, $W_0 = \{a^{\mathcal{I}_0} \mid a \in ob(\mathcal{A})\}$. We construct $h$ as the union of maps $h_m$, $m \geq 0$, where each $h_m$ is defined on $W_m$ and has the following properties: $h_{m+1}(w) = h_m(w)$, for all $w \in W_m$, and

(a$_m$) for every $w \in W_m$, if $w \in B^{\mathcal{I}_0}$ then $h_m(w) \in B^{\mathcal{I}}$, for each $B \in Bcon(\mathcal{T})$;

(b$_m$) for all $u, v \in W_m$, if $(u, v) \in R^{\mathcal{I}_0}$ then $(h_m(u), h_m(v)) \in R^{\mathcal{I}}$, for each $R \in role^{\pm}(\mathcal{T})$.

For the basis of induction, we set $h_0(a_i^{\mathcal{I}_0}) = a_i^{\mathcal{I}}$, for $a_i \in ob(\mathcal{A})$. Property (a$_0$) follows then from (72) and (b$_0$) from **(ABox)** of Remark 5.15.

For the induction step, suppose that $h_m$ has already been defined for $W_m$, $m \geq 0$. Set $h_{m+1}(w) = h_m(w)$ for all $w \in W_m$. Consider an arbitrary $v \in W_{m+1} \setminus W_m$. By **(forest)**, there is a unique $u \in W_m$ such that $(u, v) \in E_a$, for some $\mathfrak{T}_a$. Let $\ell_a(u, v) = S$. Then, by **(copy)**, $cp(v) = inv(ds)$. By **(role)**, $u \in (\exists S)^{\mathcal{I}_0}$ and, by (a$_m$), $h_m(u) \in (\exists S)^{\mathcal{I}}$, which means that there is $w \in \Delta^{\mathcal{I}}$ with $(h_m(u), w) \in S^{\mathcal{I}}$. Set $h_{m+1}(v) = w$. As $cp(v) = inv(ds)$ and $(\exists inv(S))^{\mathcal{I}_0} \neq \emptyset$, it follows from (73) that if $v \in B^{\mathcal{I}_0}$ then $w' \in B^{\mathcal{I}}$ whenever we have $w' \in (\exists inv(S))^{\mathcal{I}}$. As $w \in (\exists inv(S))^{\mathcal{I}}$, we obtain (a$_{m+1}$) for $v$. To show (b$_{m+1}$), we notice that, by **(role)**, we have $(w, v) \in R^{\mathcal{I}_0}$, for some $w \in W_{m+1}$, just in two cases: either $w \in W_{m+1} \setminus W_m$, and then $w = v$ and $Id \sqsubseteq_{\mathcal{T}}^* R$, or $w \in W_m$, and then $w = u$ and $S \sqsubseteq_{\mathcal{T}}^* R$. In the former case, $(h_{m+1}(v), h_{m+1}(v)) \in R^{\mathcal{I}}$ because $Id^{\mathcal{I}}$ is the identity relation by **(role)**. In the latter case, we have $(u, v) \in S^{\mathcal{I}_0}$; hence $(h_{m+1}(u), h_{m+1}(v)) \in S^{\mathcal{I}}$ and, as $S \sqsubseteq_{\mathcal{T}}^* R$, $(h_{m+1}(u), h_{m+1}(v)) \in R^{\mathcal{I}}$. □

Assume now that, in the query $\mathsf{q}(\vec{x}) = \exists \vec{y}\, \varphi(\vec{x}, \vec{y})$, we have $\vec{y} = y_1, \dots, y_k$, and $\varphi$ is a quantifier-free formula. Our next lemma shows that in this case to check whether $\mathcal{I}_0 \models \mathsf{q}(\vec{a})$ it suffices to consider only the points of depth $\leq m_0$ in $\Delta^{\mathcal{I}_0}$, for some $m_0$ that does not depend on $|\mathcal{A}|$.

**Lemma 7.4** *Let $m_0 = k + |role^{\pm}(\mathcal{T})|$. If $\mathcal{I}_0 \models \exists \vec{y}\, \varphi(\vec{a}, \vec{y})$ then there is an assignment $\mathfrak{a}_0$ in $W_{m_0}$ (i.e., $\mathfrak{a}_0(y_i) \in W_{m_0}$ for all i) such that $\mathcal{I}_0 \models^{\mathfrak{a}_0} \varphi(\vec{a}, \vec{y})$.*

**Proof** Suppose that $\mathcal{I}_0 \models^{\mathfrak{a}} \varphi(\vec{a}, \vec{y})$, for some assignment $\mathfrak{a}$ in $\Delta^{\mathcal{I}_0}$, and that there is $y_i$, $1 \leq i \leq k$, with $\mathfrak{a}(y_i) \notin W_{m_0}$. Let $Y$ be the minimal subset of $\vec{y}$ that contains $y_i$ and every $y$ such that either $P(y', y)$ or $P(y, y')$ is a subformula of $\varphi$, for some $y' \in Y$ and some role name $P$. Let $y_j \in Y$ be such that there is $m > |role^{\pm}(\mathcal{T})|$ with $\mathfrak{a}(y_j) \in W_m$ and $\mathfrak{a}(y) \notin W_{m-1}$ for all $y \in Y$ (for convenience, $W_{-1} = \emptyset$ as before). Clearly, such an $m$ exists: $\mathfrak{a}(y_i) \notin W_{m_0}$, $Y$ has at most $k$ variables and, by **(forest)**, relations $P^{\mathcal{I}_0}$ can connect a point in $W_n \setminus W_{n-1}$ only with a point in $W_{n+1} \setminus W_{n-2}$, for $n \geq 1$. Let $w = \mathfrak{a}(y_j)$ be a point in $\mathfrak{T}_a$. As $w \in W_m \setminus W_{m-1}$, we have $cp(w) = dr$, for some $R \in role^{\pm}(\mathcal{T})$. As there are at most $|role^{\pm}(\mathcal{T})|$ distinct labels in each labeled tree $\mathfrak{T}_a$ and in view of **(copy)**, for each point $u$ of depth $> |role^{\pm}(\mathcal{T})|$, there is a point $u'$ of depth $\leq |role^{\pm}(\mathcal{T})|$ in the same $\mathfrak{T}_a$ such that $cp(u) = cp(u')$; by **(iso)**, the trees generated by $u$ and $u'$ are isomorphic. So, there is an isomorphism $g$ from the labeled tree generated by $w$ (which contains all $\mathfrak{a}(y)$, for $y \in Y$) onto the labeled tree generated by some point of depth $\leq |role^{\pm}(\mathcal{T})|$ in $\mathfrak{T}_a$. Define a new assignment $\mathfrak{a}_Y$ by taking $\mathfrak{a}_Y(y) = g(\mathfrak{a}(y))$ if $y \in Y$ and $\mathfrak{a}_Y(y) = \mathfrak{a}(y)$ otherwise. By **(copy)**, **(concept)** and **(role)** we then have $\mathcal{I}_0 \models^{\mathfrak{a}_Y} \varphi(\vec{a}, \vec{y})$ and $\mathfrak{a}_Y(y) \in W_{m_0}$, for each $y \in Y$. If $\mathfrak{a}_Y(y_j) \notin W_{m_0}$ for some $j$, we repeat the described construction. After at most $k$ iterations we shall obtain an assignment $\mathfrak{a}_0$ required by the lemma. ❏

To complete the proof of Theorem 7.1, we encode the problem '$\mathcal{K} \models \mathsf{q}(\vec{a})$?' as a model checking problem for first-order formulas. In precisely the same way as in Section 6.1, we fix a signature that contains unary predicates $A$, $\overline{A}$, for each concept name $A$, and binary predicates $P$, $\overline{P}$, for each role name $P$, and then represent the ABox $\mathcal{A}$ of $\mathcal{K}$ as a first-order model $\mathfrak{A}_{\mathcal{A}}$ with domain $ob(\mathcal{A})$. Now we define a first-order formula $\varphi_{\mathcal{T},\mathsf{q}}(\vec{x})$ in the above signature such that (i) $\varphi_{\mathcal{T},\mathsf{q}}(\vec{x})$ depends on $\mathcal{T}$ and $\mathsf{q}$ but not on $\mathcal{A}$, and (ii) $\mathfrak{A}_{\mathcal{A}} \models \varphi_{\mathcal{T},\mathsf{q}}(\vec{a})$ iff $\mathcal{I}_0 \models \mathsf{q}(\vec{a})$.

We begin by defining formulas $\psi_B(x)$, for $B \in Bcon(\mathcal{T})$, that describe the types of the elements of $ob(\mathcal{A})$ in the model $\mathcal{I}_0$ in the following sense (see also (72)):

$$\mathfrak{A}_{\mathcal{A}} \models \psi_B[a_i] \quad \text{iff} \quad a_i^{\mathcal{I}_0} \in B^{\mathcal{I}_0}, \qquad \text{for } B \in Bcon(\mathcal{T}) \text{ and } a_i \in ob(\mathcal{A}). \qquad (74)$$

These formulas are defined as the 'fixed-points' of sequences $\psi_B^0(x), \psi_B^1(x), \dots$ of formulas with one free variable, where

$$\psi_B^0(x) = \begin{cases} A(x), & \text{if } B = A, \\ E_q R^{\mathcal{T}}(x), & \text{if } B = {\geq}\, q\, R, \end{cases}$$

$$\psi_B^i(x) = \psi_B^0(x) \quad \vee \bigvee_{B_1 \sqcap \cdots \sqcap B_k \sqsubseteq B \in \mathsf{ext}(\mathcal{T})} \left( \psi_{B_1}^{i-1}(x) \wedge \cdots \wedge \psi_{B_k}^{i-1}(x) \right), \quad \text{for } i \geq 1,$$

and $E_q R^{\mathcal{T}}(x)$ is given by (46). (As in Section 6.1, to simplify presentation we use $\mathsf{ext}(\mathcal{T})$ instead of $\mathcal{T}$.) It should be clear that if there is some $i$ such that, for all $B \in Bcon(\mathcal{T})$, $\psi_B^i(x) \equiv \psi_B^{i+1}(x)$ (i.e., every $\psi_B^i(x)$ is equivalent to $\psi_B^{i+1}(x)$ in first-order logic), then $\psi_B^i(x) \equiv \psi_B^j(x)$ for every $B \in Bcon(\mathcal{T})$ and $j \geq i$. So the minimum such $i$ does not exceed $N = |Bcon(\mathcal{T})|$, and we set $\psi_B(x) = \psi_B^N(x)$.

Next we introduce sentences $\theta_{B,dr}$, for $B \in Bcon(\mathcal{T})$ and $dr \in dr(\mathcal{T})$, that describe the types of elements in $dr(\mathcal{T})$ in the following sense (see also (73)):

$$\mathfrak{A}_{\mathcal{A}} \models \theta_{B,dr} \quad \text{iff} \quad w \in B^{\mathcal{I}_0}, \text{ for } B \in Bcon(\mathcal{T}) \text{ and each } w \in \Delta^{\mathcal{I}_0} \text{ with } cp(w) = dr. \qquad (75)$$

(By **(concept)**, this definition is correct.) These sentences are defined similarly to $\psi_B(x)$. Namely, for each $B \in Bcon(\mathcal{T})$ and each $dr \in dr(\mathcal{T})$, we inductively define a sequence

$\theta_{B,dr}^0, \theta_{B,dr}^1, \dots$ by taking

$$\theta_{B,dr}^0 = \rho_{B,dr}^0 \qquad \text{and} \qquad \theta_{B,dr}^i = \rho_{B,dr}^i \ \ \vee \bigvee_{B_1 \sqcap \cdots \sqcap B_k \sqsubseteq B \in \mathsf{ext}(\mathcal{T})} \left( \theta_{B_1,dr}^{i-1} \wedge \cdots \wedge \theta_{B_k,dr}^{i-1} \right), \ \text{for } i \geq 1,$$

where $\rho_{B,dr}^i = \bot$, for all $i \geq 0$, whenever $B \neq \exists R$ and

$$\rho_{\exists R,dr}^0 = \exists x \, \psi_{\exists inv(R)}(x) \qquad \text{and} \qquad \rho_{\exists R,dr}^i = \bigvee_{ds \, \in \, dr(\mathcal{T})} \theta_{\exists inv(R),ds}^{i-1}, \qquad \text{for } i \geq 1.$$

It should be clear that there is $i \leq |role^{\pm}(\mathcal{T})| \cdot N$ such that $\theta_{B,dr}^i \equiv \theta_{B,dr}^{i+1}$, for all $B \in Bcon(\mathcal{T})$ and $dr \in dr(\mathcal{T})$. So we set $\theta_{B,dr} = \theta_{B,dr}^{|role^{\pm}(\mathcal{T})| \cdot N}$.

Now we consider the directed graph $G_{\mathcal{T}} = (V_{\mathcal{T}}, E_{\mathcal{T}})$, where $V_{\mathcal{T}}$ is the set of all equivalence classes $[R]$, $[R] = \{R' \mid R \equiv_{\mathcal{T}}^* R'\}$, such that $\exists R$ is not empty in *some* model of $\mathcal{T}$, and $E_{\mathcal{T}}$ is the set of all pairs $([R_i], [R_j])$ such that

**(path)** $\mathcal{T} \models \exists inv(R_i) \sqsubseteq \geq q \, R_j$ and either $inv(R_i) \not\sqsubseteq_{\mathcal{T}}^* R_j$ or $q \geq 2$,

and $R_j$ has no proper sub-role satisfying **(path)**. We have $([R_i], [R_j]) \in E_{\mathcal{T}}$ iff, for any ABox $\mathcal{A}'$, whenever the minimal untangled model $\mathcal{I}_0$ of $(\mathcal{T}, \mathcal{A}')$ contains a copy $w$ of $inv(dr_i')$, for $R_i' \in [R_i]$, then $w$ is connected to a copy of $inv(dr_j')$, for $R_j' \in [R_j]$, by all relations $S$ with $R_j \sqsubseteq_{\mathcal{T}}^* S$.

Recall now that we are given a query $\mathsf{q}(\vec{x}) = \exists \vec{y} \, \varphi(\vec{x}, \vec{y})$, where $\varphi$ is a quantifier-free positive formula and $\vec{y} = y_1, \dots, y_k$. Let $\Sigma_{\mathcal{T}, m_0}$ be the set of all paths in the graph $G_{\mathcal{T}}$ of length $\leq m_0$. More precisely,

$$\Sigma_{\mathcal{T}, m_0} = \{\varepsilon\} \cup \big\{ ([R_1], [R_2], \dots, [R_n]) \mid 1 \leq n \leq m_0, ([R_j], [R_{j+1}]) \in E_{\mathcal{T}}, \ \text{for } 1 \leq j < n \big\}.$$

For $\sigma, \sigma' \in \Sigma_{\mathcal{T}, m_0}$ and a role $R \in role^{\pm}(\mathcal{T})$, we write $\sigma \xrightarrow{R} \sigma'$ if one of the following three conditions is satisfied: (i) $\sigma = \sigma'$ and $Id \sqsubseteq_{\mathcal{T}}^* R$, (ii) $\sigma.[S] = \sigma'$ or (iii) $\sigma = \sigma'.[inv(S)]$, for some role $S$ with $S \sqsubseteq_{\mathcal{T}}^* R$.

Let $\Sigma_{\mathcal{T}, m_0}^k$ be the set of all $k$-tuples of the form $\vec{\sigma} = (\sigma_1, \dots, \sigma_k)$, $\sigma_i \in \Sigma_{\mathcal{T}, m_0}$. Intuitively, when evaluating the query $\exists \vec{y} \varphi(\vec{x}, \vec{y})$ over $\mathcal{I}_0$, each bound, or non-distinguished, variable $y_i$ is mapped to a point $w$ in $W_{m_0}$. However, the first-order model $\mathfrak{A}_{\mathcal{A}}$ does not contain the points from $W_{m_0} \setminus W_0$, and to represent them, we use the following 'trick.' By **(forest)**, every point $w$ in $W_{m_0}$ is uniquely determined by the pair $(a, \sigma)$, where $a^{\mathcal{I}_0}$ is the root of the tree $\mathfrak{T}_a$ containing $w$, and $\sigma$ is the sequence of labels $\ell_a(u, v)$ on the path from $a^{\mathcal{I}_0}$ to $w$. It follows from the unraveling procedure and **(path)** that $\sigma \in \Sigma_{\mathcal{T}, m_0}$. So, in the formula $\varphi_{\mathcal{T}, \mathsf{q}}$ we are about to define we assume that the $y_i$ range over $W_0$ and represent the first component of the pairs $(a, \sigma)$, whereas the second component is encoded in the $i$th member of $\vec{\sigma}$ (these $y_i$ should not be confused with the $y_i$ in the original query $\mathsf{q}$, which range over all of $W_{m_0}$). In order to treat arbitrary terms $t$ occurring in $\varphi(\vec{x}, \vec{y})$ in a uniform way, we set $t^{\vec{\sigma}} = \varepsilon$, if $t = a \in ob(\mathcal{A})$ or $t = x_i$, and $t^{\vec{\sigma}} = \sigma_i$, if $t = y_i$ (the distinguished variables $x_i$ and the object names $a$ are mapped to $W_0$ and do not require the second component of the pairs).

Given an assignment $\mathfrak{a}_0$ in $W_{m_0}$ we denote by $split(\mathfrak{a}_0)$ the pair $(\mathfrak{a}, \vec{\sigma})$, where $\mathfrak{a}$ is an assignment in $\mathfrak{A}_{\mathcal{A}}$ and $\vec{\sigma} = (\sigma_1, \dots, \sigma_k) \in \Sigma_{\mathcal{T}, m_0}^k$ are such that

- for each distinguished variable $x_i$, $\mathfrak{a}(x_i) = a$ with $a^{\mathcal{I}_0} = \mathfrak{a}_0(x_i)$;

- for each bound variable $y_i$, $\mathfrak{a}(y_i) = a$ and $\sigma_i = ([R_1], \ldots, [R_n])$, $n \le m_0$, with $a^{\mathcal{I}_0}$ being the root of the tree containing $\mathfrak{a}_0(y_i)$ and $R_1, \ldots, R_n$ being the sequence of labels $\ell_a(u, v)$ on the path from $a^{\mathcal{I}_0}$ to $\mathfrak{a}_0(y_i)$.

Not every pair $(\mathfrak{a}, \vec{\sigma})$, however, corresponds to an assignment in $W_{m_0}$ because some paths in $\vec{\sigma}$ may not exist in our $\mathcal{I}_0$: $G_{\mathcal{T}}$ represents possible paths in *all* models for the fixed TBox $\mathcal{T}$ and varying ABox. As follows from the unraveling procedure, a point in $W_{m_0} \setminus W_0$ corresponds to $a \in ob(\mathcal{A})$ and $\sigma = ([R], \ldots) \in \Sigma_{\mathcal{T}, m_0}$ iff $a$ does not have enough $R$-witnesses in $\mathcal{A}$, i.e., iff $\mathfrak{A}_{\mathcal{A}} \models \neg\psi^0_{\ge q\,R}[a] \land \psi_{\ge q\,R}[a]$, for some $q \in Q^R_{\mathcal{T}}$. Thus, for every $(\mathfrak{a}, \vec{\sigma})$ with $\vec{\sigma} = (\sigma_1, \ldots, \sigma_k)$, there is an assignment $\mathfrak{a}_0$ in $W_{m_0}$ with $split(\mathfrak{a}_0) = (\mathfrak{a}, \vec{\sigma})$ iff $\mathfrak{A}_{\mathcal{A}} \models^{\mathfrak{a}} \eta^{\vec{\sigma}}(\vec{y})$, where

$$\eta^{(\sigma_1, \ldots, \sigma_k)}(y_1, \ldots, y_k) \;=\; \bigwedge_{\substack{1 \le i \le k \\ \sigma_i \ne \varepsilon}} \bigvee_{q \in Q^{R_i}_{\mathcal{T}}} \left( \neg\psi^0_{\ge q\,R_i}(y_i) \;\land\; \psi_{\ge q\,R_i}(y_i) \right)$$

and each $R_i$, for $1 \le i \le k$ with $\sigma_i \ne \varepsilon$, is such that $\sigma_i = ([R_i], \ldots)$.

We define now, for every $\vec{\sigma} \in \Sigma^k_{\mathcal{T}, m_0}$, concept name $A$ and role name $R$,

$$A^{\vec{\sigma}}(t) \;=\; \begin{cases} \psi_A(t), & \text{if } t^{\vec{\sigma}} = \varepsilon, \\ \theta_{A, inv(ds)}, & \text{if } t^{\vec{\sigma}} = \sigma'.[S], \text{ for some } \sigma' \in \Sigma_{\mathcal{T}, m_0}, \end{cases}$$

$$R^{\vec{\sigma}}(t_1, t_2) \;=\; \begin{cases} R^{\mathcal{T}}(t_1, t_2), & \text{if } t^{\vec{\sigma}}_1 = t^{\vec{\sigma}}_2 = \varepsilon, \\ (t_1 = t_2), & \text{if } t^{\vec{\sigma}}_1 \xrightarrow{R} t^{\vec{\sigma}}_2 \text{ and either } t^{\vec{\sigma}}_1 \ne \varepsilon \text{ or } t^{\vec{\sigma}}_2 \ne \varepsilon, \\ \bot, & \text{otherwise}, \end{cases}$$

where $R^{\mathcal{T}}(y_1, y_2)$ is given by (47). We claim that, for every assignment $\mathfrak{a}_0$ in $W_{m_0}$ and $(\mathfrak{a}, \sigma) = split(\mathfrak{a}_0)$,

$$\mathcal{I}_0 \models^{\mathfrak{a}_0} A(t) \quad \text{iff} \quad \mathfrak{A}_{\mathcal{A}} \models^{\mathfrak{a}} A^{\vec{\sigma}}(t), \qquad \text{for all concept names } A \text{ and terms } t, \tag{76}$$

$$\mathcal{I}_0 \models^{\mathfrak{a}_0} R(t_1, t_2) \quad \text{iff} \quad \mathfrak{A}_{\mathcal{A}} \models^{\mathfrak{a}} R^{\vec{\sigma}}(t_1, t_2), \qquad \text{for all roles } R \text{ and terms } t_1, t_2. \tag{77}$$

For $A(a)$, $A(x_i)$ or $A(y_i)$ with $\sigma_i = \varepsilon$ the claim follows from (74). For $A(y_i)$ with $\sigma_i = \sigma'.[S]$, by **(copy)**, we have $cp(\mathfrak{a}(y_i)) = inv(dr)$, for some $R \in [S]$; the claim then follows from (75). For $R(y_{i_1}, y_{i_2})$ with $\sigma_{i_1} = \sigma_{i_2} = \varepsilon$, the claim follows from **(ABox)**. Let us consider the case of $R(y_{i_1}, y_{i_2})$ with $\sigma_{i_2} \ne \varepsilon$: we have $\mathfrak{a}_0(y_{i_2}) \notin W_0$ and thus, by **(role)**, $\mathcal{I}_0 \models^{\mathfrak{a}_0} R(y_{i_1}, y_{i_2})$ iff

- $\mathfrak{a}_0(y_{i_1})$, $\mathfrak{a}_0(y_{i_2})$ are in the same tree $\mathfrak{T}_a$, for $a \in ob(\mathcal{A})$, i.e., $\mathfrak{A}_{\mathcal{A}} \models^{\mathfrak{a}} (y_{i_1} = y_{i_2})$,

- and either $(\mathfrak{a}_0(y_{i_1}), \mathfrak{a}_0(y_{i_2})) \in E_a$ and then $\ell_a(\mathfrak{a}_0(y_{i_1}), \mathfrak{a}_0(y_{i_2})) = S$ for some $S \sqsubseteq^*_{\mathcal{T}} R$, or $(\mathfrak{a}_0(y_{i_2}), \mathfrak{a}_0(y_{i_1})) \in E_a$ and then $\ell_a(\mathfrak{a}_0(y_{i_2}), \mathfrak{a}_0(y_{i_1})) = S$ for some $inv(S) \sqsubseteq^*_{\mathcal{T}} R$, or $\mathfrak{a}_0(y_{i_1}) = \mathfrak{a}_0(y_{i_2})$ and then $Id \sqsubseteq^*_{\mathcal{T}} R$, i.e., $\sigma_{i_1} \xrightarrow{R} \sigma_{i_2}$.

Other cases are similar and left to the reader.

Finally, let $\varphi^{\vec{\sigma}}(\vec{x}, \vec{y})$ be the result of attaching the superscript $\vec{\sigma}$ to each atom of $\varphi$ and

$$\varphi_{\mathcal{T}, \mathfrak{q}}(\vec{x}) \;=\; \exists \vec{y} \bigvee_{\vec{\sigma} \in \Sigma^k_{\mathcal{T}, m_0}} \left( \varphi^{\vec{\sigma}}(\vec{x}, \vec{y}) \;\land \eta^{\vec{\sigma}}(\vec{y}) \right).$$

As follows from (76)–(77), for every assignment $\mathfrak{a}_0$ in $W_{m_0}$, we have $\mathcal{I}_0 \models^{\mathfrak{a}_0} \varphi(\vec{x}, \vec{y})$ iff $\mathfrak{A}_\mathcal{A} \models^{\mathfrak{a}} \varphi^{\vec{\sigma}}(\vec{x}, \vec{y})$ for $(\mathfrak{a}, \sigma) = split(\mathfrak{a}_0)$. For the converse direction notice that, if $\mathfrak{A}_\mathcal{A} \models^{\mathfrak{a}} \eta^{\vec{\sigma}}(\vec{y})$ then there is an assignment $\mathfrak{a}_0$ in $W_{m_0}$ with $split(\mathfrak{a}_0) = (\mathfrak{a}, \vec{\sigma})$.

Clearly, $\mathfrak{A}_\mathcal{A} \models \varphi_{\mathcal{T},\mathsf{q}}(\vec{a})$ iff $\mathcal{I}_0 \models \mathsf{q}(\vec{a})$, for every tuple $\vec{a}$. We also note that, for every pair of tuples $\vec{a}$ and $\vec{b}$ of object names in $ob(\mathcal{A})$, $\varphi^{\vec{\sigma}}(\vec{a}, \vec{b})$ is a positive existential sentence with inequalities, and so is domain-independent.[10] It is also easily seen that, for each $\vec{b}$, $\eta^{\vec{\sigma}}(\vec{b})$ is domain-independent. It follows from the minimality of $\mathcal{I}_0$ that $\varphi_{\mathcal{T},\mathsf{q}}(\vec{a})$ is domain-independent, for each tuple $\vec{a}$ of object names in $ob(\mathcal{A})$.

Finally, note that the resulting query contains $\leq |role^\pm(\mathcal{T})|^{k \cdot (k + |role^\pm(\mathcal{T})|)}$ disjuncts. □

## 8. *DL-Lite* without the Unique Name Assumption

In this section, unless otherwise stated, we assume that the interpretations *do not* respect the UNA, that is, we may have $a_i^\mathcal{I} = a_j^\mathcal{I}$ for distinct object names $a_i$ and $a_j$. The consequence relation $\models^{\mathsf{noUNA}}$ refers to the class of such interpretations.

Description logics without the UNA are usually extended with additional equality and inequality constraints of the form:

$$a_i \approx a_j \qquad \text{and} \qquad a_i \not\approx a_j,$$

where $a_i, a_j$ are object names. Their semantics is quite obvious: we have $\mathcal{I} \models a_i \approx a_j$ iff $a_i^\mathcal{I} = a_j^\mathcal{I}$, and $\mathcal{I} \models a_i \not\approx a_j$ iff $a_i^\mathcal{I} \neq a_j^\mathcal{I}$. The equality and inequality constraints are supposed to belong to the ABox part of a knowledge base. It is to be noted, however, that reasoning with equalities is LogSpace-reducible to reasoning without them:

**Lemma 8.1** *For every KB $\mathcal{K} = (\mathcal{T}, \mathcal{A})$, one can construct in LogSpace in the size of $\mathcal{A}$ a KB $\mathcal{K}' = (\mathcal{T}, \mathcal{A}')$ without equality constraints such that $\mathcal{I} \models \mathcal{K}$ iff $\mathcal{I} \models \mathcal{K}'$, for every interpretation $\mathcal{I}$.*

**Proof** Let $G = (V, E)$ be the undirected graph with

$$V = ob(\mathcal{A}), \qquad E = \big\{ (a_i, a_j) \mid a_i \approx a_j \in \mathcal{A} \text{ or } a_j \approx a_i \in \mathcal{A} \big\}$$

and $[a_i]$ the set of all vertices of $G$ that are reachable from $a_i$. Define $\mathcal{A}'$ by removing all the equality constraints from $\mathcal{A}$ and replacing every $a_i$ with $a_j \in [a_i]$ with minimal $j$. Note that this minimal $j$ can be computed in LogSpace: just enumerate the object names $a_j$ with respect to the order of their indexes $j$ and check whether the current $a_j$ is reachable from $a_i$ in $G$. It remains to recall that reachability in undirected graphs is SLogSpace-complete and that SLogSpace = LogSpace (Reingold, 2008). □

As we mentioned in Section 5.3, the logics of the form $DL\text{-}Lite_\alpha^{\mathcal{H}}$ do not 'feel' whether we adopt the UNA or not. With this observation and Lemmas 5.17, 5.18 and 8.1 at hand, we obtain the following result as a consequence of Theorem 5.13:

---

10. A query $\mathsf{q}(\vec{x})$ is said to be domain-independent in case $\mathfrak{A}_\mathcal{A} \models^{\mathfrak{a}} \mathsf{q}(\vec{x})$ iff $\mathfrak{A} \models^{\mathfrak{a}} \mathsf{q}(\vec{x})$, for each $\mathfrak{A}$ such that the domain of $\mathfrak{A}$ contains $ob(\mathcal{A})$, and $A^\mathfrak{A} = A^{\mathfrak{A}_\mathcal{A}}$ and $P^\mathfrak{A} = P^{\mathfrak{A}_\mathcal{A}}$, for all concept and role names $A$ and $P$.

**Theorem 8.2** *With or without the UNA, for combined complexity,* (i) *satisfiability of DL-Lite$_{bool}^{\mathcal{H}}$ KBs is* NP-*complete;* (ii) *satisfiability of DL-Lite$_{horn}^{\mathcal{H}}$ KBs is* P-*complete; and* (iii) *satisfiability of DL-Lite$_{krom}^{\mathcal{H}}$ and DL-Lite$_{core}^{\mathcal{H}}$ KBs is* NLogSpace-*complete. These results hold even if the KBs contain role disjointness,* (a)*symmetry,* (ir)*reflexivity and transitivity constraints, equalities and inequalities.*

On the other hand, from Corollary 6.2 and Lemmas 5.17, 5.18 and 8.1 we can derive the following:

**Theorem 8.3** *Without the UNA, satisfiability and instance checking for DL-Lite$_{bool}^{\mathcal{H}}$ KBs are in* AC$^0$ *for data complexity. These problems are also in* AC$^0$ *if the KBs contain role disjointness,* (a)*symmetry and* (ir)*reflexivity constraints and inequalities. However, they are* LogSpace-*complete if the KBs may contain equalities, and* NLogSpace-*complete if role transitivity constraints are allowed.*

We also note that our complexity results (Corollary 5.12, Theorems 6.5, 6.6 and 6.7) for the logics *DL-Lite$_{\alpha}^{\mathcal{HF}}$* and *DL-Lite$_{\alpha}^{\mathcal{HN}}$* do not depend on the UNA.

In this section, we analyze the combined and data complexity of reasoning in the logics of the form *DL-Lite$_{\alpha}^{(\mathcal{HF})}$* and *DL-Lite$_{\alpha}^{(\mathcal{HN})}$* (as well as their fragments) without the UNA. The obtained and known results are summarized in Table 2 on page 17.

## 8.1 *DL-Lite$_{\alpha}^{(\mathcal{HN})}$*: Arbitrary Number Restrictions

The following theorem shows that the interaction between number restrictions and the possibility of identifying objects in the ABox results in a higher complexity.

**Theorem 8.4** *Without the UNA, satisfiability of DL-Lite$_{core}^{\mathcal{N}}$ KBs (even without equality and inequality constraints) is* NP-*hard for both combined and data complexity.*

**Proof** The proof is by reduction of the following variant of the 3SAT problem—called *monotone one-in-three 3SAT*—which is known to be NP-complete (Garey & Johnson, 1979): given a *positive* 3CNF formula

$$\varphi \;\; = \;\; \bigwedge_{k=1}^{n} \big(a_{k,1} \vee a_{k,2} \vee a_{k,3}\big),$$

where each $a_{k,j}$ is one of the propositional variables $a_1, \ldots, a_m$, decide whether there is an assignment for the variables $a_j$ such that *exactly one variable* is true in each of the clauses in $\varphi$. To encode this problem in the language of *DL-Lite$_{core}^{\mathcal{N}}$*, we need object names $a_i^k$, for $1 \leq k \leq n$, $1 \leq i \leq m$, and $c_k$ and $t^k$, for $1 \leq k \leq n$, role names $S$ and $P$, and concept names $A_1, A_2, A_3$. Let $\mathcal{A}_{\varphi}$ be the ABox containing the following assertions:

$$S(a_i^1, a_i^2), \;\; \ldots, \;\; S(a_i^{n-1}, a_i^n), \;\; S(a_i^n, a_i^1), \qquad \text{for } 1 \leq i \leq m,$$
$$S(t^1, t^2), \;\; \ldots, \;\; S(t^{n-1}, t^n), \;\; S(t^n, t^1),$$
$$P(c_k, t^k), \qquad \text{for } 1 \leq k \leq n,$$
$$P(c_k, a_{k,j}^k), \;\; A_j(a_{k,j}^k), \qquad \text{for } 1 \leq k \leq n, \;\; 1 \leq j \leq 3,$$

and let $\mathcal{T}$ be the TBox with the following axioms:

$$A_1 \sqsubseteq \neg A_2, \quad A_2 \sqsubseteq \neg A_3, \quad A_3 \sqsubseteq \neg A_1, \quad \geq 2\,S \ \sqsubseteq \ \bot, \quad \geq 4\,P \ \sqsubseteq \ \bot.$$

Clearly, $(\mathcal{T}, \mathcal{A}_\varphi)$ is a *DL-Lite*$^{\mathcal{N}}_{core}$ KB and $\mathcal{T}$ does not depend on $\varphi$ (so that we cover both combined and data complexity). We claim that the answer to the monotone one-in-three 3SAT problem is positive iff $(\mathcal{T}, \mathcal{A}_\varphi)$ is satisfiable without the UNA.

($\Leftarrow$) Suppose $\mathcal{I} \models (\mathcal{T}, \mathcal{A}_\varphi)$. Define an assignment $\mathfrak{a}$ of the truth values $\mathbf{f}$ and $\mathbf{t}$ to propositional variables by taking $\mathfrak{a}(a_i) = \mathbf{t}$ iff $(a_i^1)^{\mathcal{I}} = (t^1)^{\mathcal{I}}$. Our aim is to show that $\mathfrak{a}(a_{k,j}) = \mathbf{t}$ for exactly one $j \in \{1, 2, 3\}$, for each $k$, $1 \leq k \leq n$. For all $j \in \{1, 2, 3\}$, we have $(c_k^{\mathcal{I}}, (a_{k,j}^k)^{\mathcal{I}}) \in P^{\mathcal{I}}$. Moreover, $(a_{k,i}^k)^{\mathcal{I}} \neq (a_{k,j}^k)^{\mathcal{I}}$ for $i \neq j$. As $c_k^{\mathcal{I}} \in (\leq 3\,P)^{\mathcal{I}}$ and $(c_k^{\mathcal{I}}, (t^k)^{\mathcal{I}}) \in P^{\mathcal{I}}$, we then must have $(a_{k,j}^k)^{\mathcal{I}} = (t^k)^{\mathcal{I}}$ for some unique $j \in \{1, 2, 3\}$. It follows from functionality of $S$ that, for each $1 \leq k \leq n$, we have $(a_{k,j}^1)^{\mathcal{I}} = (t^1)^{\mathcal{I}}$ for exactly one $j \in \{1, 2, 3\}$.

($\Rightarrow$) Let $\mathfrak{a}$ be an assignment satisfying the monotone one-in-three 3SAT problem. Take some $a_{i_0}$ with $\mathfrak{a}(a_{i_0}) = \mathbf{t}$ (clearly, such an $i_0$ exists, for otherwise $\mathfrak{a}(\varphi) = \mathbf{f}$) and construct an interpretation $\mathcal{I} = (\Delta^{\mathcal{I}}, \cdot^{\mathcal{I}})$ by taking:

- $\Delta^{\mathcal{I}} = \{y_k, z^k \mid 1 \leq k \leq n\} \cup \{x_i^k \mid \mathfrak{a}(a_i) = \mathbf{f},\ 1 \leq i \leq m, 1 \leq k \leq n\}$,

- $c_k^{\mathcal{I}} = y_k$ and $(t^k)^{\mathcal{I}} = z^k$, for $1 \leq k \leq n$,

- $(a_i^k)^{\mathcal{I}} = \begin{cases} x_i^k, & \text{if } \mathfrak{a}(a_i) = \mathbf{f}, \\ z^k, & \text{if } \mathfrak{a}(a_i) = \mathbf{t}, \end{cases}$ for $1 \leq i \leq m,\ 1 \leq k \leq n$,

- $S^{\mathcal{I}} = \{((a_i^1)^{\mathcal{I}}, (a_i^2)^{\mathcal{I}}), \ldots, ((a_i^{n-1})^{\mathcal{I}}, (a_i^n)^{\mathcal{I}}), ((a_i^n)^{\mathcal{I}}, (a_i^1)^{\mathcal{I}}) \mid 1 \leq i \leq m\}$,

- $P^{\mathcal{I}} = \{(c_k^{\mathcal{I}}, (t^k)^{\mathcal{I}}), (c_k^{\mathcal{I}}, (a_{k,1}^k)^{\mathcal{I}}), (c_k^{\mathcal{I}}, (a_{k,2}^k)^{\mathcal{I}}), (c_k^{\mathcal{I}}, (a_{k,3}^k)^{\mathcal{I}}) \mid 1 \leq k \leq n\}$.

It is readily checked that $\mathcal{I} \models (\mathcal{T}, \mathcal{A}_\varphi)$. $\qquad\square$

In fact, the above lower bound is optimal:

**Theorem 8.5** *Without the UNA, satisfiability of DL-Lite*$^{\mathcal{N}}_\alpha$, *DL-Lite*$^{(\mathcal{HN})}_\alpha$ *and DL-Lite*$^{(\mathcal{HN})^+}_\alpha$ *KBs with equality and inequality constraints is* NP-*complete for both combined and data complexity and any* $\alpha \in \{core, krom, horn, bool\}$.

**Proof** The lower bound is immediate from Theorem 8.4, and the matching upper bound can be proved by the following non-deterministic algorithm. Given a *DL-Lite*$^{(\mathcal{HN})^+}_{bool}$ KB $\mathcal{K} = (\mathcal{T}, \mathcal{A})$, we

- guess an equivalence relation $\sim$ over $ob(\mathcal{A})$;

- select in each equivalence class $a_i/_\sim$ a representative, say $a_i$, and replace every occurrence of $a \in a_i/_\sim$ in $\mathcal{A}$ with $a_i$;

- fail if the equalities and inequalities are violated in the resulting ABox—i.e., if it contains $a_i \not\approx a_i$ or $a_i \approx a_j$, for $i \neq j$;

- otherwise, remove the equality and inequality constraints from the ABox and denote the result by $\mathcal{A}'$;

- use the NP satisfiability checking algorithm for $DL\text{-}Lite_{bool}^{(\mathcal{HN})^+}$ to decide whether the KB $\mathcal{K}' = (\mathcal{T}, \mathcal{A}')$ is consistent under the UNA.

Clearly, if the algorithm returns 'yes,' then $\mathcal{I}' \models \mathcal{K}'$, for some $\mathcal{I}'$ respecting the UNA, and we can construct a model $\mathcal{I}$ of $\mathcal{K}$ (not necessarily respecting the UNA) by extending $\mathcal{I}'$ with the following interpretation of object names: $a^{\mathcal{I}} = a_i^{\mathcal{I}'}$, whenever $a_i$ is the representative of $a/_\sim$ ($\mathcal{I}$ coincides with $\mathcal{I}'$ on all other symbols). Conversely, if $\mathcal{I} \models \mathcal{K}$ then we take the equivalence relation $\sim$ defined by $a_i \sim a_j$ iff $a_i^{\mathcal{I}} = a_j^{\mathcal{I}}$. Let $\mathcal{I}'$ be constructed from $\mathcal{I}$ by removing the interpretations of all object names that are not representatives of the equivalence classes for $\sim$. It follows that $\mathcal{I}'$ respects the UNA and $\mathcal{I}' \models \mathcal{K}'$, so the algorithm returns 'yes.' ❏

## 8.2 $DL\text{-}Lite_\alpha^{(\mathcal{HF})}$: Functionality Constraints

Let us consider now $DL\text{-}Lite_{bool}^{(\mathcal{HF})^+}$ and its fragments. The following lemma shows that for these logics reasoning without the UNA can be reduced in polynomial time in the size of the ABox to reasoning under the UNA.

**Lemma 8.6** *For every $DL\text{-}Lite_{bool}^{(\mathcal{HF})^+}$ KB $\mathcal{K} = (\mathcal{T}, \mathcal{A})$ with equality and inequality constraints, one can construct in polynomial time in $|\mathcal{A}|$ a $DL\text{-}Lite_{bool}^{(\mathcal{HF})^+}$ KB $\mathcal{K}' = (\mathcal{T}, \mathcal{A}')$ such that $\mathcal{A}'$ contains no equalities and inequalities and $\mathcal{K}$ is satisfiable without the UNA iff $\mathcal{K}'$ is satisfiable under the UNA.*

**Proof** In what follows by *identifying $a_j$ with $a_k$ in $\mathcal{A}$* we mean replacing each occurrence of $a_k$ in $\mathcal{A}$ with $a_j$. We construct $\mathcal{A}'$ by first identifying $a_j$ with $a_k$, for each $a_j \approx a_k \in \mathcal{A}$, and removing the equality from $\mathcal{A}$, and then exhaustively applying the following procedure to $\mathcal{A}$:

- if $\geq 2\,R \sqsubseteq \bot \in \mathcal{T}$ and $R(a_i, a_j), R(a_i, a_k) \in \mathsf{Cl}_{\mathcal{T}}^{\mathbf{e}}(\mathcal{A})$, for distinct $a_j$ and $a_k$, then identify $a_j$ with $a_k$ (recall that a functional $R$ cannot have transitive sub-roles and thus $\mathsf{Cl}_{\mathcal{T}}^{\mathbf{e}}(\mathcal{A})$ is enough).

If the resulting ABox contains $a_i \not\approx a_i$, for some $a_i$, then, clearly, $\mathcal{K}$ is not satisfiable, so we add $A(a_i)$ and $\neg A(a_i)$ to the ABox, for some concept name $A$. Finally, we remove all inequalities from the ABox and denote the result by $\mathcal{A}'$. It should be clear that $\mathcal{A}'$ is computed from $\mathcal{A}$ in polynomial time and that, without the UNA, $\mathcal{K}$ is satisfiable iff $\mathcal{K}'$ is satisfiable. So it suffices to show that $\mathcal{K}'$ is satisfiable without the UNA iff it is satisfiable under the UNA. The implication ($\Leftarrow$) is trivial.

($\Rightarrow$) Observe that the above procedure ensures that

$$q_{R,a}^{\mathbf{e}} \leq 1, \qquad \text{for each } R \text{ with } \geq 2\,S \sqsubseteq \bot \in \mathcal{T}, R \sqsubseteq_{\mathcal{T}}^* S \text{ and } a \in ob(\mathcal{A}')$$

(see page 37 for definitions). Let $\mathcal{K}''$ be the $DL\text{-}Lite_{bool}^{(\mathcal{HN})^-}$ KB provided by Lemma 5.17 for $\mathcal{K}'$. It follows from the above property and the proofs of Lemma 5.14 and Corollary 5.16 that if $\mathcal{K}''$ is satisfiable without the UNA then $(\mathcal{K}'')^{\ddagger e}$ is satisfied in a first-order model with some constants interpreted by the same domain element. As $(\mathcal{K}'')^{\ddagger e}$ is a universal first-order sentence containing no equality, it is satisfiable in a first-order model such that all constants are interpreted by distinct elements. It follows from the proofs of Lemma 5.14 and Corollary 5.16 that this first-order model can be unraveled into a model $\mathcal{J}$ for $\mathcal{K}''$ respecting the UNA. By Lemma 5.17, $\mathcal{J}$ is a model of $\mathcal{K}'$. ❏

The reduction above cannot be done better than in P, as shown by the next theorem:

**Theorem 8.7** *Without the UNA, satisfiability of DL-Lite$_{core}^{\mathcal{F}}$ KBs (even without equality and inequality constraints) is P-hard for both combined and data complexity.*

**Proof** The proof is by reduction of the entailment problem for Horn-CNF (as in the proof of Theorem 6.7). Let

$$\varphi \;=\; \bigwedge_{k=1}^{n} \big( a_{k,1} \wedge a_{k,2} \rightarrow a_{k,3} \big) \;\wedge\; \bigwedge_{l=1}^{p} a_{l,0}$$

be a Horn-CNF formula, where each $a_{k,j}$ and each $a_{l,0}$ is one of the propositional variables $a_1, \ldots, a_m$ and $a_{k,1}$, $a_{k,2}$, $a_{k,3}$ are all distinct, for each $k$, $1 \le k \le n$. To encode the P-complete problem '$\varphi \models a_i$?' in the language of *DL-Lite$_{core}^{\mathcal{F}}$* we need object names $t$, $a_i^k$, for $1 \le k \le n$, $1 \le i \le m$, and $f_k$ and $g_k$, for $1 \le k \le n$, and role names $P$, $Q$, $S$ and $T$. The ABox $\mathcal{A}$ contains the following assertions

$$S(a_i^1, a_i^2), \;\; \ldots, \;\; S(a_i^{n-1}, a_i^n), \;\; S(a_i^n, a_i^1), \qquad \text{for } 1 \le i \le m,$$
$$P(a_{k,1}^k, f_k), \;\; P(a_{k,2}^k, g_k), \;\; Q(g_k, a_{k,3}^k), \;\; Q(f_k, a_{k,1}^k), \qquad \text{for } 1 \le k \le n,$$
$$T(t, a_{l,0}^1), \qquad \text{for } 1 \le l \le p,$$

and the TBox $\mathcal{T}$ asserts that all of the roles are functional:

$$\ge 2\, P \sqsubseteq \bot, \qquad \ge 2\, Q \sqsubseteq \bot, \qquad \ge 2\, S \sqsubseteq \bot \qquad \text{and} \qquad \ge 2\, T \sqsubseteq \bot.$$

Clearly, $\mathcal{K} = (\mathcal{T}, \mathcal{A})$ is a *DL-Lite$_{core}^{\mathcal{F}}$* KB and $\mathcal{T}$ does not depend on $\varphi$. We claim that $\varphi \models a_j$ iff $(\mathcal{T}, \mathcal{A} \cup \{\neg T(t, a_j^1)\})$ is not satisfiable without the UNA. To show this, it suffices to prove that $\varphi \models a_j$ iff $\mathcal{K} \models^{\mathsf{noUNA}} T(t, a_j^1)$.

($\Rightarrow$) Suppose $\varphi \models a_j$. Then we can derive $a_j$ from $\varphi$ using the following inference rules:

- $\varphi \models a_{l,0}$ for each $l$, $1 \le l \le p$;

- if $\varphi \models a_{k,1}$ and $\varphi \models a_{k,2}$, for some $k$, $1 \le k \le n$, then $\varphi \models a_{k,3}$.

We show that $\mathcal{K} \models^{\mathsf{noUNA}} T(t, a_j^1)$ by induction on the length of the derivation of $a_j$ from $\varphi$. The basis of induction is trivial. So assume that $a_j = a_{k,3}$, $\varphi \models a_{k,1}$, $\varphi \models a_{k,2}$, for some $k$, $1 \le k \le n$, and that $\mathcal{K} \models^{\mathsf{noUNA}} T(t, a_{k,1}^1) \wedge T(t, a_{k,2}^1)$. Suppose also that $\mathcal{I} \models \mathcal{K}$. Since $T$ is functional, we have $(a_{k,1}^1)^{\mathcal{I}} = (a_{k,2}^1)^{\mathcal{I}}$. Since $S$ is functional, $(a_{k,1}^{k'})^{\mathcal{I}} = (a_{k,2}^{k'})^{\mathcal{I}}$, for all $k'$, $1 \le k' \le n$, and in particular, for $k' = k$. Then, since $P$ is functional, $f_k^{\mathcal{I}} = g_k^{\mathcal{I}}$, from which, by functionality of $Q$, $(a_{k,3}^k)^{\mathcal{I}} = (a_{k,1}^k)^{\mathcal{I}}$. Finally, since $S$ is functional, $(a_{k,3}^{k'})^{\mathcal{I}} = (a_{k,1}^{k'})^{\mathcal{I}}$, for all $k'$, $1 \leq k' \leq n$, and in particular, for $k' = 1$. Thus, $\mathcal{I} \models T(t, a_j^1)$ and therefore $\mathcal{K} \models^{\mathsf{noUNA}} T(t, a_j^1)$.

($\Leftarrow$) Suppose that $\varphi \not\models a_j$. Then there is an assignment $\mathfrak{a}$ such that $\mathfrak{a}(\varphi) = \mathbf{t}$ and $\mathfrak{a}(a_j) = \mathbf{f}$. Construct an interpretation $\mathcal{I}$ by taking

- $\Delta^{\mathcal{I}} = \{x_i^k \mid \mathfrak{a}(a_i) = \mathbf{f},\ 1 \leq k \leq n, 1 \leq i \leq m\} \cup \{z^k, u_k, v_k \mid 1 \leq k \leq n\} \cup \{w\}$,

- $(a_i^k)^{\mathcal{I}} = \begin{cases} x_i^k, & \text{if } \mathfrak{a}(a_i) = \mathbf{f}, \\ z^k, & \text{if } \mathfrak{a}(a_i) = \mathbf{t}, \end{cases}$  for $1 \leq k \leq n$ and $1 \leq i \leq m$,

- $t^{\mathcal{I}} = w$,  $T^{\mathcal{I}} = \{(w, z^1)\}$,

- $S^{\mathcal{I}} = \{((a_i^1)^{\mathcal{I}}, (a_i^2)^{\mathcal{I}}), \dots, ((a_i^{n-1})^{\mathcal{I}}, (a_i^n)^{\mathcal{I}}), ((a_i^n)^{\mathcal{I}}, (a_i^1)^{\mathcal{I}}) \mid 1 \leq i \leq m\}$,

- $f_k^{\mathcal{I}} = u_k$  and  $g_k^{\mathcal{I}} = \begin{cases} v_k, & \text{if } \mathfrak{a}(a_{k,2}) = \mathbf{f}, \\ u_k, & \text{if } \mathfrak{a}(a_{k,2}) = \mathbf{t}, \end{cases}$  for $1 \leq k \leq n$,

- $P^{\mathcal{I}} = \{((a_{k,1}^k)^{\mathcal{I}}, f_k^{\mathcal{I}}), ((a_{k,2}^k)^{\mathcal{I}}, g_k^{\mathcal{I}}) \mid 1 \leq k \leq n\}$,

- $Q^{\mathcal{I}} = \{(g_k^{\mathcal{I}}, (a_{k,3}^k)^{\mathcal{I}}), (f_k^{\mathcal{I}}, (a_{k,1}^k)^{\mathcal{I}}) \mid 1 \leq k \leq n\}$.

It is readily checked that $\mathcal{I} \models \mathcal{K}$ and $\mathcal{I} \not\models T(t, a_j^1)$, and so $\mathcal{K} \not\models^{\mathsf{noUNA}} T(t, a_j^1)$.  ❑

The above result strengthens the NLogSpace lower bound for instance checking in $DL\text{-}Lite_{core}^{\mathcal{F}}$ proved by Calvanese et al. (2008).

**Corollary 8.8**  *Without the UNA, satisfiability of $DL\text{-}Lite_\alpha^{\mathcal{F}}$, $DL\text{-}Lite_\alpha^{(\mathcal{HF})}$ and $DL\text{-}Lite_\alpha^{(\mathcal{HF})^+}$ KBs, $\alpha \in \{core, krom, horn\}$, with equalities and inequalities is P-complete for both combined and data complexity.*

*Without the UNA, satisfiability of $DL\text{-}Lite_{bool}^{\mathcal{F}}$, $DL\text{-}Lite_{bool}^{(\mathcal{HF})}$ and $DL\text{-}Lite_{bool}^{(\mathcal{HF})^+}$ KBs with equalities and inequalities is NP-complete for combined complexity and P-complete for data complexity.*

**Proof**  The upper bounds follow from Lemma 8.6 and the corresponding upper bounds for the UNA case. The NP lower bound for combined complexity is obvious and the polynomial lower bounds follow from Theorem 8.7.  ❑

## 8.3 Query Answering: Data Complexity

The P and coNP upper bounds for query answering without the UNA follow from the results for Horn-$\mathcal{SHIQ}$ (Hustadt et al., 2005; Eiter et al., 2008) and $\mathcal{SHIQ}$ (Ortiz et al., 2006, 2008; Glimm et al., 2007), respectively (see the discussion at the beginning of Section 7). We present here the following result:

**Theorem 8.9**  *Without the UNA, positive existential query answering for $DL\text{-}Lite_{horn}^{\mathcal{H}}$ KBs with role disjointness, (a)symmetry, (ir)reflexivity constraints and inequalities is in $AC^0$ for data complexity. This problem is LogSpace-complete if, additionally, equalities are allowed in the KBs.*

**Proof** The proof follows the lines of the proof of Theorem 7.1 and uses the observation that models without the UNA give no more answers than their untangled counterparts. More precisely, let KB $\mathcal{K}' = (\mathcal{T}', \mathcal{A}')$ be as above. Suppose that it is *consistent*. Let $\mathsf{q}(\vec{x})$ be a positive existential query in the signature of $\mathcal{K}'$. Given $\mathcal{K}'$, Lemma 5.17 provides us with a KB $\mathcal{K}$. It is easy to see that $\mathcal{K}$ is a *DL-Lite*$_{horn}^{\mathcal{H}}$ KB extended with inequality constraints. The following is an analogue of Lemma 7.2, which also allows us to get rid of those inequalities:

**Lemma 8.10** *For every tuple $\vec{a}$ of object names in $\mathcal{K}'$, we have $\mathcal{K}' \models^{\mathsf{noUNA}} \mathsf{q}(\vec{a})$ iff $\mathcal{I} \models \mathsf{q}(\vec{a})$ for all untangled models $\mathcal{I}$ of $\mathcal{K}$ (respecting the UNA).*

**Proof** ($\Rightarrow$) Suppose that $\mathcal{K}' \models^{\mathsf{noUNA}} \mathsf{q}(\vec{a})$ and $\mathcal{I}$ is an untangled model of $\mathcal{K}$. As $\mathcal{I}$ respects the UNA, by Lemma 5.17 and in view of satisfiability of $\mathcal{K}'$, which ensures that (44) holds, we then have $\mathcal{I} \models \mathcal{K}'$ and therefore, $\mathcal{I} \models \mathsf{q}(\vec{a})$.

($\Leftarrow$) Suppose $\mathcal{I}' \models \mathcal{K}'$. We construct an interpretation $\mathcal{J}'$ respecting the UNA as follows. Let $\Delta^{\mathcal{J}'}$ be the disjoint union of $\Delta^{\mathcal{I}'}$ and $ob(\mathcal{A})$. Define a function $h\colon \Delta^{\mathcal{J}'} \to \Delta^{\mathcal{I}'}$ by taking $h(a) = a^{\mathcal{I}'}$, for each $a \in ob(\mathcal{A})$, and $h(w) = w$, for each $w \in \Delta^{\mathcal{I}'}$, and let

$$a^{\mathcal{J}'} = a, \qquad A^{\mathcal{J}'} = \left\{ u \mid h(u) \in A^{\mathcal{I}'} \right\} \qquad \text{and} \qquad P^{\mathcal{J}'} = \left\{ (u, v) \mid (h(x), h(v)) \in P^{\mathcal{I}'} \right\},$$

for each object, concept and role name $a$, $A$, $P$. Clearly, $\mathcal{J}'$ respects the UNA and $\mathcal{J}' \models \mathcal{K}'$. It also follows that $h$ is a homomorphism.

By Lemma 5.17, there is a model $\mathcal{I}$ of $\mathcal{K}$ with the same domain as $\mathcal{J}'$ that coincides with $\mathcal{J}'$ on all symbols in $\mathcal{K}'$. As $\mathcal{I} \models \mathsf{q}(\vec{a})$, we must then have $\mathcal{J}' \models \mathsf{q}(\vec{a})$, and since $h$ is a homomorphism, $\mathcal{I}' \models \mathsf{q}(\vec{a})$. Therefore, $\mathcal{K}' \models^{\mathsf{noUNA}} \mathsf{q}(\vec{a})$ as required. ❑

The remaining part of the proof is exactly as in Theorem 7.1 (since now we may assume that $\mathcal{K}$ is a *DL-Lite*$_{horn}^{\mathcal{H}}$ KB containing no inequality constraints).

LogSpace-completeness for the case with equalities follows from Lemma 8.1. ❑

## 9. Conclusion

In this article, we investigated the boundaries of the 'extended *DL-Lite* family' of description logics by providing a thorough and comprehensive understanding of the interaction between various *DL-Lite* constructs and their impact on the computational complexity of reasoning. We studied 40 different logics, classified according to five mutually orthogonal features: (1) the presence or absence of role inclusion assertions, (2) the form of the allowed concept inclusion assertions, distinguishing four main logical groups called *core*, *Krom*, *Horn*, and *Bool*, (3) the form of the allowed numeric constraints, ranging from none, to global functionality constraints only, and to arbitrary number restrictions, (4) the presence or absence of the unique name assumption (and equalities and inequalities between object names, if this assumption is dropped), and (5) the presence or absence of standard role constraints such as role disjointness, role symmetry, asymmetry, reflexivity, irreflexivity and transitivity. For all of the resulting logics, we studied the combined and data complexity of KB satisfiability and instance checking, as well as the data complexity of answering positive existential queries.

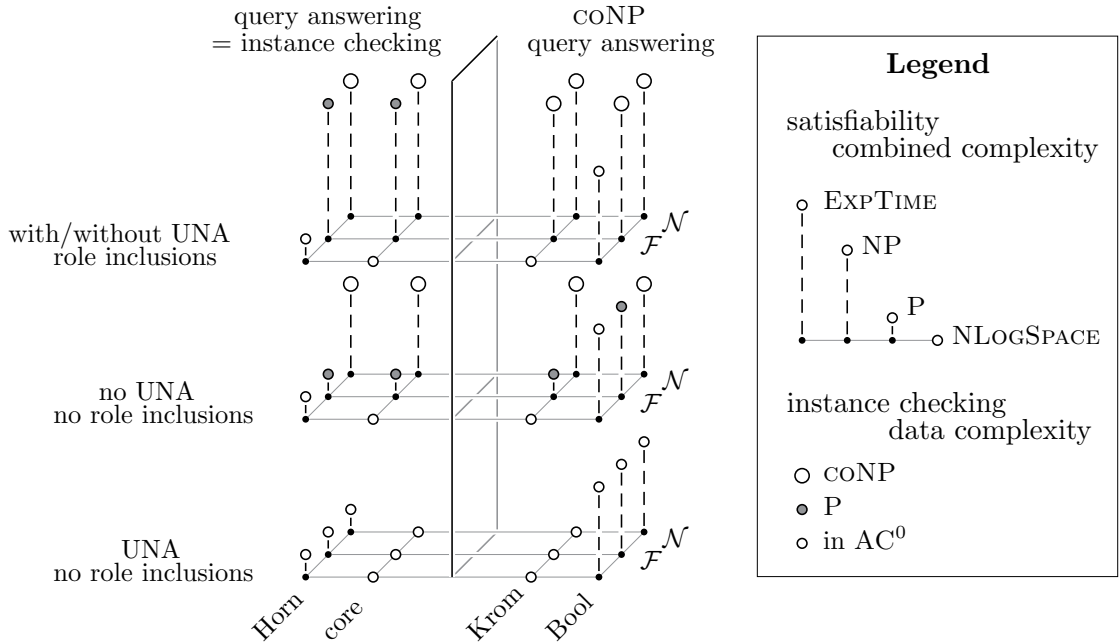

Figure 6: Complexity of basic *DL-Lite* logics.

The obtained tight complexity results are illustrated in Figure 6, where the combined complexity of satisfiability is represented by the height of vertical dashed lines, while the data complexity of instance checking by the size and color of the circle on top of these lines (recall that satisfiability and instance checking are reducible to the complement of each other). The data complexity of query answering for the core and Horn logics, shown on the left-hand side of the separating vertical plane, coincides with the data complexity of instance checking; for the Krom and Bool logics, shown on the right-hand side of the plane, query answering is always data-complete for CoNP. The upper layer shows the complexity of logics with role inclusions, in which case it does not depend on whether we adopt the UNA or not. The middle and the lower layers deal with the logics without role inclusions when the UNA is dropped or adopted, respectively. In each of these layers, the twelve languages are arranged in the 4 × 3 grid: one axis shows the type of concepts inclusions allowed (Horn, core, Krom, Bool), while the other the type of number restrictions (none, global functionality $\mathcal{F}$ or arbitrary $\mathcal{N}$). Some observations are in order:

- Under the UNA but without role inclusions, number restrictions do not increase the complexity of reasoning, which depends only on the form of concept inclusions allowed.

- On the other hand, without any form of number restrictions, the logics can have role inclusions and are insensitive to the UNA; again, the complexity is determined by the shape of concept inclusions only.

- In either of the above cases, instance checking is in $\text{AC}^0$ for data complexity, which means that the problems are first-order rewritable.

- Without UNA adopted and without either disjunctions or role inclusions, functionality leads to P-completeness of instance checking for data complexity, which suggests its reducibility to Datalog.

- For data complexity, there is no difference between the *core* and *Horn* logics, and between the *Krom* and *Bool* ones, which means that the core and Krom logics can be extended with conjunctions on the left-hand side of concept inclusions 'for free.'

Finally, for the logics $DL\text{-}Lite_\alpha^{(\mathcal{HF})}$ and $DL\text{-}Lite_\alpha^{(\mathcal{HN})}$ with both (qualified) number restrictions and role inclusions, whose interaction is restricted by conditions $(\mathbf{A_1})$–$(\mathbf{A_3})$, the complexity of reasoning always coincides with the complexity of the fragments $DL\text{-}Lite_\alpha^{\mathcal{F}}$ and, respectively, $DL\text{-}Lite_\alpha^{\mathcal{N}}$ without role inclusions, no matter whether we adopt the UNA or not.

Role disjointness, symmetry and asymmetry constraints can be added to any of the languages without changing their complexity. In fact, the $DL\text{-}Lite_\alpha^{(\mathcal{HN})}$ and $DL\text{-}Lite_\alpha^{(\mathcal{HF})}$ logics contain all of the above types of constraints together with role reflexivity and irreflexivity. We conjecture that (ir)reflexivity constraints can be added to all other logics without affecting their complexity. However, if we extend any $DL\text{-}Lite$ logic with role transitivity constraints, then the combined complexity of satisfiability remains the same, while instance checking and query answering become data-hard for NLogSpace. And the addition of equality between object names—which only makes sense if the UNA is dropped—leads to an increase from membership in $AC^0$ to LogSpace-completeness for data complexity; all other results remain unchanged.

The list of DL constructs considered in this paper is far from being complete. For example, it would be of interest to analyze the impact of nominals, role chains and Boolean operators on roles on the computational behavior of the $DL\text{-}Lite$ logics. Another interesting and practically important problem is to investigate in depth the interaction between various constructs with the aim of pushing restrictions like $(\mathbf{A_1})$–$(\mathbf{A_3})$ as far as possible.

One of the main ideas behind the $DL\text{-}Lite$ logics was to provide efficient access to large amounts of data through a high-level conceptual interface. This is supposed to be achieved by representing the high-level view of the information managed by the system as a $DL\text{-}Lite$ TBox $\mathcal{T}$, the data stored in a relational database as an ABox $\mathcal{A}$, and then rewriting positive existential queries to the knowledge base $(\mathcal{T}, \mathcal{A})$ as standard first-order queries to the database represented by $\mathcal{A}$. Such an approach is believed to be viable because, for a number of $DL\text{-}Lite$ logics, the query answering problem is in $AC^0$ for data complexity; cf. Theorems 7.1, 8.9 and Figure 6. The first-order rewriting technique has been implemented in various system, notably in QuOnto (Acciarri et al., 2005; Poggi et al., 2008b), which can query, relying on ontology-to-relational mappings, data stored in any standard relational database management system, and in Owlgres (Stocker & Smith, 2008), which can access an ABox stored in a Postgres database (though, to the best of our knowledge, the latter implementation is incomplete for conjunctive query answering). It is to be noted, however, that the size of the rewritten query can be substantially larger than the size of the original query, which can cause problems even for a very efficient database query engine.

For a positive existential query $\mathsf{q}$ and TBox $\mathcal{T}$, there are two major sources of high complexity of the first-order formula $\varphi_{\mathcal{T}, \mathsf{q}}$ in the proof of Theorem 7.1: (i) the formulas $\psi_B(x)$ computing whether an ABox object is an instance of a concept $B$ (and the formulas $\theta_{R,dr}$ computing whether objects with outgoing $R$-arrows are instances of $B$), and (ii) the disjunction over the paths $\vec{\sigma}$ in the graph $G_{\mathcal{T}}$. In the case of $DL\text{-}Lite_{core}^{(\mathcal{HN})}$, the size of $\psi_B(x)$ is linear in $|\mathcal{T}|$, while for $DL\text{-}Lite_{horn}^{(\mathcal{HN})}$ it can become exponential (however, various optimizations are possible). The size of the disjunction in (ii) is exponential in the number of non-distinguished variables in q. One way of removing source (i) would be to extend the given database (ABox) $\mathcal{A}$ by precomputing the Horn closure of the ABox with respect to the TBox and storing the resulting data in a supplementary database. This approach is advocated by Lutz et al. (2008) for querying databases via the description logic $\mathcal{EL}$. It could also be promising for the Horn fragments of expressive description logics such as $\mathcal{SHIQ}$ (Hustadt et al., 2005; Hustadt, Motik, & Sattler, 2007)—containing $DL\text{-}Lite_{horn}^{\mathcal{HF}}$ as a sub-language—for which the data complexity of instance checking (Hustadt et al., 2005, 2007) and conjunctive query answering is polynomial (Eiter et al., 2008). The disadvantage of using a supplementary database is the necessity to update it every time the ABox is changed. It would be interesting to investigate this alternative approach for $DL\text{-}Lite$ logics and compare it with the approach described above. Another important problem is to characterize those queries for which the disjunction in (ii) can be represented by a formula of polynomial size.

As the unique name assumption is replaced in OWL by the constructs `sameAs` and `differentFrom` (i.e., $\approx$ and $\not\approx$), a challenging problem is to investigate possible ways of dealing with equality (inequality does not require any special treatment as shown in the proof of Lemma 8.10). Although reasoning with equality is LogSpace-reducible to reasoning without it (cf. Lemma 8.1), we lose the property of first-order rewritability, and computing equivalence classes under $\approx$ may be too costly for real-world applications.

$DL\text{-}Lite$ logics are among those few examples of DLs for which usually very complex 'non-standard' reasoning problems—such as checking whether one ontology is a conservative extension of another one with respect to a given signature $\Sigma$ (Kontchakov et al., 2008), computing minimal modules of ontologies with respect to $\Sigma$ (Kontchakov et al., 2009) or uniform interpolants (Wang, Wang, Topor, & Pan, 2008)—can be supported by practical reasoning tools. However, only first steps have been made in this direction, and more research is needed in order to include these reasoning problems and tools into the standard OWL toolkit. It would be also interesting to investigate the unification problem for $DL\text{-}Lite$ logics (Baader & Narendran, 2001).

Finally, there exist certain parallels between the Horn logics of the $DL\text{-}Lite$ family, $\mathcal{EL}$, Horn-$\mathcal{SHIQ}$ and the first-order language of tuple and equality generating dependencies, TGDs and EGDs, used in the theory of databases (see, e.g., Gottlob & Nash, 2008). Further investigations of the relationships between these logics may lead to a deeper understanding of the role description logics can play in the database framework.

## Acknowledgments

This research has been partially supported by FET project TONES (Thinking ONtologiES), funded within the EU 6th Framework Programme under contract FP6-7603, and by the large-scale integrating project (IP) OntoRule (ONTOlogies meet Business RULEs ONtologiES), funded by the EC under ICT Call 3 FP7-ICT-2008-3, contract number FP7-231875. We thank the referees for their constructive criticism, comments, and suggestions.